\documentclass[intlimits,twoside,a4paper]{article}

\usepackage[cp1251]{inputenc}
\usepackage{multirow}
\usepackage[eqsecnum]{cmpj3}

\usepackage{bm}

\newcommand{\bea}{\begin{eqnarray}}
\newcommand{\eea}{\end{eqnarray}}
\newcommand{\ba}{\begin{array}}
\newcommand{\ea}{\end{array}}
\newcommand{\edc}{\end{document}}
\newcommand{\bc}{\begin{center}}
\newcommand{\ec}{\end{center}}
\newcommand{\be}{\begin{equation}}
\newcommand{\ee}{\end{equation}}

\newtheorem{thm}{Theorem}[section]

\newtheorem{defin}[thm]{Definition}

\newtheorem{rem}{Remark}[section]

\articletype{Review}

\issue{2022}{25}{3}{32501}
\doinumber{10.5488/CMP.25.32501}

\title[Phase diagrams of Ising and Potts]
{Phase diagrams of lattice models on Cayley tree and chandelier network: a review}

\author[H. Ak\i n]{H. Ak\i n\refaddr{label1}\refaddr{label2}\thanks{Corresponding author: \email{hakin@ictp.it; akinhasan25@gmail.com}.}}
\addresses{
	\addr{label1}The Abdus Salam International Centre for Theoretical Physics
	(ICTP), Strada Costiera, 11, I - 34151 Trieste, Italy
\addr{label2} Department of Mathematics,
Faculty of Arts and Sciences, Harran University, 63290, \c{S}anl\i urfa,
Turkey
}


\date{Received May 11, 2022, in final form July 27, 2022}

\begin{document}
	
	\maketitle
% ----------------------------------------------------------------
\begin{abstract}
The main purpose of this review paper is to give systematically
all the known results on phase diagrams corresponding to lattice
models (Ising and Potts) on Cayley tree (or Bethe lattice) and
chandelier networks. A detailed survey of  various modelling
applications of lattice models is reported. By using Vannimenus's
approach, the recursive equations of Ising and Potts models
associated to a given Hamiltonian on the Cayley tree are presented
and analyzed. The corresponding phase diagrams with programming
codes in  different programming languages are plotted. To
detect the phase transitions in the modulated phase, we
investigate in detail the actual variation of the wave-vector $q$
with temperature and the Lyapunov exponent associated with the
trajectory of our current recursive system. We determine the
transition between commensurate ($C$) and incommensurate ($I$) phases
by means of the Lyapunov exponents, wave-vector, and strange
attractor for a comprehensive comparison. We survey the dynamical
behavior of the Ising model on the chandelier network. We examine
the phase diagrams of the Ising model corresponding to a given
Hamiltonian on a new type of ``Cayley-tree-like lattice'', such as
{\it triangular, rectangular, pentagonal chandelier networks
(lattices)}. Moreover, several open problems are discussed.

%\textbf{PACS}: 05.70.Fh;  05.70.Ce; 75.10.Hk.\\
\printkeywords
\end{abstract}

\section{Introduction}
In order to understand the existing laws in nature, we need to
learn the dynamic behavior of atoms and molecules contained in the
objects that we can see and be born with. In this framework, we must
determine the behavior of many atoms and molecules at microscopic
levels. This issue is one of the current research topics of
statistical mechanics. To fully understand a well-organized system
with a high level of structure, it becomes imperative to create a
mathematical framework. With the help of this mathematical
framework, we investigate large clusters of microscopic entities
by means of statistical methods and probability theory~\cite{Baxter}. 

In statistical physics, one of the interesting examples of lattice
models with competing interactions is the Axial
Next-Nearest-Neighbor Ising model (or the ANNNI model) on Bethe
lattices. Vannimenus~\cite{Vannimenus} studied the Ising model
with competing nearest-neighbor and next-nearest-neighbor
interactions on Cayley tree and showed that the phase diagram
contains a modulated phase as obtained for similar models on
periodic lattices $(d>1)$. Furthermore, he determined that the
wave-vector has a ``devil's staircase'' behavior (see
\cite{MTA1985a,Yokoi,Aubry1982} for details). This model was first
examined on the order two Cayley tree and then on the Cayley tree of the order
$k$ ($k>2$) \cite{UGAT2012IJMPC}. Much of the work so far has
been done for Hamiltonians with interaction nearest neighbors
\cite{BleherG,CMCS,Silca-C,Marques}. Investigation of phase
diagrams of different lattice models has been the focus of
attention of many researchers
\cite{Peierls,Pirogov-Sinai1975,Pirogov-Sinai1976}.

In the present review paper, we study the structure of phase
diagrams by means of the limit behavior of nonlinear dynamical
systems corresponding to certain latice models. We derive the
recursive equations associated to Potts models, which is a
generalization of the Ising model, on the Cayley tree. The
recursive equations for certain Ising and Potts models are derived
\cite{GMP2008,GTA2005}. Then, iteration equations are obtained with
the help of derived equations. We analyse the dynamical
systems by considering them to certain initial values in the iteration
equations $({{x}^{(n)}},y_{1}^{(n)},y_{2}^{(n)},y_{3}^{(n)})\in
{{\mathbb{R}}^{4}}$. We study transitions between the phases by
linearizing the nonlinear equations around a given fixed point of
the transition curves in the phase diagrams (see
\cite{GU2011a,GTAU2011a}). The transition lines between the phases
is an open problem for the Potts models.

One of the most important issues of statistical mechanics is the
phase transition phenomenon in the equilibrium system
\cite{Minlos,Monroe1992,Simon,Sinai,Berker,Yang-Lee,Yasar}.
Lattice models are cartoons created to uncover different
aspects of the basic statistical mechanics, especially when the
phase transition phenomenon and spontaneous symmetry are broken
\cite{Yokoi,Monroe2003}. The simplest of all these models is the
Ising model, originally introduced as a ferromagnetic model. This
model has many applications in the fields such as physics, machine
learning, chemistry, biology, medicine, computer science, and even in
sociology \cite{LLS,Schelling}. Ernst studied the one-dimensional
model proposed by his adviser in 1925 and is today called the
Ising model \cite{Georgi,Inawashiro,Inawashiro-T1983}. He
suggested that there is no phase transition in one dimension
lattice $\mathbb{Z}$ or even in any dimension. Later, Ising's
suggestion was proven wrong. Onsager exactly solved the phase
transition problem in two dimensions in 1944 \cite{Onsager1944}
(see \cite{Wu} for details). Many studies have demonstrated that
the phase transition occurs
\cite{Dobrushin,G2002,GAT2007,GR2009b,GAUT2011a,GAUT2011b,Monroe1994,Tragtenberg}.
Since 1925, the Ising model has become one of the best-studied
models of statistical mechanics. The Ising model and its coherent
invariance in criticality had long been conjectured by physicists.

In 1981, Vannimenus  \cite{Vannimenus} obtained a modulated phase
diagram for the Ising model with nearest and prolonged nearest
interactions. Vannimenus proved that the multi-critical Lifshitz
point is formed at zero temperature. From that time on, many
researchers
\cite{GMP2008,GU2011a,GTAU2011a,Monroe1992,GAUT2011b,GTA2009a}
have been interested in the phase diagrams and Lifshitz points for
Ising and Potts models with the nearest and next
nearest-interaction on the Cayley tree. For more information on
the Potts model, see Wu's work \cite{Wu}
(\cite{G2002,GR2000,G1990,GR2006,G2004}). Inawashiro et al.
\cite{Inawashiro,Inawashiro-T1983} determined an intermediate
range in addition to paramagnetic, ferromagnetic, and
antiferromagnetic phases by considering the Ising spins with the
nearest neighbor and next-nearest neighbor interactions on a
Cayley tree.

Mariz et al. \cite{MTA1985a} generalized these results by adding
the external magnetic field to the one-level next nearest
neighbor. In \cite{AGUT2012ACTA}, we investigated new phase
diagrams corresponding to the Ising model with competing ternary
and binary interactions on a Cayley tree of arbitrary order. We
then computed the Lyapunov exponents and described the modulated
phases associated to the Ising model with competing interactions
on a Cayley tree of arbitrary order \cite{UGAT2012IJMPC}.

The Potts model was originally proposed by Domb \cite{Potts1952}
(see \cite{GAT2007}). In previous years, the Potts model did not
attract much interest compared to the Ising model. However, in
recent years, the Potts model has been very versatile in terms of
mathematics and physics, because it has a rich structure. Potts
model has many applications in chemistry \cite{Berker,Elliott},
biology \cite{GG1992,Merks-Glazier,OGR2003}, sociology
\cite{LLS,Schelling} and computer science~\cite{CMCS,BKYZ,Won},
statistical mechanics \cite{ZinatiCodello2017,Akin-Ulusoy-2022}.

In order to determine whether the Potts model with competing
interactions has a phase transition or not, Ganikhodjaev et al.
\cite{GTA2005,GR2009b,GMM2006,GR2009a,GR2005a,GR2005b} considered
the approach of using recursive equations for partition functions.
In 2008, Ganikhodjaev et al. \cite{GMP2008} obtained the phase
diagrams of the competitive Potts model with nearest neighbor and
next nearest neighbor interactions in a Cayley tree of the order two.
They showed that the diagram consists of six phases:
ferromagnetic, paramagnetic, modulated, antiphase and
paramodulated, all meeting at the multicritical point
$(T=0,p=1/3)$. %Also, they have reported on a new phase which we
%denote as paramodulated, found at low temperatures and
%characterized by zero average magnetization lying inside the
%modulated phase. Such a phase, inherent in the Potts model has no
%analogues in the Ising setting.
In \cite{TGUA2013ACTA}, we studied the phase diagrams associated
to the Potts model with competing nearest-neighbor, prolonged
next-nearest neighbor and two level triple neighbor interactions
on a Cayley tree of 3rd-order. Later in their work, by means
of the Visual Basic computer programming language, Ganikhodjaev et
al. \cite{GTA2009a} plotted the phase diagrams associated to the
competing Potts model with nearest-neighbor interactions,
prolonged next-nearest-neighbor interactions and one-level
next-nearest-neighbor interactions on a Cayley tree.% Also, they
%have proved that as soon as the same-level interaction is nonzero,
%the paramagnetic phase found at high temperatures for $J_o=0$
%disappears, while Ising model does not obtain such property (see
%\cite{Vannimenus,Inawashiro,Inawashiro-T1983,MTA1985a,GMP2008}).
%At vanishing temperature, we have plotted the phase diagram for
%all values and signs of $J_1$, $J_p$ and $J_o$.
%We have seen that several interesting properties were exhibited
%for typical values of $J_o/J_1$ and at finite temperatures.

In our recent papers, we  generalized the results obtained in
\cite{Vannimenus,MTA1985a,GMP2008,Inawashiro,Inawashiro-T1983} to
Potts and Ising models with competing interactions on the Cayley
tree of arbitrary order. However, the exact results are obtained
for Ising systems. Since it is difficult and complicated to derive
the iterative equations for Potts models on a Cayley tree of
arbitrary order, only the phase diagrams for certain $k$ values
are examined~\cite{UGAT2012IJMPC,AGUT2012ACTA,UGTA2010AIP,GU2011PhysicaA,GATU2013JPhysConfSer}.

%Within the last decades, types and number of Gibbs measures
%associated to the Ising and the Potts models on Cayley tree have
%been extensively studied by many reseachers
%\cite{Ro2005,Ro2008,Ro2006,RR2008,RR2009,RS2006,AkinT2011CMP} and
%\cite{GR2000,G1990,GR2006,G2004,GMM2006,GR2009a,GR2005a,GR2005b,GAUT2011a,GAUT2011b,G1990a,Won}.
%In \cite{ART2011JSP}, in addition to the established Gibbs
%measures on order-$k_0$ Cayley tree, we have determined a class of
%new Gibbs measures associated to the Ising model on a Cayley tree
%of higher order $k>k_0$. Ganikhodjaev and Rozikov \cite{GR2009a}
%have shown that the Ising model on the two-order Cayley tree has
%phase transitions for some parameters, as well as a full
%determination of periodic and non-periodic Gibbs measures for four
%mutually interacting (outer space, nearest neighbor, double and
%triple neighborhoods)
%\cite{Georgi,Kindermann,Preston,Ro2005,Ro2008,Ro2006,RR2008,RR2009,RS2006}.
%%One of the objectives of this paper was the exact determination
%%of the periodic and non-periodic Gibbs measures of the Potts and
%%Ising models associated with different Hamiltonians.
%Recently, many authors have proved the existence of phase
%transition taking int account the recurrence equations associated
%with the Ising-Vanniminus model on two and three order Cayley
%trees
%\cite{GTA2009a,GAUT2011b,ART2011JSP,Akin2016,Akin2017,Utkir-GM2013,RAU2014}.
%Ak\i n \cite{Akin2017-arXiv} has analytically studied the
%recurrence equations corresponding to an Ising model with three
%competing interactions on a order three Cayley tree.

Nonlinear iterative equations related to the rich world of
dynamical systems is one of the main concerns of statistical
physics on the Cayley tree. It is also an up-to-date issue that
has been extensively studied \cite{CMCS,GAT2007,GR2009a}. We
observe the behavior of the phase diagrams by using numerical
calculations for these nonlinear equations. We will review the
references in this area as precisely or completely as possible.

The paper is organized as follows.

In section \ref{PRELIMINARIES}, the definitions and preliminaries
are given.

In section \ref{Dynamical behavior of Ising model on CT}, we comprehensively study
the phase diagrams of Ising models on the Cayley tree with
competing interactions  (see
\cite{Vannimenus,MTA1985a,Inawashiro,Inawashiro-T1983} and
references therein). Ising model corresponding to the Hamiltonian
with different binary and ternary neighborhood interactions on a
Cayley tree of 2{nd} order is surveyed. The results of the
works obtained by Vannimenus \cite{Vannimenus}, Mariz et al.
\cite{MTA1985a} and Inawashiro et al.
\cite{Inawashiro,Inawashiro-T1983} are generalized. The phase
diagrams of corresponding nonlinear dynamical systems are plotted
by creating new computer codes. We examine the variation of the
wave vector $q$ versus desired temperatures in the modulated
phases and the Lyapunov exponent corresponding to the trajectory
of the given iterative systems in all their
aspects. %We investigate the variation of wave-vector analysis  and
%the behaviors of the Lyapunov exponent corresponding the given
%Hamiltonian.
We study  the strange attractors in detail
\cite{GATU2013JPhysConfSer}. We present a few typical figures of
strange attractors for some finite temperatures. Several
interesting properties of the phases are studied for typical
values of the parameters. We numerically show  that the
paramodulated and the para-ferro transitions are continuous. By
linearizing the given system around the fixed point, we obtain the
transition lines (see \cite{Vannimenus} for details).

In section~\ref{Dynamical behavior of Potts model}, the nonlinear
dynamical systems of the Potts models corresponding to the given
Hamiltonians on a Cayley tree of 2{nd} order are analyzed by
means of the computer programs and some relevant phase diagrams
are plotted. For the Potts models on the Cayley tree, the
transition lines are obtained from stability conditions and
special points in the phase diagrams corresponding to the Potts
models are analyzed by numerical iterations. In addition, the wave
vectors versus temperature for some critical points in the
modulated phases are plotted. %The Potts models associated to the
%Hamiltonians are generalized to the Cayley tree of arbitrary
%order.
The phase diagrams of the generalized nonlinear dynamical systems
are plotted. Open problems about phase diagrams of Potts models
are presented.

%%In Section \ref{Cayley-like lattice}, we have investigated the
%%dynamical behavior of Ising model on Chandelier networks (Cayley-like
%%lattice). We have studied the phase diagrams for the Ising model
%%on a Cayley tree-like lattice, a new lattice type called {\it
%%triangular, Rectangular, pentagonal Chandeliers}, with competing
%%nearest-neighbor interactions $J_1$, prolonged
%%next-nearest-neighbor interactions $J_p$ and one-level
%%next-nearest-neighbor quinary interactions $J^{(5)}_{l_1}$.

In section \ref{Cayley-like lattice}, we introduce chandelier
networks such as {\it triangular, rectangular, pentagonal
chandelier networks}. We study the dynamical behavior of the given
Ising models on chandelier networks. We deal with the phase
diagrams for the Ising model on a Cayley tree-like lattice, a new
type of chandelier lattice, with nearest neighbor interactions
$J_1$, prolonged next nearest neighbor interactions $J_p$, and
one-step quinary nearest neighbor interactions $J^{(5)}_{l_1}$. We
show that the phase diagrams associated to Ising model on
chandelier networks contain some multicritical Lifshitz points
that are at nonzero temperature, as well as many modulated new
phases \cite{UGAT2012IJMPC}. At vanishing temperature, the phase
diagram is completely determined for all values and signs of the
parameters $J_1,J_p$ and $J^{(5)}_{l_1}$. In the modulated phase,
the variation of the wave vector versus temperature is also
analyzed for some given critical points. We explain the
similarities and differences between chandelier networks and
Cayley tree. We present some open problems about the Ising models
defined on the chandelier networks.

In section \ref{CONCLUSIONS-RECOMMENDATIONS}, the results obtained
in the present review article are summarized. We compare the
obtained results, especially phase diagrams of Ising and Potts
models.

In section \ref{Openproblems}, we focus on the open problems that
arise in the development of the theory of Ising and Potts models
on Cayley trees and chandelier networks.

Note that, for the sake of completeness, we  give the system
of equations and some figures obtained in the articles we
published before, if needed.

\section{\textbf{Preliminaries}}\label{PRELIMINARIES}

\subsection{Cayley tree}
A Cayley tree of the order $k$ ($k >1$) is a weave pattern in which
$(k + 1)$ edges from each vertice point extend infinitely as shown
in figure~\ref{cayley-tree-k=3} ($k=3$). A Cayley tree (or Bethe
lattice \cite{Ostilli}) is a graph that is connected and contains
no circuits (see \cite{Utkir-GM2013,Rozikov-Kitap} for details).
Let $\Gamma^k = (V,L)$ be a semi-infinite Cayley tree of the order
$k\geqslant 1$ with the root $x^0$ (whose each vertex has exactly $k+1$
edges, except for the root $x^0$, which has $k$ edges). Here, $V$
is the set of vertices and $L$ is the set of edges. The vertices
$x$ and $y$ are called {\it nearest neighbors} and they are
denoted by $\ell=\langle{x,y}\rangle$ if there exists only one
edge connecting them. A collection of the pairs
$\langle{x,x_1}\rangle,\dots,\langle{x_{d-1},y}\rangle$ is called
a {\it path} from the point $x$ to the point $y$. The distance
$d(x,y), x,y\in V$, on the Cayley tree, is the length (the number
of edges) of the
shortest path connecting $x$ with $y$.\\
The sphere of radius $n$ on $V$ is represented by
$$
W_n=\{x\in V: d(x,x^{(0)})=n \}
$$
and the ball of radius $n$ is denoted by
$$
V_n=\{x\in V: d(x,x^{(0)})\leqslant n \}.
$$
For any $x\in W_n$, the set of direct successors of the vertex $x$
is defined by
$$
S(x)=\{y\in W_{n+1}: d(x,y)=1 \}.
$$
It is easy to see that this type of tree is the special case of
uniformly bounded tree \cite{Dang2015}.
\begin{defin} For $K\subseteq V$ a
spin configuration $\sigma_{K}$ on $K$ is defined by a function
$$
x\in K\rightarrow \sigma_{K}(x)\in \Phi=\{-1,+1\}.
$$
The set of all configurations on $K$ is denoted by $\Phi^{K}.$
\end{defin}

\begin{figure} [!htbp]\label{cayley-tree-k=3}
\centering
\includegraphics[width=50mm]{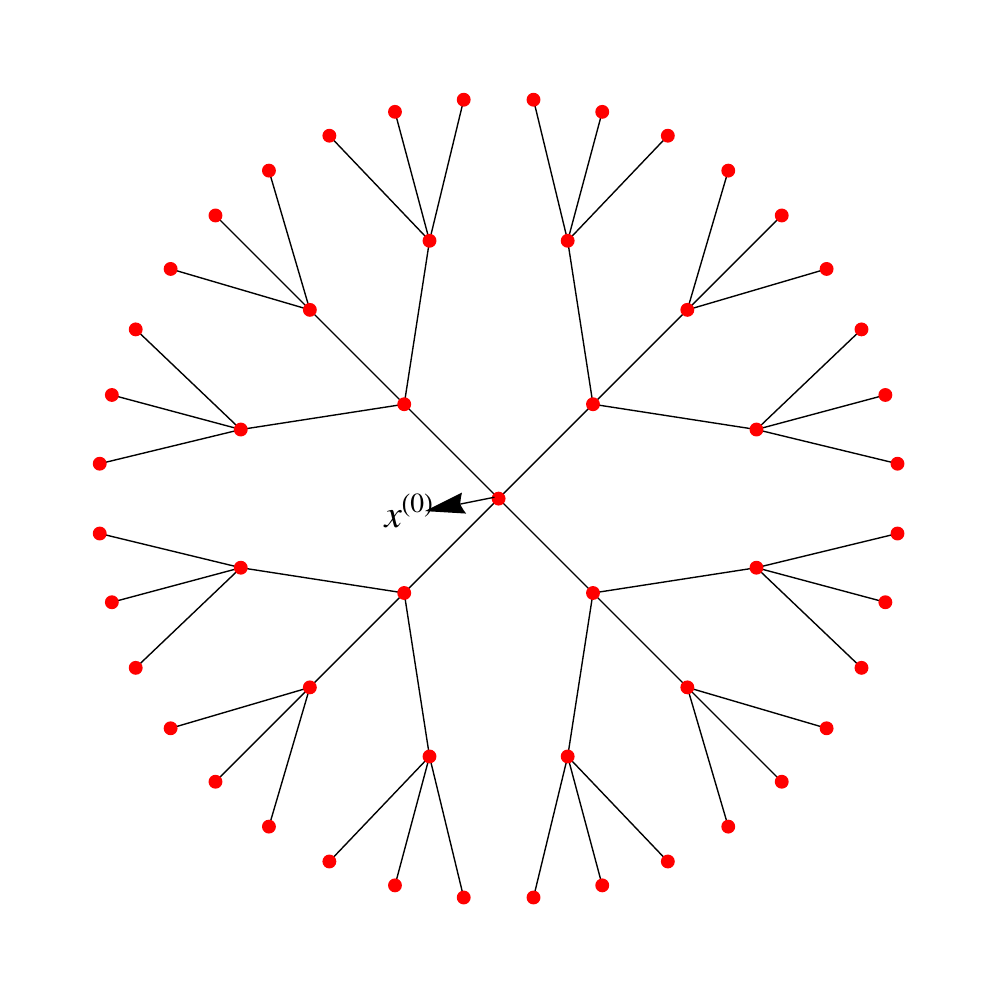}
\caption{(Colour online) Cayley tree of order $k=3$.}\label{cayley-tree-k=3}
\end{figure}
The fixed vertex $x^{(0)}$ is called the 0th level and the
vertices in $W_n$ are called the $n$th level. For the sake of
simplicity, we put $|x| = d(x, x^{(0)}), x \in V$
\cite{GAUT2011Chaos}.

Let us now define the neighborhoods that represent mutually
competing interactions that we will use throughout the paper.
\begin{defin}\label{neighborhoods}
Hereafter, we  use the following definitions for
neighborhoods.
\begin{enumerate}
    \item For $x,y\in V$, the vertices $x$ and $y$ are called {\it
\textbf{nearest-neighbors (NN)}} if there exists an edge $\ell\in
L$ connecting them, which is denoted by $\ell=\langle x,y\rangle$.
    \item Two vertices $x,y\in V$ are called {\it \textbf{the
next-nearest-neighbors (NNN)}} if there exists a vertex $z \in V$
such that $x, z$ and $y, z$ are NN, that is if $d(x,y)=2$.
    \item The vertices $x$ and $y$ are called {\it \textbf{prolonged
next-nearest-neighbors}} if $|x|\neq |y|$ and $d(x,y)=2$, and it is
denoted by $\rangle x,y\langle$.
    %\item The vertices $x,y$ and $z$ are called {\it \textbf{ternary prolonged
%    next-nearest-neighbors}} if $|x|\neq |y|$ and is denoted by
%    $\widetilde{>x,y<}$.
    \item The next-nearest-neighbor vertices $x,y\in V$ that are not
prolonged are called {\it \textbf{one-level
next-nearest-neighbors}} if $|x|=|y|$ and are denoted by
$\widetilde{\rangle x,y\langle}$.

\item For $x\in W_n,y\in S(x)$, $z\in S(y)$ ($x\in W_n,y\in
W_{n+1}$, $z\in W_{n+2}$) and $n\in \mathbb{Z}_+$, the triple of
vertices $x,y,z$ is called {\it \textbf{prolonged ternary
next-nearest-neighbor (PTNNN)}} if there exist two adjacent edges
$\ell_1=\langle x,y\rangle,\ell_2=\langle y,z\rangle\in L$ and is
denoted by $\widetilde{\langle x,y,z\rangle}$.

\item Three vertices $x, y$ and $z$ are called a
\emph{\textbf{triple of neighbors}} if $\langle x, y\rangle$,
$\langle y, z\rangle$ are nearest neighbors. They are denoted by
$\langle  x, y, z \rangle.$
%\item The triple of vertices $x,y,z\in V$ that are not prolonged
%is called {\it \textbf{two-level}} since $|x|=|z|$ and is denoted
%by $\bar{<x,y,z>}$.
\end{enumerate}
\end{defin}
%\begin{defin}\label{coupling}
%In this paper, we will consider Hamiltonian with \textbf{competing
%nearest-neighbor interactions} $J_1$, \textbf{prolonged
%next-nearest-neighbor interactions} $J_p$ and \textbf{one-level
%next-nearest neighbors interaction} $J_o$, \textbf{ternary
%prolonged  next nearest-neighbor interactions} $J_{t_p}$.  The
%coefficients $J_1,J_p,J_o,J_{t_p}\in \mathbf{R}$ are called
%coupling constant.
%\end{defin}
%Note that the joule, symbol $J$, is a derived unit of energy in
%the International System of Units.

\subsection{Ising model}\label{Ising model}
As stated in the introduction, the Ising model is a model
introduced to explain the concept of ferromagnetism. The model can
be derived from quantum mechanical considerations, by various
conjectures and rough conclusions.

In 1925, Ising \cite{Ising-1925} incorrectly concluded that there
is no phase transition in all dimensions. In 1936, Peierls
\cite{Peierls-1936} unexpectedly proved that the Lenz-Ising model
undergoes a phase transition  in 2 dimensions. Ising only solved the
one-dimensional statistical problem and demonstrated that his
model does not act like a ferromagnetic body \cite{Peierls-1936}.
There are several reasons why the Ising model has received great
attention from physicists, mathematicians, and scientists working
in other fields. Despite its simplicity, the Ising model is a
usable and interesting model (see \cite{ART2011JSP}).

As stated in the  \cite{Kindermann}, Ising discussed only
magnetic interpretation, but since his first definition in 1925,
the same model has been applied to a number of other physical and
biological systems such as gases, binary alloys, and cellular
structures. A sociologically oriented application was proposed by
Weidlich~\cite{Weidlich-1971}. We  give only one example here.
Let us take the Hamiltonian
\begin{equation}\label{Ising-Ham}
E_{\lambda}=-\sum _{i,j} I_{ij}\sigma _i\sigma _j-\mu H\sum _i
\sigma _i,
\end{equation}
where $E_{\lambda}$ represents the total energy of a certain spin
configuration $\lambda(\sigma_l, \sigma_2,\ldots ,\sigma_n)$ of
$n$ magnets (where every $\sigma_i$ has a given value, $+ 1$ or $- 1$)
is assumed to depend on the external field $H$ and on spin-spin
interaction parameters $I_{ij}$.  In the simplest case, $I_{ij}$
is defined by
$$ I_{ij}=\left\{
\begin{array}{cc}
 I>0, & \text{for}\ (i,j)=\text{nearest\ neighbours\ in\ the lattice}, \\
 0, & \text{otherwise}.
\end{array}
\right.
$$
Considering the Hamiltonian \eqref{Ising-Ham}, Weidlich
\cite{Weidlich-1971} examined an application of the Ising model in
sociology. Weidlich \cite{Weidlich-1971} considers a group of
people. The state of each of these individuals is chosen to be
either liberal (downward) or conservative (upward). Here, the total
energy $E_{\lambda}$ is called the tension. The first expression
in \eqref{Ising-Ham} is the tension arising from the interactions
of people. The external field (the second expression) may be the
liberal or conservative state of government. According to Weidlich
\cite{Weidlich-1971}, the tension is lowest when all people agree
with each other and with the government. Of course, in such an
application it may be necessary to ignore special neighborhoods
and regular grid restrictions.

%\newpage
\section{The dynamical behavior of the Ising model on a Cayley tree}\label{Dynamical behavior of Ising model on CT}
In this section, we deal with the phase diagrams of dynamical
systems associated with the Ising models corresponding to the
Hamiltonians on a given Cayley tree with certain interactions. So
far, this topic has been extensively studied due to the emergence
of non-trivial magnetic orders (see
\cite{Vannimenus,MTA1985a,Inawashiro,Inawashiro-T1983} and some 
references therein).

\subsubsection{Vannimenus-Ising model with competing interactions
on a Cayley tree}\label{Vannimenus-Ising model} Let us consider
the Hamiltonian
\begin{equation}\label{Vhm}
H(\sigma)=-{{J}_{p}}\sum\limits_{\rangle x,y\langle}{\sigma
(x)\sigma (y)}-J\sum\limits_{\langle x,y\rangle}{\sigma (x)\sigma
(y)},
\end{equation}
where $\rangle x,y\langle$ represents the prolonged next nearest
neighbor interaction and $\langle x,y\rangle$ represents the
nearest neighbor interaction. We  refer to this model as the
Vannimenus-Ising model for short.

By considering the Hamiltonian \eqref{Vhm}, in addition to the
paramagnetic and ferromagnetic phases in the drawn phase diagram,
Vannimenus \cite{Vannimenus} also detected a new modulated phase,
(see figure~\ref{van-phase-D}). In this section, we  explain
in detail how the phase diagrams of dynamical systems and phase
variations are determined.
\begin{figure} [!t]
%	\label{vannimenus-tanim}
\centering
\includegraphics[width=55mm]{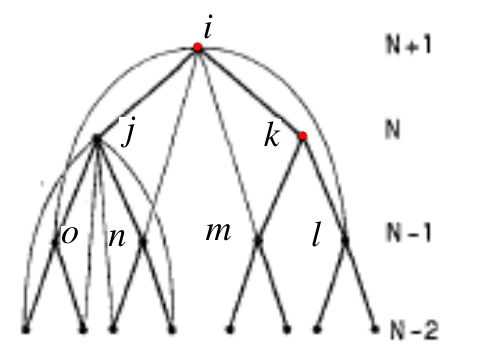}
\caption{(Colour online) Four successive generations of a Cayley tree (two
branches), with nearest-neighbor (heavy lines) and prolonged
next-nearest neighbor (thin lines)
interactions.}\label{vannimenus-tanim}
\end{figure}
Vannimenus \cite{Vannimenus} considered a standard approach,
which is to write down recursive equations relating the partition
function of an $N$-generation tree to the partition functions of
its subsystems with $(N- 1)$ generations (see figure
\ref{vannimenus-tanim}). To derive the recursive equations, he
built the tree in the reverse direction, starting from its surface
and moving towards its root. For a sufficiently large system,
Vannimenus considered only the fastest growing terms.
%Vannimenus have considered the standard approach consisting in
%writing down recurrence equations relating the partition function
%of an $N$-generation tree to the partition functions of its
%subsystems containing $(N- 1)$ generations (see Figure
%\ref{vannimenus-tanim}). In order to derive recursive equations,
%he have builded up the tree in the reverse direction, starting
%from its surface and going towards its root. For a large enough
%system, Vannimenus considered only the fastest growing terms.

For the partial partition functions 
%$ Z^{(N+1)}\left(
%\begin{array}{ccc}
%  & i &  \\
% k &  & j
%\end{array}
%\right)$, 
$ Z^{(N+1)}\left(
\begin{smallmatrix}
	& i &  \\
	k &  & j
\end{smallmatrix}
\right)$
there are a priori eight different partial partition
functions $Z^{(N+1)}$, where $i,j,k\in \{-1,+1\}$. Therefore, one
can obtain the following equation to derive the partial partition
functions (see figure~\ref{vannimenus-tanim}):
\begin{eqnarray}\label{ppf-Vann}
Z^{(N+1)}\left(
\begin{array}{ccc}
  & i &  \\
 j &  & k
\end{array}
\right)&=&\sum _{l,m,n,o\in \{-1,+1\}} \re^{[\beta Ji(k+j)+\beta
J_pi(o+n+m+l)]}\\\nonumber &\times& Z^{(N)}\left(
\begin{array}{ccc}
& j &   \\
o &  & n
\end{array}
\right)Z^{(N)}\left(
\begin{array}{ccc}
 & k &  \\
 m &  & l
\end{array}
\right).
\end{eqnarray}

After simplifying, taking into account eight partial partition
functions \eqref{ppf-Vann}, Vannimenus \cite{Vannimenus} derived
the following recursive equations:
\begin{eqnarray}\label{recurent}
\left\{
\begin{array}{l}
 x^{(n+1)}=\frac{1}{a^2D}[(1+b^2x^{(n)})^2+(y_1^{(n)}+b^2y_2^{(n)})^2], \\
 y_1^{(n+1)}=\frac{2}{D}[b^2y_1^{(n)}+y_2^{(n)}](x^{(n)}+b^2), \\
 y_2^{(n+1)}=-\frac{2}{a^2D}[b^2y_2^{(n)}+y_1^{(n)}](x^{(n)}b^2+1),
\end{array}
\right.
\end{eqnarray}
where
$D=\left(b^2+x^{(n)}\right)^2+\left(y_2^{(n)}+b^2y_1^{(n)}\right)^2
$ and $a=\re^{J/T}, b=\re^{J_p/T}.$\\
Furthermore, the average magnetization $m$ of the $N{\text{th}}$ generation is
given by
\[
m=\frac{2(y_1^{(n)}+y_2^{(n)})(x^{(n)}+1)}{(1+x^{(n)})^2+(y_2^{(n)}+y_1^{(n)}){}^2}.
\]
\subsection{How do we plot graphs of the phase diagrams of the Ising model?}
A phase diagram of a dynamical system describes the structure of
the phases, the stability of the phases, the transitions from one
phase to another, and the corresponding transition lines
\cite{Inawashiro,Inawashiro-T1983,G2011Chaos}. In previous works
\cite{Vannimenus,UGAT2012IJMPC,AUT2010AIP}, corresponding
recurrence equations were analyzed using various computer codes,
and thus the mathematical and physical interpretations of these systems
of nonlinear equations were established. To run these programs,
the phase diagrams were plotted using high performance computers.
One examines the limit $(x^{(n)},\,\,
y_{1}^{(n)},\,\,y_{2}^{(n)})\rightarrow (x^{*},\,\, y_{1}^{*},\,\, y_{2}^{*})$
as $n\rightarrow\infty$.

%The recursive equations \eqref{recurent} provide us (numerically)
%with the phase diagram in $(T/J,-J_p/J)$ space. Starting from
%random initial conditions (with $y_1^{(1)},y_2^{(1)}\neq 0$), one
%iterates the recurrence relations \eqref{recurent} and observes
%their behavior after a large number of iterations.

The recursive equations \eqref{recurent} give us (numerically) the
phase diagram in $(T/J,-J_p/J)$-space. By obtaining the initial
conditions (with $y_1^{(1)},\,\,y_2^{(1)}\neq 0$), we obtain the next
terms in the recurrence equations~\eqref{recurent}, so that we can
observe the behavior of these equations by performing a large
number of iterations. As a result of these iterations, we try to
reach the fixed points $(x^*,\,\,y_1^*,\,\,y_2^*)$ of these iterative
equations. Then, taking this fixed point as new initial data, one
can construct a phase (or Gibbs measure) that corresponds to a
paramagnetic phase if $y_1^*=0,\,\,y_2^*=0$ or to a ferromagnetic
phase if $y_1^*,\,\,y_2^*\neq 0.$ Secondly, the system may be
periodic with period $p$ and  taking into account the sequence
$(x^{(k)},\,\,y_1^{(k)},\,\,y_2^{(k)})$ with $k=0,\,\,1,\,\,\ldots,\,\,p-1$  as new
initial data, one can construct a phase. 
Here, the case $p=2$ corresponds to the antiferromagnetic phase and
the case $p=4$ corresponds to the so-called antiphase, denoted for compactness
by~$ \left\langle 2 \right\rangle  $. It corresponds to an appropriate phase with a given
period. Finally, the system can remain aperiodic, which is
associated to an incommensurable phase. It is difficult to
numerically distinguish between a non-periodic state and a state
with a very long period (see
\cite{Vannimenus,UGAT2012IJMPC,Inawashiro,Inawashiro-T1983,G2011Chaos,AUT2010AIP}
for details).

If these iterations are based on the resulting values and
$\lim\limits_{n\rightarrow \infty } x^{(n)}\neq
0,\,\,\lim\limits_{n\rightarrow \infty }
y_1^{(n)}=0,\lim\limits_{n\rightarrow \infty } y_2^{(n)}=0$, then
the diagram shows a paramagnetic phase region (shortly,
\textbf{P}) (high symmetric phase). If at least one of the limits
$\lim\limits_{n\rightarrow \infty } y_1^{(n)}$ and
$\lim\limits_{n\rightarrow \infty} y_2^{(n)}$ is non-zero, then the
relevant phase corresponds to the ferromagnetic~(\textbf{F})
region in our phase diagram. Otherwise, if there are no limit
values, i.e., after a certain iteration, the terms of the index are
not always the same, then the terms of these sequences will either
change periodically or randomly. In such cases, for example, if
the terms after 100000 terms are the same as the two periods, then
we have expressed this region with period 2, briefly \textbf{P}2.
If it is 3 periods, then we have expressed the region with period
3, briefly \textbf{P}3. Thus, we have termed the modulated phase
(\textbf{M}) in the phase regions in the other case, which is
called the periodic phase, up to period 12 (\textbf{P}12).

Let us consider the equation
\begin{equation}\label{Periodicfixedpoint}
\lim_{n\rightarrow \infty }
(x^{(n)},y_1^{(n)},y_2^{(n)})=\lim_{n\rightarrow \infty }
(x^{(n+p)},y_1^{(n+p)},y_2^{(n+p)})=(x^{(*)},y_1^{(*)},y_2^{(*)}).
\end{equation}
The nonnegative integer $p$ is called the period of convergent
sequence $(x^{(n)},y_1^{(n)},y_2^{(n)})$. For simplicity, let
$F^{(n)}=(x^{(n)},y_1^{(n)},y_2^{(n)})$.

To illustrate this process graphically, we have expressed
each phase with distinct colours. %Thus, a different drawing was
%made from the phase diagrams drawn up to now in the literature.
We have fixed the corresponding colouring of the phase diagrams in
 table \ref{colors}.

\begin{center}
\begin{table}
  \centering
  \caption{(Colour online) The colours of the phase diagrams.}\label{colors}
  \vspace{2mm}
\begin{tabular}{|c|c|c|}\hline
Periodic point &colour& Kind of Phases \\\hline $F^{(n)}
\rightarrow (x^{*}, 0, 0)$ & White & %\textcolor[rgb]{1.0,1.0,1.0}{'PARAMAGNETIC'}
\textquotedblleft PARAMAGNETIC"
\\\hline
 $F^{(n)}
\rightarrow (x^{*}, y_1^{*},y_2^{*})$ &Red & \textcolor[rgb]{1.0,
0.0, 0.0}{\textquotedblleft FERROMAGNETIC"}\\\hline &Mangotango&
\textcolor[rgb]{1.0, 0.51, 0.26}{\textquotedblleft
PARAMODULATED"}\\\hline $\left.
\begin{array}{l}
 F^{(n)} \\
 F^{(n+2)}
\end{array}
\right\}\rightarrow (x^{(*)},y_1^{(*)},y_2^{(*)})$& Yellow &
\textcolor[rgb]{1.0, 1.0, 0.0}{\textquotedblleft PERIOD 2"}
\\\hline $\left.
\begin{array}{l}
 F^{(n)} \\
 F^{(n+3)}
\end{array}
\right\}\rightarrow (x^{(*)},y_1^{(*)},y_2^{(*)})$&Lime
&\textcolor[rgb]{0.0, 1.0, 0.0}{\textquotedblleft PERIOD
3"}\\\hline $\left.
\begin{array}{l}
 F^{(n)} \\
 F^{(n+4)}
\end{array}
\right\}\rightarrow (x^{(*)},y_1^{(*)},y_2^{(*)})$&Aqua  &
\textcolor[rgb]{0.0, 1.0, 1.0}{\textquotedblleft PERIOD 4"}
\\\hline $\left.
\begin{array}{l}
 F^{(n)} \\
 F^{(n+5)}
\end{array}
\right\}\rightarrow (x^{(*)},y_1^{(*)},y_2^{(*)})$&Blue  &
\textcolor[rgb]{0.0, 0.0, 1.0}{\textquotedblleft PERIOD 5"}
\\\hline
$\left.
\begin{array}{l}
 F^{(n)} \\
 F^{(n+6)}
\end{array}
\right\}\rightarrow (x^{(*)},y_1^{(*)},y_2^{(*)})$&Fuchsia &
\textcolor[rgb]{1.0, 0.0, 1.0}{\textquotedblleft PERIOD
6"}\\\hline $\left.
\begin{array}{l}
 F^{(n)} \\
 F^{(n+7)}
\end{array}
\right\}\rightarrow (x^{(*)},y_1^{(*)},y_2^{(*)})$&3DDkShadow &
\textcolor[rgb]{0.54, 0.47, 0.36}{\textquotedblleft PERIOD
7"}\\\hline $\left.
\begin{array}{l}
 F^{(n)} \\
 F^{(n+8)}
\end{array}
\right\}\rightarrow (x^{(*)},y_1^{(*)},y_2^{(*)})$& Maroon% (maroon(html/css))
& \textcolor[rgb]{0.5, 0.0, 0.0}{\textquotedblleft PERIOD
8"}\\\hline $\left.
\begin{array}{l}
 F^{(n)} \\
 F^{(n+9)}
\end{array}
\right\}\rightarrow (x^{(*)},y_1^{(*)},y_2^{(*)})$& Green
&\textcolor[rgb]{0.0, 0.5, 0.0}{\textquotedblleft PERIOD 9"
}\\\hline $\left.
\begin{array}{l}
 F^{(n)} \\
 F^{(n+10)}
\end{array}
\right\}\rightarrow (x^{(*)},y_1^{(*)},y_2^{(*)})$& Olive &
\textcolor[rgb]{0.5, 0.5, 0.0}{\textquotedblleft PERIOD
10"}\\\hline $\left.
\begin{array}{l}
 F^{(n)} \\
 F^{(n+11)}
\end{array}
\right\}\rightarrow (x^{(*)},y_1^{(*)},y_2^{(*)})$& Teal &
\textcolor[rgb]{0.0, 0.5, 0.5}{\textquotedblleft PERIOD
11"}\\\hline $\left.
\begin{array}{l}
 F^{(n)} \\
 F^{(n+p)}
\end{array}
\right\}\rightarrow (x^{(*)},y_1^{(*)},y_2^{(*)})$& Purple ($p>
11$) &\textcolor[rgb]{0.5, 0.0, 0.5}{\textquotedblleft
MODULATED"}\\\hline
\end{tabular}
\end{table}
\end{center}
%%\begin{table}[!h]
%%\caption{Comparison of percentages.}
%%\begin{tabular}{lclclclclc}
%%\hline \hline
%%Mode &  Var  &  Cum\\
%%\hline
%%{}       & EF   & CHF    & EF2   & CHF2\\
%%1   &  17.5 & 19.1   & 17.5  & 19.1\\
%%2   &  11.8 & 12.7   & 29.3  &  31.9\\
%%3   &  6.6  &  5.6         & 35.9    &  37.4\\
%%\hline
%%\end{tabular}
%%\end{table}
\begin{figure} [!htbp]\label{van-phase-D}
\centering
\includegraphics[width=55mm]{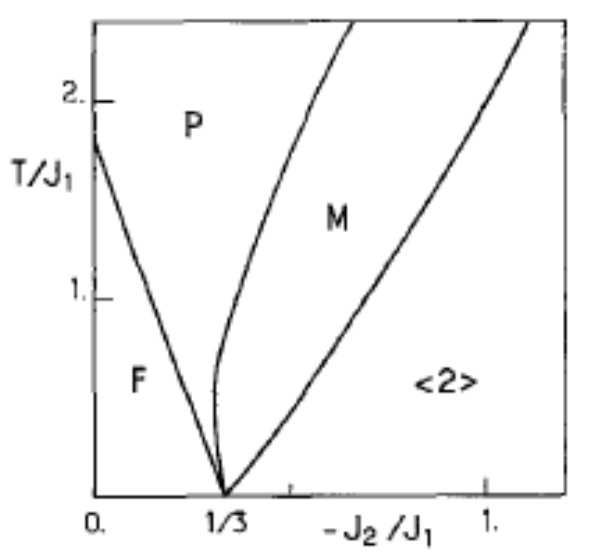}
\caption{The phase diagram of Vannimenus model on the rectangular
region $[0,1.3]\times [0,2.3]$ (see
\cite{Vannimenus}).}\label{van-phase-D}
\end{figure}

\begin{figure} [!htbp]
	%\label{van-phase-D1}
\centering
\includegraphics[width=45mm]{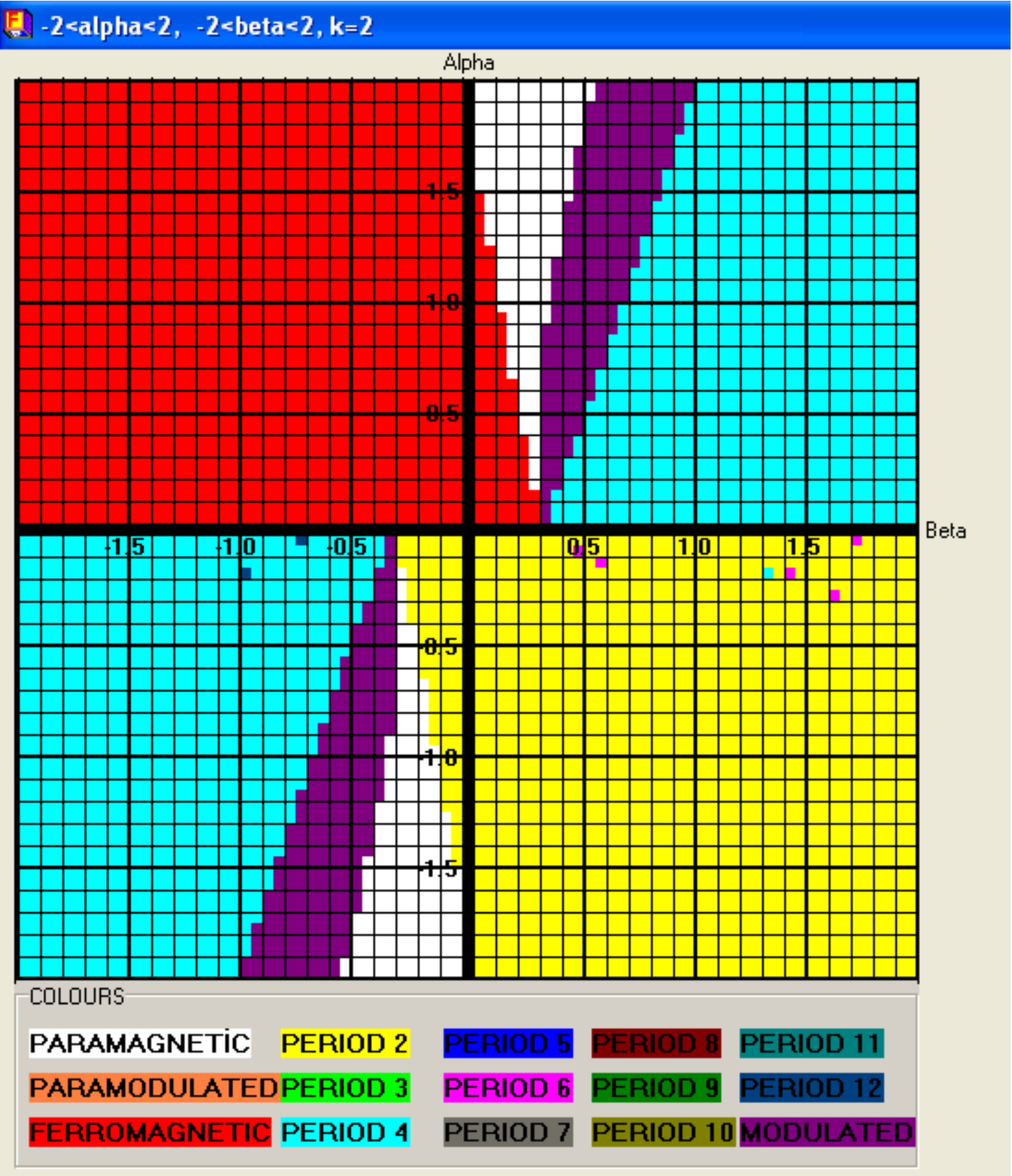}
\caption{(Colour online) The phase diagram of Vannimenus model on
the rectangular region $[-2,2]\times [-2,2]$ (see
\cite{GAUT2011a}).}\label{van-phase-D1}
\end{figure}

In \cite{GU2011a}, the generalization of the Vannimenus's study on
a Cayley tree of arbitrary order was introduced. In these works,
the phase diagrams of two of the simplest Ising models were
examined in detail. The phase diagram in \cite{Vannimenus} was
plotted for the first region only. In the study by Vannimenus, the
phase diagram was investigated. The phase diagram of the same model was
first tried in Visual Basic, then $C^{++}$ and later Delphi
programming languages. The best result was obtained with the code
written in Delphi programming language (see figure
\ref{van-phase-D1}). %Previous work was verified in experiments.
%Thus, we have successfully scanned the source and verified the
%results.
Vannimenus \cite{Vannimenus} studied the Ising model in
conjunction with the Hamiltonian~\eqref{Vhm} on the Cayley tree of the
order two. In  figure \ref{van-phase-D} plotted by Vannimenus
\cite{Vannimenus}, the phase diagram of the dynamical system
corresponding to the Ising model with competing interactions
includes ferromagnetic~(\textbf{F}), paramagnetic (\textbf{P}),
modulated (\textbf{M}), and  periodic (+ + -- --) structure ($\left\langle 
2 \right\rangle  $, in short). Considering the same Hamiltonian, we analyzed
the phase diagrams associated with the Ising model on a Cayley
tree of arbitrary order \cite{UGAT2012IJMPC}. We accurately
plotted the phase diagrams \cite{UGAT2012IJMPC}. We characterized
the phases by the sequence of stable points of the recursion
relations. We examined the effect of the order of the Cayley tree
on the phase diagrams \cite{UGAT2012IJMPC,AGUT2012ACTA}.
%When the pictures
%in Figure \ref{Ising-phase-D2} are compared, we can say that the
%second one is of better quality (see Figures
%\ref{Ising-phase-D2}).
%Vannimenus \cite{Vannimenus} studied the Ising model associated
%with the Hamiltonian \eqref{Vhm} on the Cayley tree of order two.
%In Figure \ref{van-phase-D} plotted by Vannimenus
%\cite{Vannimenus}, the phase diagram of the Ising model with
%competing interactions contains ferromagnetic (\textbf{F}),
%paramagnetic (\textbf{P}), modulated (\textbf{M}) and  the
%periodic (+ + - -) structure ($<2>$).
%
\begin{figure} [!htbp]
	%\label{Ising-phase-D2}
\centering
\includegraphics[width=41mm]{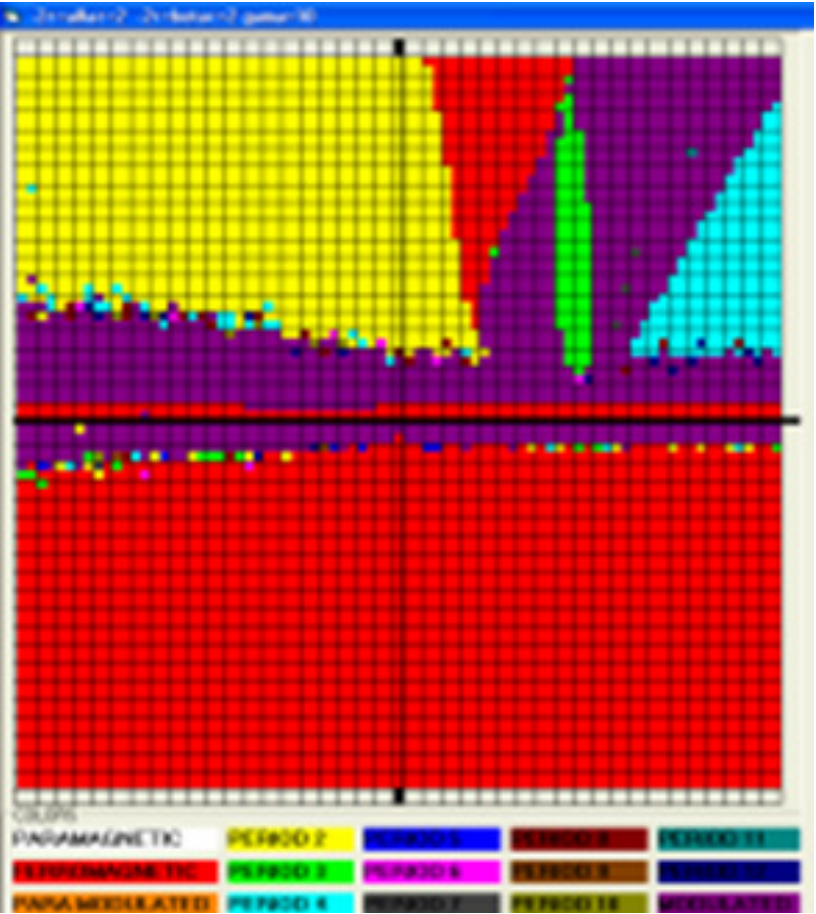}\ \ \ \ \ \ \ \ \ \ \
\ \ \ \ \
\includegraphics[width=40mm]{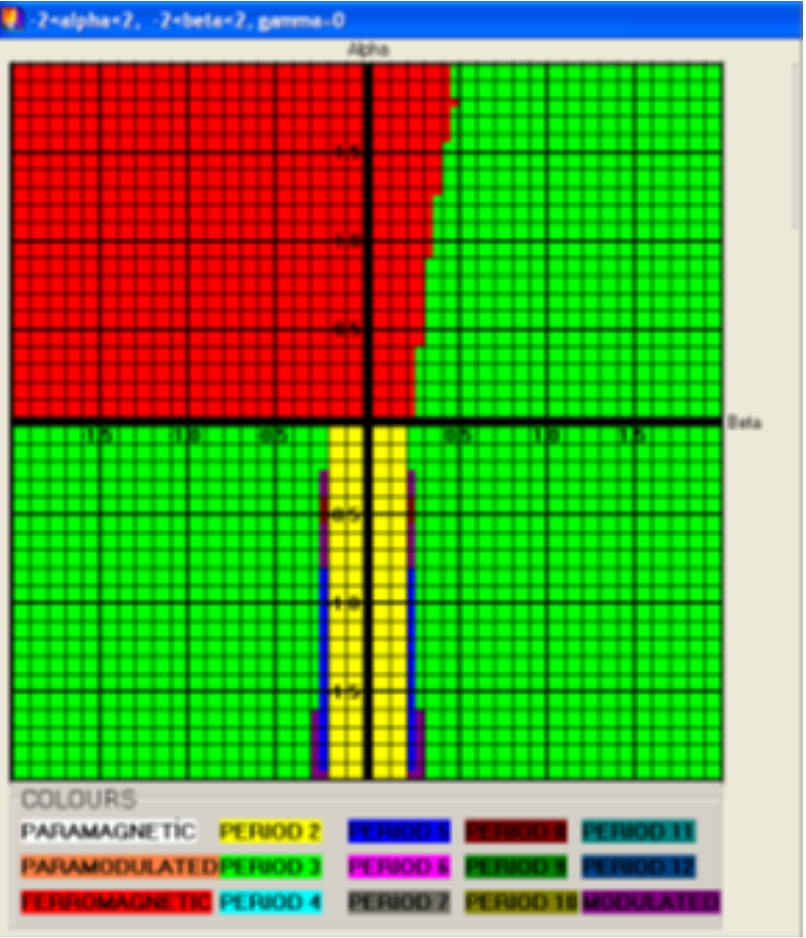}
\caption{(Colour online) The phase diagrams on the rectangular
region $[-2,2]\times [-2,2]$ of some given Ising
models.}\label{Ising-phase-D2}
\end{figure}
%Using the same Hamiltonian, we obtained the Ising model on an
%arbitrary order Cayley tree. Our results are completely
%overlapping. Taking into account a new method to plot the phase
%diagrams, we can exactly picture the phase diagrams. When we
%compare the first two images below, we can say that the second one
%is better quality (see Figures \ref{Ising-phase-D2}).
Looking at the phase diagram to the right of  figure~\ref{Ising-phase-D2}, the image quality obtained is better than
that on the left of figure~\ref{Ising-phase-D2}. The phase
diagram was drawn using the $C^{++}$ programming language on the
left. The figure on the right was drawn using the Delphi
Programming language. %%The second picture has a clearer picture quality.
%The exchange of wave vectors, the presence of critical points at
%non-zero points, and the fact that the phase diagrams are in color
%separated by regions are among our other findings
%Note that one of our other goals is to obtain a general
%mathematical theory for the phase diagrams corresponding to the
%dynamic systems. However, it is not easy to develop this theory.
In addition, the mean magnetization equations for each level of
the above-mentioned Ising and Potts models were derived
\cite{Vannimenus,Yesilleten}.
\begin{rem}
It should be noted that we have used Vannimenus' approach to plot
phase diagrams.
\end{rem}
%\newpage
\subsection{The phase diagrams of Ising model with NN and PTNNN
interactions}\label{PhaseDiagrams} In this subsection, for
convenience we consider one of the simplest Ising models (see
\cite{GAUT2011b,AGUT2012ACTA}). Note that this subsection is based
on  \cite{AGUT2012ACTA}.

Let us consider the following Hamiltonian
\begin{equation}\label{Prolonged-hm}
H(\sigma )=-{{J}_{{{t}_{p}}}}\sum\limits_{ {\mathop{\rangle
x,y,z\langle}}\,}{}\sigma (x)\sigma (y)\sigma
(z)-{{J}_{1}}\sum\limits_{\langle x,y\rangle}\sigma (x)\sigma (y),
\end{equation}
where $\mathop{\rangle x,y,z\langle}$ represents competing
\textbf{prolonged ternary next nearest-neighbor} (PTNNN) and
$\langle x,y\rangle$ represents competing
\textbf{nearest-neighbor} (NN), where
${{J}_{{{t}_{p}}}},{{J}_{1}}\in\mathbb{R}$ are coupling constants.

\begin{figure} [!htbp]
	%\label{cayley-level2}
\centering
\includegraphics[width=65mm]{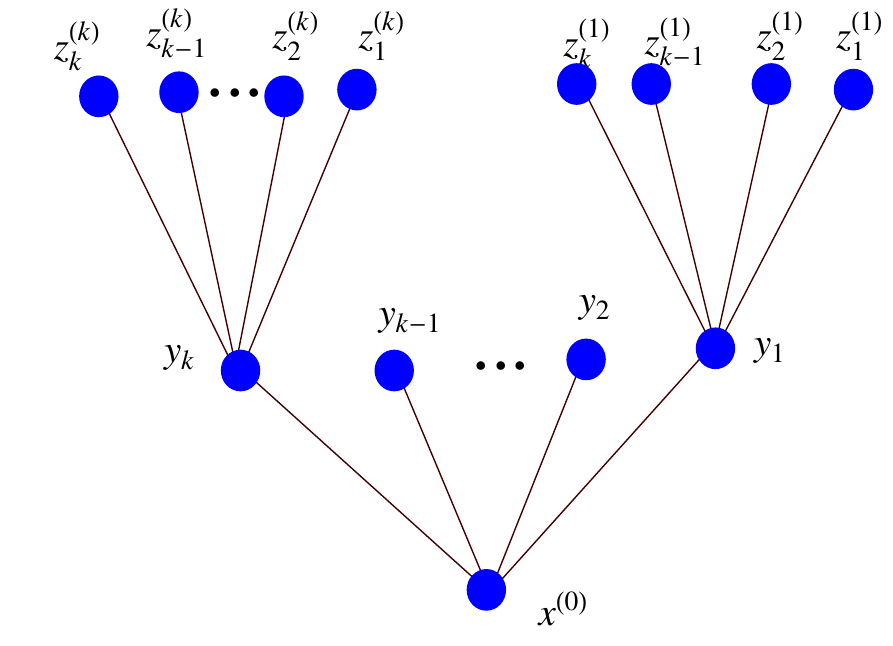}
\caption{(Colour online)
Two generations of arbitrary order $k>1$ semi-infinite Cayley tree
$\Gamma^{k}$ (branching ratio is finite
$k$).}\label{cayley-level2}
\end{figure}

%In order to obtain the recurrent equations, we consider the
%relation of the partition function on ${{V}_{n}}$ to the partition
%function on subsets of ${{V}_{n-1}}$. Given the initial conditions
%on ${{V}_{1}}$, the recurrence equations indicate how their
%influence propagates down the tree.
Here, we  analyze the dynamical behavior and the phase diagram
of the Ising model corresponding to the Hamiltonian
\eqref{Prolonged-hm}. We investigate the relationship between the
partition function on $V_n$ and the partition function on subsets
of $V_{n-1}$ to obtain the recurrent equations. The recurrence
equations show how their impact propagates along the tree for a
given initial condition on $V_1$.\\
Let
\[
\sigma _{S}^{+}(x^{(0)})=\left(
\begin{matrix}
   \sigma ({{y}_{k}}),\cdots ,\sigma ({{y}_{2}}),\sigma ({{y}_{1}})  \\
  + \\
\end{matrix} \right),
\]
be a configuration in $\{-1,+1\}^{B_1(x^{(0)})}$ (see figure
\ref{cayley-level2}), where $B_1(x^{(0)})=\{y\in
S(x^{(0)})\}\cup\{x^{(0)}\}$.\\
Let $m$ represent the number of spins down ($\sigma (y_{i})=-1$)
at the first level $W_1$, where $0\leqslant m\leqslant k$. Then, at the
first level $W_1$, $(k-m)$ is the number of spins up ($\sigma
(y_{i})=+1$).
%
%Let $m$ be the number of spins down, i.e., $\sigma ({{y}_{i}})=-1$
%on the first level ${{W}_{1}}$, where $0\leqslant m\leqslant k$. Then
%$(k-m)$ is the number of spins up, i.e., $\sigma ({{y}_{i}})=+1$
%on the first level ${{W}_{1}}$.

In figure~\ref{cayley-level2}, the root of the lattice that
emanates $k$ edges of $\Gamma^{k}$ is a fixed vertex $x^{(0)}$,
therefore $y_j \in S(x^{(0)}),\,\,z_i^{(j)} \in S(y_j)$. Denote a
configuration in $\{-1,+1\}^{B_1(y_{i})}$ by
\[
\sigma _{S}^{+}(y_{i})=\left(\begin{matrix}
   \sigma (z_{k}^{(i)}),\cdots ,\sigma (z_{2}^{(i)}),\sigma (z_{1}^{(i)})\\
   +  \\
\end{matrix} \right).
\]
Let the $m$ spins on $\{-1,+1\}^{B_1(S(y_{i}))}$ (the second level
$W_{2}$) be spins down, i. e., $\sigma (z_{j}^{(i)})=-1$, where
$0\leqslant m \leqslant k.$ Let the remaining number of $(k-m)$ of them be
spin up i. e., $\sigma (z_{j}^{(i)})=+1$. Let
\[
\sigma _{S}^{-}(x^{(0)})=\left(
\begin{matrix}
   \sigma ({{y}_{k}}),\ldots ,\sigma ({{y}_{2}}),\sigma ({{y}_{1}})\\
   -  \\
\end{matrix} \right),
\]
be a configuration in $\{-1,+1\}^{S(x^{(0)})}$. Let $m$ be the
number of spins down, i. e., $\sigma (y_{i})=-1$ at the first level
${{W}_{1}}$, where $0\leqslant m\leqslant k.$ Considering the vertices in
the subsets $W_1$ and $W_2$ of the graph $\Gamma^{k}$, we can
similarly construct the following configuration. Let
\[\sigma _{S}^{-}({{y}_{i}})=\left(
\begin{matrix}
   \sigma (z_{k}^{(i)}),\cdots ,\sigma (z_{2}^{(i)}),\sigma (z_{1}^{(i)})  \\
   - \\
\end{matrix} \right),
\]
be a configuration in $\{-1,+1\}^{B_1(y_{i})}$ (see figure \ref{cayley-level2}).
Let $m$ be the number of spins down $\sigma (z_{j}^{(i)})=-1$ at
the second level ${{W}_{2}}$, where $0 \leqslant m \leqslant k.$

For clarity, denote the configuration of the set
$\Phi^{B_1(x^{(0)})}$ by
\[S_{m}^{(k-m)}(\sigma ({{x}^{(0)}}))=\left( \begin{matrix}
   \overbrace{++\cdots +}^{k-m}\overbrace{--\cdots -}^{m}  \\
   \sigma (x^{(0)})  \\
\end{matrix} \right).\]
Let
\[
{{Z}^{(n)}}\left( \begin{matrix}
   {{i}_{1}},{{i}_{2}},\ldots ,{{i}_{k}}  \\
   {{i}_{0}}  \\
\end{matrix} \right)={{Z}^{n}}=\sum\limits_{{\sigma_{n}}\in {{\Omega }^{{{V}_{n}}}}}{\exp (-\beta H({{\sigma }_{n}}))},
\]
be the partition function on ${{V}_{n}}$, where the spin in the
root $x^{(0)}$ is ${{i}_{0}}$ and the $k$ spins in the proceeding
ones are ${{i}_{1}},{{i}_{2}},\ldots ,{{i}_{k}}$. A priori, there
are $2^{k+1}$ partition functions $Z^{(n)}$. However, it is
plausible to assume that the different branches are comparable.

To make the paper more readable, we use shorter notations, as
shown below. As a result, we can show that there are only four
independent variables, namely
%There are a priori $2^{k+1}$ different $Z^{(n)}$ to consider. It
%is reasonable, though, to assume that the different branches are
%equivalent, as is usually done for models on trees. We will use
%shorter notations to increase the readability of the article as
%follows, therefore we can show that there are only four
%independent variables, namely
\[
{{z}_{1}}={{Z}^{(n)}}\left( \begin{matrix}
   +,+,\ldots ,+  \\
   +  \\
\end{matrix} \right), \quad {{z}_{2}}={{Z}^{(n)}}\left( \begin{matrix}
   -,-,\ldots ,-  \\
   +  \\
\end{matrix} \right),\] \[{{z}_{3}}={{Z}^{(n)}}\left( \begin{matrix}
   +,+,\ldots ,+  \\
   -  \\
\end{matrix} \right), \quad {{z}_{4}}={{Z}^{(n)}}\left( \begin{matrix}
   -,-,\ldots ,-  \\
   -  \\
\end{matrix} \right).\]
Then, arbitrary partial partition function ${{Z}^{(n)}}\left(
\begin{matrix}
   {{i}_{1}},{{i}_{2}},\ldots ,{{i}_{k}}  \\
   {{i}_{0}}  \\
\end{matrix} \right)$ is a combination of ${{z}_{1}},{{z}_{2}},{{z}_{3}},{{z}_{4}}$.
For instance, for $\sigma ({{i}_{0}})=+$,  the partial partition
function can be written by
\[ {{Z}^{(n)}}\left( \begin{matrix}
   {{i}_{1}},{{i}_{2}},\ldots ,{{i}_{k}}  \\
   +  \\
\end{matrix} \right)=\left( \begin{matrix}
   \overbrace{++\ldots +}^{k-m}\overbrace{--\cdots -}^{m}  \\
   +  \\
\end{matrix} \right)=z_{1}^{\frac{m}{k}}z_{2}^{\frac{k-m}{k}},
\]
where $m$ is the number of spins down on ${{W}_{1}}.$ Similarly,
for $\sigma ({{i}_{0}})=-1$, the partial partition function can be
written by
\[
{{Z}^{(n)}}\left( \begin{matrix}
   {{i}_{1}},{{i}_{2}},\ldots ,{{i}_{k}}  \\
   -  \\
\end{matrix} \right)=\left( \begin{matrix}
   \overbrace{++\cdots +}^{k-m}\overbrace{--\cdots -}^{m}  \\
  - \\
\end{matrix} \right)=z_{3}^{\frac{m}{k}}z_{4}^{\frac{k-m}{k}}.
\]
%where $m$ is the number of spins up on ${{W}_{1}}.$
By adjusting the variables ${{u}_{i}}=\sqrt[k]{z_{i}}$, one gets
$$
\left\{
\begin{array}{l}
 u_1^{'}=a\left(bu_1+b^{-1}u_2\right)^k, \\
 u_2^{'}=a^{-1}\left(b^{-1}u_3+bu_4\right)^k, \\
 u_3^{'}=a^{-1}\left(b^{-1}u_1+bu_2\right)^k, \\
 u_4^{'}=a\left(bu_3+b^{-1}u_4\right)^k,
\end{array}
\right.
$$
%\begin{eqnarray*}
%u_{1}^{'}&=&a{{(b{{u}_{1}}+{{b}^{-1}}{{u}_{2}})}^{k}};\\
%u_{2}^{'}&=&{{a}^{-1}}{{({{b}^{-1}}{{u}_{3}}+b{{u}_{4}})}^{k}};\\
%u_{3}^{'}&=&{{a}^{-1}}{{({{b}^{-1}}{{u}_{1}}+b{{u}_{2}})}^{k}};\\
%u_{4}^{'}&=&a{{(b{{u}_{3}}+{{b}^{-1}}{{u}_{4}})}^{k}},
%\end{eqnarray*}
where $a=\re^{J_{1}/T}$, $b=\re^{J_{t_{p}}}/T)$ and the variables
$u_{i}^{'}$ represent the ${{Z}^{(n+1)}}$.

Using the partial partition functions $u_{i}$, we obtain the total
partition function in the form;
\[
Z^{(n)}=(u_1+u_2)^k+(u_3+u_4)^k.
\]
To plot the phase diagrams, we consider the following reduced
variables
\begin{equation}\label{partition1aa}
\left\{
\begin{array}{l}
 x=\frac{u_2+u_3}{u_1+u_4}, \\
 y_1=\frac{u_1-u_4}{u_1+u_4}, \\
 y_2=\frac{u_2-u_3}{u_1+u_4}.
\end{array}
\right.
\end{equation}
The variable $x$ in the system \eqref{partition1aa} is simply a
measure of the frustration of nearest-neighbor bonds, not an order
parameter like $y_1,y_2$
%The variable $x$ is just a measure of the frustration of the
%nearest-neighbor bonds and is not an order parameter like
%${{y}_{1}},{{y}_{2}}$
(see \cite{Vannimenus} for details). By substituting the variables
in the equation \eqref{partition1aa} as follows:
$$
u_1=\frac{(1+y_1)A}{2},\quad u_2=\frac{(x+y_2)A}{2},\quad u_3=\frac{(x-y_2)A}{2}, \quad u_4=\frac{(1-y_1)A}{2},
$$
where $A=u_1+u_4$,  and if we replace the variables $u_{i}^{'}$ in
the system \eqref{partition1aa}, we derive the recurrence
equations:
\begin{equation}\label{partition1ab}
\left\{
\begin{array}{l}
{{x}^{'}}=\frac{1}{{{a}^{2}}D}[(x-{{y}_{2}}+{{b}^{2}}(1-{{y}_{1}}{{))}^{k}}+{{(1+{{y}_{1}}+{{b}^{2}}(x+{{y}_{2}}))}^{k}}],\\
y_{1}^{'}=\frac{1}{D}[({{b}^{2}}(1+{{y}_{1}})+x+{{y}_{2}}{{)}^{k}}-{{({{b}^{2}}(x-{{y}_{2}})+1-{{y}_{1}})}^{k}}],\\
y_{2}^{'}=\frac{1}{{{a}^{2}}D}[(x-{{y}_{2}}+{{b}^{2}}(1-{{y}_{1}}{{))}^{k}}-{{(1+{{y}_{1}}+{{b}^{2}}(x+{{y}_{2}}))}^{k}}],
\end{array}
\right.
\end{equation}
where $
D=\left(x+y_{2}+b^{2}(1+y_{1})\right)^{k}+\left(b^{2}(x-y_{2})+1-y_{1}\right)^{k}.
$\\
For the sake of completeness we  restate the method studied in
\cite{Vannimenus,GAUT2011Chaos}. With a suitable computer code, we
calculate the consecutive terms of the recurrence equations
\eqref{partition1ab} numerically, taking into account the rules
given in table \ref{colors}. Thus, with this method, we plot the
phase diagram corresponding to the recursive relations
\eqref{partition1ab} in the space
$(-{{J}_{{{t}_{p}}}}/{{J}_{1}},T/{{J}_{1}})$. Assume
$(-{{J}_{{{t}_{p}}}}/{{J}_{1}})=\beta $, $T/{{J}_{1}}=\alpha $ and
respectively $b=\exp(-{{\alpha }^{-1}}\beta )$, $a=\exp({{\alpha
}^{-1}})$.

By substituting the initial conditions
$$
\left\{
\begin{array}{l}
{{x}^{(1)}}=\frac{{{b}^{2k}}+{{a}^{2k}}}{a((ab{{)}^{2k}}+1)},\\
y_{1}^{(1)}=\frac{{{(ab)}^{2k}}-1}{{{(ab)}^{2k}}+1},\\
y_{2}^{(1)}=\frac{{{b}^{2k}}-{{a}^{2k}}}{{{a}^{2}}((ab{{)}^{2k}}+1)},
\end{array}
\right.
$$
under boundary condition $\bar{\sigma}^{(n)}(V\backslash
V_{n})\equiv 1$ in the recursive equations \eqref{partition1ab},
we obtain the second and third terms, respectively, so we run the
computer code and numerically calculate the terms after the $n$th
term of this sequence for natural number $n$ large
enough (see \cite{Vannimenus,AGUT2012ACTA,GanMohd2016} for details). %we iterates the recurrence relations \eqref{partition1a}
%and observes their behavior after a large number of iterations.

In the simplest sense, as explained earlier, a fixed point
$({{x}^{*}},y_{1}^{*},y_{2}^{*})$ is obtained. In other words, for
sufficiently large natural number $n$, the other terms of this
sequence are equal or become equal after a certain period.
%In the simplest situation a fixed point
%$({{x}^{*}},y_{1}^{*},y_{2}^{*})$ is reached.
For each binary value $(-J_{{t}_{p}}/J_{1},T/J_{1})\in
[-2,2]\times [-2,2]$ selected in the two-dimensional plane, we
obtain a fixed point $y_{1}^{*}=0,y_{2}^{*}=0$, then kind of the
phase becomes paramagnetic phase and point
$(-J_{{t}_{p}}/J_{1},T/J_{1})$ is marked with white or we obtain a
ferromagnetic phase if $y_{1}^{*},y_{2}^{*}\ne 0$ taking into
account the average magnetization equation.%From formula of average magnetization
%follows that a situation where $y_{1}^{*},y_{2}^{*}\ne 0$ but
%$m=0$ can not occur.
\begin{figure} [!htbp]
	%\label{prolonged-tenary1}
\centering
\includegraphics[width=44mm]{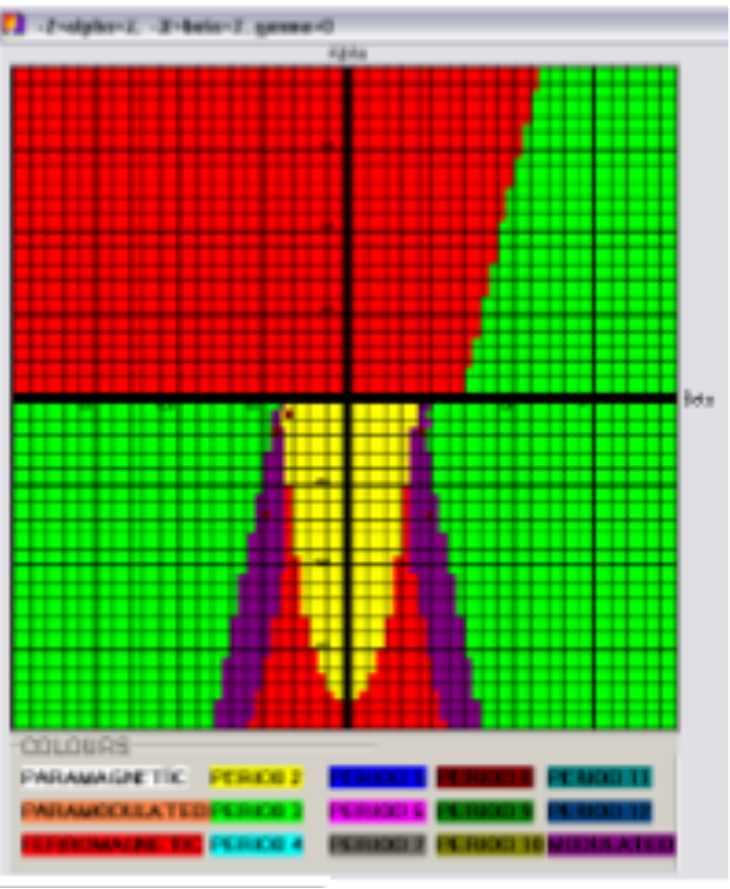}\ \ \ \ \ \ \ \
\includegraphics[width=46mm]{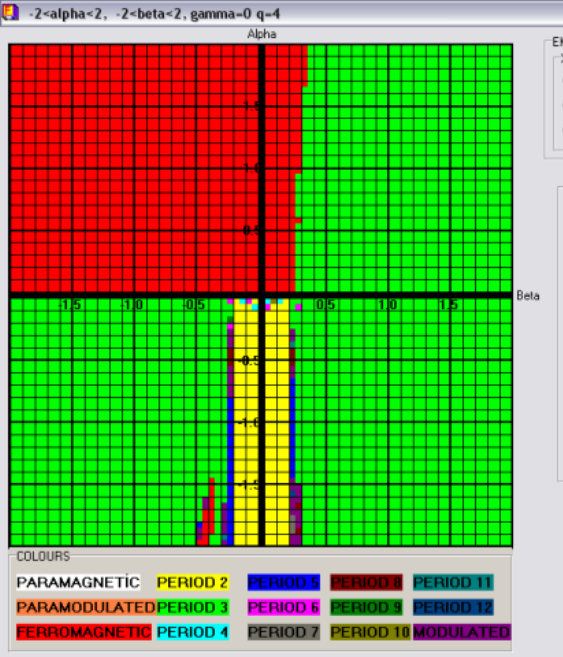}\label{fazk4}
\caption{(Colour online)  Relevant phase diagrams of the model
\eqref{Prolonged-hm} for $k=2$ (left-hand figure) and $4$ (right-hand
figure), respectively (see
\cite{AGUT2012ACTA}).}\label{prolonged-tenary1}
\end{figure}

Figures in \ref{prolonged-tenary1} illustrate the resulting
phase diagrams for various values of $k$. The phase diagram for
$J_1>0$ and $k=2$ incorporates a ferromagnetic phase and a novel
phase with period $3$ (see \cite{GAUT2011b,AGUT2012ACTA}). The
phase diagram of the recursive system in \eqref{partition1ab}
contains interesting images for competing interactions
$(J_{1}<0)$. The phase diagram in this case is extremely detailed,
with at least four multicritical Lifshitz points, two of which are
at zero temperature and the other two at nonzero temperature. The
key novelty is the existence of multicritical Lifshitz points that
are not at zero temperature, as opposed to zero temperature in
prior publications by Vannimenus \cite{Vannimenus} and Mariz
\cite{MTA1985a}. It is worth noting that the paramagnetic phase
has vanished in this situation. The role of $k$ is clearly
apparent on the phase diagrams.

Note that the recurrence equations \eqref{partition1ab} and phase
diagrams in this subsection are based on the work~\cite{AGUT2012ACTA}.
\subsection{Variation of wave-vectors}\label{variation of wavevectors}

Here, we examine in detail the variation of the wave vector with
temperature in the modulated phase region. The variation of the
wave vector shows narrow commensurate in the incommensurate
regions \cite{GU2011PhysicaA}. An incommensurate structure with a
given wave-vector is represented by a fixed defect concentration
\cite{Aubry1982}. Furthermore, the wave-vector of the modulated structure
(i.e. the defect concentration) varies continuously with piecewise
constant parts at each commensurate value (the resulting curve is
called a devil's staircase) \cite{Aubry1982}.

In this subsection, we  study the behavior of the average
magnetization in a period or the dynamical magnetization as a
function of the reduced temperature. This investigation allows us
to characterize the nature of the transition. The average
magnetization $m$ associated with the model \eqref{Prolonged-hm}
for the $n$th generation is given by
\allowdisplaybreaks
\begin{eqnarray}
m&=&\frac{Z_{+}^{(n)}-Z_{-}^{(n)}}{Z_{+}^{(n)}+Z_{-}^{(n)}}=\frac{\sum\limits_{i_0,i_1,\ldots,i_{k}\in
\{-1,1\}}{i_0{{Z}^{(n)}}\left( \begin{matrix}
   {{i}_{1}},{{i}_{2}},\cdots ,{{i}_{k}}  \\
   i_0  \\
\end{matrix} \right)}}{\sum\limits_{i_0,i_1,\ldots,i_{k}\in
\{-1,1\}}{{{Z}^{(n)}}\left( \begin{matrix}
   {{i}_{1}},{{i}_{2}},\cdots ,{{i}_{k}}  \\
   i_0  \\
\end{matrix} \right)}} \nonumber \\
&=&\frac{{{(1+x+{{y}_{1}}+{{y}_{2}})}^{k}}-{{(1+x-{{y}_{1}}-{{y}_{2}})}^{k}}}
{{{(1+x+{{y}_{1}}+{{y}_{2}})}^{k}}+{{(1+x-{{y}_{1}}-{{y}_{2}})}^{k}}}
\label{wavevectorIsing1a}
\end{eqnarray}
and we obtain the magnetization of the root $x^{(0)}$ as
\[
 \left\langle {{\sigma}_{0}} \right\rangle =\underset{n\to \infty }{\mathop{\lim
 }}\,\frac{Z_{+}^{(n)}-Z_{-}^{(n)}}{Z_{+}^{(n)}+Z_{-}^{(n)}}.
 \]
%Here we use numerical methods to study the behaviors of the system
%given in equation \eqref{partition1a}.
Here, by using numerical calculations we investigate the behavior
of the system given in \eqref{partition1ab}.
%Lastly, we consider the variation of the wave-vector with
%temperature. A definition of the wave-vector that is convenient for
%numerical purposes is
%\[
%q=\underset{N\to \infty }{\mathop{\lim
%}}\,(\frac{1}{2}\frac{n}{N})
%\]
%where $n$ is the number of times the magnetization
%\eqref{prolonged-tenary1}  changes sign during $N$ successive
%iterations \cite{Vannimenus}.
Finally, we investigate how the wave-vector changes with temperature.
A numerically convenient definition of the wave-vector is
\[
q=\underset{N\to \infty }{\mathop{\lim
}}\,\left(\frac{1}{2}\frac{n}{N}\right),
\]
where $n$ is the number of times that the magnetization
\eqref{wavevectorIsing1a} changes the sign during $N$ successive
iterations \cite{Vannimenus}.
\begin{figure} [!htbp]
	%\label{wavevectorsP}
\centering
\includegraphics[width=60mm]{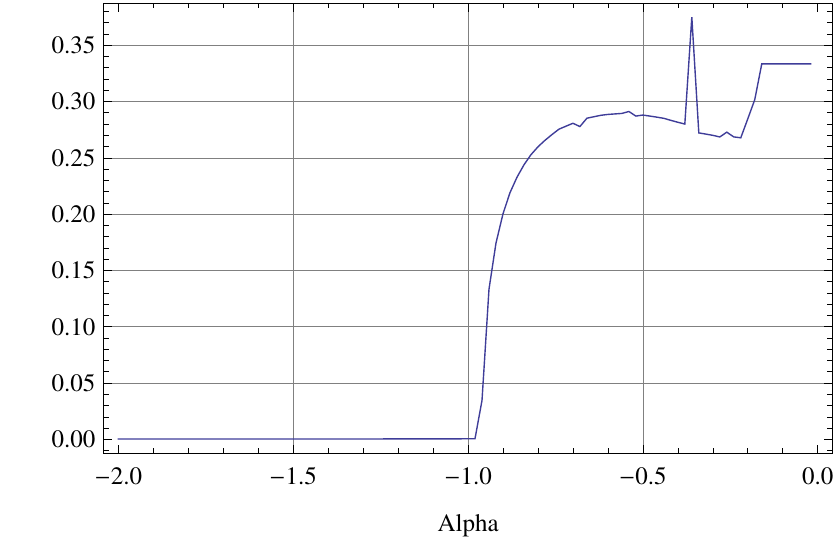}\ \ \ \
\includegraphics[width=60mm]{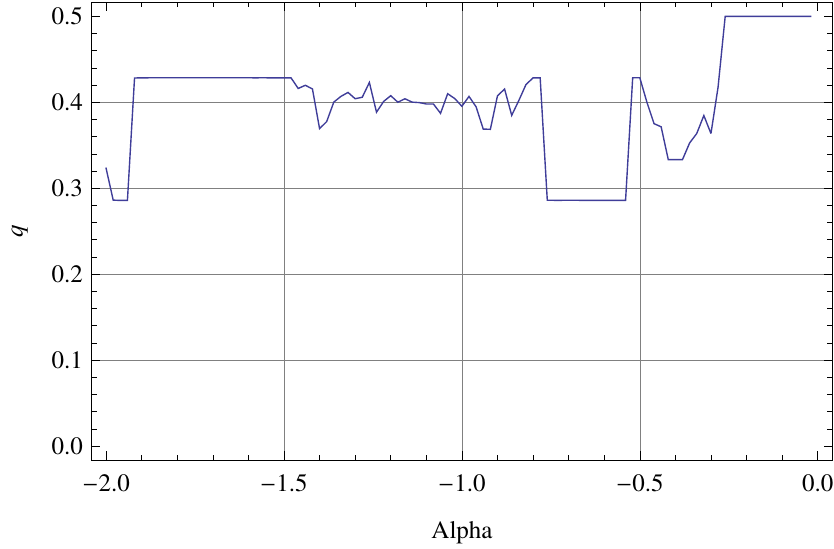}
\caption{(Colour online) Change of the wave-vectors for $\beta
=0.466$, $k=2$ (left-hand) and $\beta =-0.23$,  $k=4$ (right-hand) (see
\cite{AGUT2012ACTA}).}\label{wavevectorsP}
\end{figure}

The magnetization wave-vector varies continuously with the
temperature and the order of the Cayley tree. Figure
\ref{wavevectorsP} shows typical graphs of $q$ versus $T$. In
figure \ref{prolonged-tenary1}, for $k=2$, the point $(\beta
,\alpha )=(-0.45,-0.15)$ is the multicritical Lifshits point. The
graphs of $q$ versus $T/{{J}_{1}}$ are changing very fast around
$(-1.25,0)$. In figure \ref{prolonged-tenary1}, for $k=4$ in the
fourth region, there is a modulated phase and the island-shaped
region is quite narrow. The analysis of such general phase
diagrams is quite complicated. Therefore, the Lyapunov exponents
must be examined in detail. Moreover, for a more detailed study
of variation of the wave-vector~$q$ versus $T/{{J}_{1}}$, %we need
%to locate the main locking steps that should be present regarding
%to the general theory \cite{Vannimenus}. Because the intervals are
%very narrow, description of the distinction.
we need to find the main locking steps that should be present in
accordance with the general theory given by Vannimenus
\cite{Vannimenus}. %Because the intervals are so narrow, a
%description of the distinction is required.

In figures \ref{wavevectorsP}, we describe the variation of
wave-vectors for some values of interactions in the Ising model,
which is determined by the Hamiltonian given in
\eqref{Prolonged-hm}. As an important result in our phase
diagrams, if the Hamiltonian is added to the next nearest neighbor with the triple prolonged
(PNNN), the paramagnetic regions
disappear at  low temperatures \cite{GAUT2011b,AGUT2012ACTA}.
In the phase diagram obtained by Vannimenus, Lifshitz points
called ``multi-critical points'' are obtained at zero temperatures.
However, in our new phase diagrams, Lifshitz points appear at
non-zero temperatures. We first plotted the phase diagrams
corresponding to the Ising system of triple interaction on a
two-stage Cayley tree \cite{GAUT2011Chaos}. These phase diagrams
obtained are much richer  compared to the previous ones, and
the Lifshitz points have more different characteristics. In
\cite{GU2011PhysicaA}, the authors plotted the phase diagrams of
Ising model determined by Hamiltonian  on a Cayley tree of
arbitrary order, and thus the wave-vectors for some values were drawn.
The Ising models with three interactions have been tested to
examine the variation of wave-vectors and to plot them in three
parameters. Two-dimensional graphs have been successfully drawn
for some constants of one of these parameters. In
\cite{AGUT2012ACTA}, for particular values of the coupling
constants and $T$,  we have plotted the magnetization graphs of
the nonlinear-dynamical system.
\subsection{The study of the Lyapunov exponent}\label{Lyapunov exponent}
As stated in \cite{AGUT2012ACTA}, the stability analysis for the
ferromagnetic phase is more difficult than that of the
paramagnetic phase because the fixed point is not the same for the
whole phase. The Lyapunov exponent for the models associated to
the given Hamiltonians was calculated numerically (see
\cite{UGAT2012IJMPC}).  In addition, we demonstrated that every
region of negative $\lambda$ corresponds to the stability domain
of a given cycle~\cite{UGAT2012IJMPC}. Since the intervals are
very narrow, the description of the distinction between long-periodic
cycles (commensurate or incommensurate phases) and calculations
are difficult to perform. The answer to our problem consists of nonlinear
analysis techniques such as Lyapunov exponents corresponding to
the trajectory of the dynamical system
\cite{UGAT2012IJMPC,AGUT2012ACTA}.
%For more detailed properties of $q(T)$, this is necessary to
%locate the main locking steps that must be present according to
%the general theory \cite{Vannimenus}. These intervals may be very
%narrow, moreover the distinction between long-periodic cycles and
%truly non-periodic solutions are difficult to obtain numerically.

The positive values of Lyapunov exponents show the chaotic
structures of the phases \cite{Tome1989}. In other words, the
Lyapunov exponents of the map show the presence of chaotic
structures, associated with a strange attractor \cite{Moreira}.
Moreover, Lyapunov exponents support the existence of a region of
chaotic phases, associated with a strange attractor of the map
\cite{Moreira}. Vannimenus \cite{Vannimenus} has numerically shown
that the error bars indicate the fluctuations in the last hundred
out of 10~000 iteration steps. Lyapunov exponents measure the mean
exponential rate of convergence or divergence of trajectories
surrounding the attractors. By ensuring the accuracy of the
graphics obtained for the Lyapunov exponents of the works in the
literature, the Lyapunov exponents of the new models were drawn
\cite{UGAT2012IJMPC,AGUT2012ACTA}. Pasquini et al.~\cite{Pasquini1995} have characterized the formal expression of
the Lyapunov exponents via the convergence of the map towards a
fixed point before performing the thermodynamic limit, in a
restricted ensemble of disorder realizations. Furthermore, Pasquini et
al. \cite{Pasquini1995} determined the analytic upper and lower
bounds for the Lyapunov exponents of the product of random
matrices in the one-dimensional disordered Ising model.

Let us consider a one-dimensional map $f:\mathbb{R}\rightarrow
\mathbb{R}$. Let $x,x+\delta\in \mathbb{R}$. Under the first
iteration of the function $f$, if the distance between $f(x)$ and
$f(x+\delta)$ changes exponentially, then we have
$$
d(f(x),f(x+\delta))=\re^{\lambda}d(x,x+\delta).
$$
After $n$th iteration, one gets
\begin{equation}\label{Lyapunov-1D2}
d(f^n(x),f^n(x+\delta))=\re^{n\lambda}d(x,x+\delta).
\end{equation}
From \eqref{Lyapunov-1D2}, we have
\begin{equation*}\label{Lyapunov-1D}
\lambda =\frac{1}{n}\ln
\left(\frac{d(f^n(x),f^n(x+\delta))}{\delta}\right).
\end{equation*}
If the limit
$$
\lambda =\lim_{n\rightarrow \infty}\frac{1}{n}\ln
\left(\frac{d(f^n(x),f^n(x+\delta))}{\delta}\right)
$$
exists as $\delta\rightarrow 0$, the quantity $\lambda$ is called
Lyapunov exponent of the function $f$.

In  \cite{UGAT2012IJMPC,AGUT2012ACTA}, we were able to
establish a formal expression of the Lyapunov exponents in terms
of the approach by Vannimenus \cite{Vannimenus}.

\begin{defin} Consider a differentiable vector-valued function
$$\mathbf{F}(\mathbf{x})=({{F}_{1}}(\mathbf{x})....{{F}_{n}}(\mathbf{x})),
$$
where $\mathbf{F}:{{\mathbb{R}}^{m}}\to {{\mathbb{R}}^{n}}$ and
${{F}_{i}}:{{\mathbb{R}}^{m}}\to {{\mathbb{R}}}$. The Jacobian
matrix of $\mathbf{F}$ is a $n\times m$ matrix:
\begin{equation}
\mathbf{DF}\equiv \frac{\rd\mathbf{F}}{\rd\mathbf{x}}=
 \left( \begin{matrix}
   \frac{\rd{{F}_{1}}}{\rd{{x}_{1}}} & \cdots  & \frac{\rd{{F}_{1}}}{\rd{{x}_{m}}}  \\
   \vdots  & \vdots  & \vdots   \\
   \frac{\rd{{F}_{n}}}{\rd{{x}_{1}}} & \cdots  & \frac{\rd{{F}_{n}}}{\rd{{x}_{m}}}  \\
\end{matrix} \right).
\end{equation}
\end{defin}
In this subsection, in order to illustrate the stability of the
fixed point $({{x}^{*}},0,0)$, we consider the following
Hamiltonian (see \cite{UGAT2012IJMPC}):
\begin{equation}\label{Prolonged-hm22}
H(\sigma )=-{{J}_{{{t}_{p}}}}\sum\limits_{\overset{}
{\mathop{>x,y,z<}}\,}{}\sigma (x)\sigma (y)\sigma
(z)-{{J}_{p}}\sum\limits_{>x,y<}{\sigma (x)\sigma
(y)}-{{J}_{1}}\sum\limits_{<x,y>}{}\sigma (x)\sigma (y),
\end{equation}
where the first sum encompasses all prolonged ternary next-nearest
neighbors, the second sum encompasses all prolonged next-nearest
neighbors, and the third sum encompasses all nearest neighbors,
and $J_{t_{p}}, J_{p}, J_1\in \mathbb{R}$ are coupling constants.

Obviously, if $k=2$ and $J_{t_{p}}=0$, the Hamiltonian
\eqref{Prolonged-hm22} will be the same as the Hamiltonian
discussed by Vannimenus \cite{Vannimenus}. In the case
${{J}_{{{t}_{p}}}}\times J_1\neq 0$ and ${{J}_{p}}=0$, the model
has been investigated by Ak\i n et al.~\cite{AGUT2012ACTA}.

In \cite{UGAT2012IJMPC}, we derived the recurrent equations
associated to the Hamiltonian \eqref{Prolonged-hm22} with
${{J}_{{{t}_{p}}}}\times J_1\times{{J}_{p}}\neq 0$ as follows. For
convenience, we assume
%
%%\begin{equation}\label{IJMPC-rec2}
%%\begin{array}{rcl}
%%x^{(n+1)}&=&f(x^{(n)},y_1^{(n)},y_2^{(n)})=\frac{1}{a^2D}[(b^2 (x^{(n)}-y_2^{(n)})+ c^2(1-y_1^{(n)}))^k+(1+y_1^{(n)}+b^2 c^2(x^{(n)}+y_2^{(n)}))^k]\\[3mm]
%%y^{(n+1)}_1&=&g(x^{(n)},y_1^{(n)},y_2^{(n)})=\frac{2}{D}[(b^2 c^2(1+y_1^{(n)})+x^{(n)}+y_2^{(n)})^k-( c^2(x^{(n)}-y_2^{(n)})+b^2(1-y_1^{(n)}))^k] \\[3mm]
%%y^{(n+1)}_2&=&h(x^{(n)},y_1^{(n)},y_2^{(n)})=\frac{1}{a^2D}[(b^2(x^{(n)}-y_2^{(n)})+
%%c^2(1-y_1^{(n)}))^k-(1+y_1^{(n)}+b^2c^2(x^{(n)}+y_2^{(n)}))^k]
%%\end{array}
%%\end{equation}
%%\normalsize with
%%\begin{eqnarray*}
%%D&=&[b^2 c^2(1+y_1^{(n)})+x^{(n)}+y_2^{(n)}]^k+[
%%c^2(x^{(n)}-y_2^{(n)})+b^2(1-y_1^{(n)})]^k,
%%\end{eqnarray*}
\begin{eqnarray}\label{IJMPC-rec2a}
x_{n+1}&=&%F_{1}(x_{n},y_{n},z_{n})=
\frac{(b^2 (x_{n}-z_{n})+ c^2(1-y_{n}))^k+(1+y_{n}+b^2
c^2(x_{n}+z_{n}))^k}{a^2\left((b^2 c^2(1+y_{n})+x_{n}+z_{n})^k+(
c^2(x_{n}-z_{n})+b^2(1-y_{n}))^k\right)},\\\label{-IJMPC-rec2a}
y_{n+1}&=&%F_{2}(x_{n},y_{n},z_{n})=
\frac{2((b^2 c^2(1+y_{n})+x_{n}+z_{n})^k-(
c^2(x_{n}-z_{n})+b^2(1-y_{n}))^k)}{(b^2
c^2(1+y_{n})+x_{n}+z_{n})^k+(c^2(x_{n}-z_{n})+b^2(1-y_{n}))^k},
\\\label{IJMPC-rec2b}
z_{n+1}&=&%F_{3}(x_{n},y_{n},z_{n})=
\frac{((b^2(x_{n}-z_{n})+
c^2(1-y_{n}))^k-(1+y_{n}+b^2c^2(x_{n}+z_{n}))^k)}{a^2\left((b^2
c^2(1+y_{n})+x_{n}+z_{n})^k+(
c^2(x_{n}-z_{n})+b^2(1-y_{n}))^k\right)},\label{IJMPC-rec2c}
\end{eqnarray}
%with
%\begin{eqnarray*}
%D&=&[b^2 c^2(1+y_{n})+x_{n}+z_{n}]^k+[
%c^2(x_{n}-z_{n})+b^2(1-y_{n})]^k,
%\end{eqnarray*}
where $a=\re^{(J_1/T)}$, $b=\re^{(J_p/T)}$, $c=\re^{(J_{t_p}/T)}$,
$x_{n+1}=F_{1}(x_{n},y_{n},z_{n})$,
$y_{n+1}=F_{2}(x_{n},y_{n},z_{n})$ and
$z_{n+1}=F_{3}(x_{n},y_{n},z_{n})$ (see \cite{UGAT2012IJMPC} for
details).
%The calculation of the Lyapunov exponent can be given as follows.
%The iterative equations \eqref{IJMPC-rec2a}-\eqref{IJMPC-rec2c}
%are linearized around the successive points of the trajectory,
%giving linear iterative equations for the perturbations.

Linearizing the iterative equations
\eqref{IJMPC-rec2a}--\eqref{IJMPC-rec2c} around the subsequent
points of the trajectory yields linear iterative equations for the
perturbations. The Lyapunov exponents can be calculated in the
following way.\\
Let us consider the operator $\mathbf{F}:{{\mathbb{R}}^{3}}\to
{{\mathbb{R}}^{3}}$ defined by
\begin{equation}\label{IJMPC-rec2-fun1}\mathbf{F}(x_{n+1},y_{n+1},z_{n+1})=
[{{F}_{1}}({{x}_{n}},{{y}_{n}},{{z}_{n}}),{{F}_{2}}({{x}_{n}},{{y}_{n}},{{z}_{n}}),{{F}_{3}}({{x}_{n}},{{y}_{n}},{{z}_{n}})].
\end{equation}
Now, we  linearize the operator \eqref{IJMPC-rec2-fun1} around
the successive points of the trajectory $(x_{1},y_{1},z_{1})$,...,
$(x_{k},y_{k},z_{k})$ satisfying the recurrence equations
\label{IJMPC-rec2} for the perturbations $(\delta x_{n},\delta
y_{n},\delta z_{n})$.

From the definition of differentiability in multivariable
calculus, we have \arraycolsep=1.5pt\begin{eqnarray*}
x_{n{+}1}{-}F_1(x_k,y_k,z_k)&=&\frac{\partial
F_1(x_n,y_n,z_n)}{\partial x_n}(x_n-x_k)
+\frac{\partial F_1(x_n,y_n,z_n)}{\partial y_n}(y_n-y_k){+}\frac{\partial F_1(x_n,y_n,z_n)}{\partial z_n}(z_n-z_k),\\
{{y}_{n{+}1}}{-}{{F}_{2}}({{x}_{k}},{{y}_{k}},{{z}_{k}})&=&\frac{\partial
{{F}_{2}}({{x}_{n}},{{y}_{n}},{{z}_{n}})}{\partial
{{x}_{n}}}({{x}_{n}}{-}{{x}_{k}})+\frac{\partial
{{F}_{2}}({{x}_{n}},{{y}_{n}},{{z}_{n}})}{\partial
{{y}_{n}}}({{y}_{n}}{-}{{y}_{k}}) +\frac{\partial
{{F}_{2}}({{x}_{n}},{{y}_{n}},{{z}_{n}})}{\partial
{{z}_{n}}}({{z}_{n}}{-}{{z}_{k}}), \\
{{z}_{n+1}}{-}{{F}_{3}}({{x}_{k}},{{y}_{k}},{{z}_{k}})&=&\frac{\partial
{{F}_{3}}({{x}_{n}},{{y}_{n}},{{z}_{n}})}{\partial
{{x}_{n}}}({{x}_{n}}{-}{{x}_{k}})+\frac{\partial
{{F}_{3}}({{x}_{n}},{{y}_{n}},{{z}_{n}})}{\partial
{{y}_{n}}}({{y}_{n}}{-}{{y}_{k}})+\frac{\partial
{{F}_{3}}({{x}_{n}},{{y}_{n}},{{z}_{n}})}{\partial
{{z}_{n}}}({{z}_{n}}{-}{{z}_{k}}).
\end{eqnarray*} We obtain the
Jacobian matrix as follows:
\[
L_n=\left( \begin{matrix}
\frac{\partial {{F}_{1}}({{x}_{n}},{{y}_{n}},{{z}_{n}})}{\partial {{x}_{n}}}
& \frac{\partial {{F}_{1}}({{x}_{n}},{{y}_{n}},{{z}_{n}})}{\partial {{y}_{n}}}
& \frac{\partial {{F}_{1}}({{x}_{n}},{{y}_{n}},{{z}_{n}})}{\partial {{z}_{n}}}\\
\frac{\partial {{F}_{2}}({{x}_{n}},{{y}_{n}},{{z}_{n}})}{\partial {{x}_{n}}}
& \frac{\partial {{F}_{2}}({{x}_{n}},{{y}_{n}},{{z}_{n}})}{\partial {{y}_{n}}}
& \frac{\partial {{F}_{2}}({{x}_{n}},{{y}_{n}},{{z}_{n}})}{\partial {{z}_{n}}}\\
\frac{\partial {{F}_{3}}({{x}_{n}},{{y}_{n}},{{z}_{n}})}{\partial {{x}_{n}}}
& \frac{\partial {{F}_{3}}({{x}_{n}},{{y}_{n}},{{z}_{n}})}{\partial {{y}_{n}}}
& \frac{\partial {{F}_{3}}({{x}_{n}},{{y}_{n}},{{z}_{n}})}{\partial {{z}_{n}}}\\
\end{matrix} \right).
\]
Therefore, in a matrix form, we have
\begin{equation}\label{IJMPC-rec2-Lya}
V_{n+1}=\left(
\begin{array}{c}
 \delta x_{n+1}\\
 \delta y_{n+1}\\
 \delta z_{n+1}\\
\end{array}\right)
=L_n\left(
\begin{array}{c}
 \delta x_n\\
 \delta y_n \\
 \delta z_n \\
\end{array}\right),
\end{equation}
where the matrix $L_n$ is determined by the iteration step.

It is clear that there may be a deviation for $V_{0},V_{1},\ldots,
V_{9999}$, but after the term $V_{9999}$  the deviation will
decrease. Starting from initial conditions
\[
{{x}_{1}}=\frac{{{a}^{2k}}+{{c}^{2k}}}{a({{(bc)}^{2k}}+1)},\quad
{{y}_{1}}=\frac{{{(bc)}^{2k}}-1}{{{(bc)}^{2k}}+1},\quad
{{z}_{1}}=\frac{{{(bc)}^{2k}}-1}{{{a}^{2}}({{(bc)}^{2k}}+1)}
\]
under the boundary condition ${{\bar{\sigma }}^{(n)}}(V\backslash
{{V}_{n}})\equiv +1$, we iterate the recursive equations
\eqref{IJMPC-rec2a}--\eqref{IJMPC-rec2c} and observe the behavior
of these recursive equations after sufficiently large iterations.

Thus, the Lyapunov exponent $\lambda$ is defined by
\begin{equation}\label{Lyapunov22}
\lambda =\underset{n\to \infty }{\mathop{\lim }}\,\frac{1}{n}\log
(||{{V}_{n}}||) =\underset{n\to \infty }{\mathop{\lim
}}\,\frac{1}{n}\log \left( \sqrt{{{(\delta {{x}_{n}})}^{2}}+{{(\delta
{{y}_{n}})}^{2}}+{{(\delta {{z}_{n}})}^{2}}}\right) .
\end{equation}
When $\lambda$ is small, the incommensurate phases evolve smoothly
because the widths of the steps are very small. Though there are
tiny complete parts of the devil's staircase at the edge of each
step, they are undistinguishable \cite{Aubry1982}.
\begin{figure}[!htbp]
\centering
\includegraphics[width=55mm]{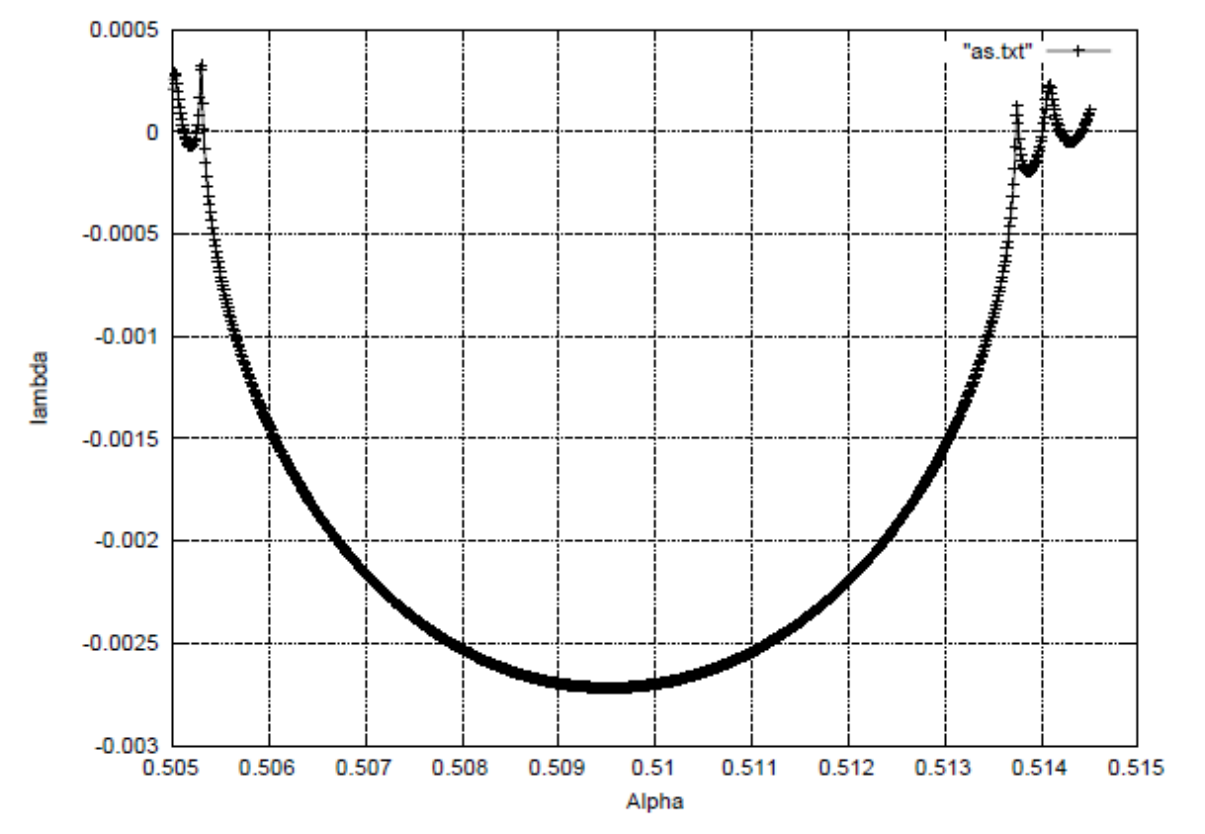}\ \ \ \ \ \
\includegraphics[width=55mm]{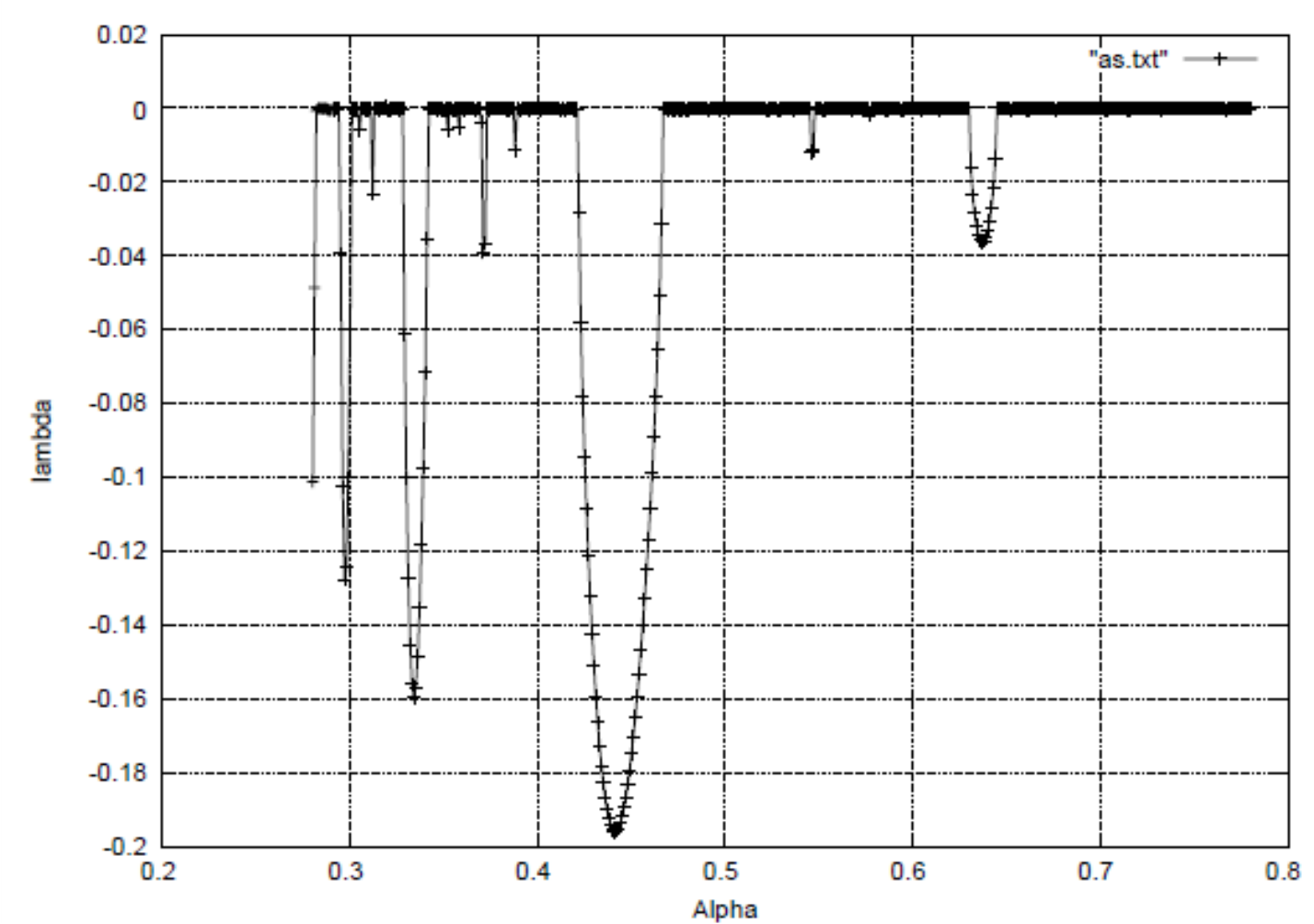}
\caption{Left-hand figure: The Lyapunov exponent $\lambda$ in the region
of the commensurate step $q=1/6$, for $\beta =0.36$ and $\gamma=0$
with $k=2$ corresponding to figure 6 given in \cite{UGAT2012IJMPC}. Right-hand figure: Graph of the Lyapunov exponent
$\lambda$: $\beta=0.23$ and $\gamma=0$ with $k=4$ corresponding to
figure~4 in \cite{UGAT2012IJMPC}.}\label{Lyapunov1}
\end{figure}

%This exponent plays for a general attractor the role that the
%logarithm of the largest eigenvalue plays for a simple fixed
%point. It is also seen numerically that $\lambda$ never takes
%significantly positive values \cite{UGAT2012IJMPC} (see Figures in
%\ref{Lyapunov1}). Figures in \ref{Lyapunov1} have been borrowed
%from the Ref. \cite{UGAT2012IJMPC}.

The logarithm of the biggest eigenvalue of the matrix plays the
same role for a generic attractor as it does for a basic fixed
point. It can also be seen numerically that $\lambda$ is never
considerably positive \cite{UGAT2012IJMPC} (see figures in
\ref{Lyapunov1}). The figures in \ref{Lyapunov1} were taken from
the reference \cite{UGAT2012IJMPC}. Non-zero Lyapunov exponent
quantifies the exponential rate of growth or  decay of the
initial perturbation obtained as a time
average with $t\rightarrow \infty$. %The vanishing of the Lyapunov
%exponent means that the set of trajectories is quasi-continuous,
%or in other terms that it has a zero frequency "phason" mode
%\cite{Vannimenus} (see Ref. \cite{Aubry1982} for details).
The Lyapunov exponent vanishing indicates that the set of
trajectories is quasi-continuous, or in other words, that it has a
zero frequency ``phason'' mode \cite{Vannimenus} (for more
information, see \cite{Aubry1982}).

The Lyapunov exponent also indicates whether an infinitesimal
change in the beginning conditions will have an infinitesimal
influence (negative exponent) or will result in a completely
different trajectory (positive exponent) \cite{Vannimenus}. For
the Ising model corresponding to the Hamiltonian
\eqref{Prolonged-hm}, the paramagnetic region in phase diagrams
disappeared, because the limit
\[
\underset{n\to \infty
}{\mathop{\lim}}\,({{x}^{(n)}},y_{1}^{(n)},y_{2}^{(n)})=({{x}^{*}},0,0)
\]
is not satisfied.
%We have seen that unlike the Potts model if the prolonged
%next-nearest neighbors $>x,y<$ is added the Hamiltonian then the
%paramagnetic region in the phase diagrams corresponding to the
%Ising model appears.

\subsection{The stability of the paramagnetic phase} 

To obtain the stability limit
surface of the paramagnetic phase, we study the stability of
paramagnetic phase. Thus, we need to linearize equations
\eqref{IJMPC-rec2a}--\eqref{IJMPC-rec2c} around the fixed point
$(x^{*}, 0, 0)$, where $x^{*}$ is given by
\begin{equation}\label{fixed22}
x^{*}=f(x^{*},0,0)=\frac{{{({{b}^{2}}x^{*}+{{c}^{2}})}^{k}}+{{(1+{{b}^{2}}{{c}^{2}}x^{*})}^{k}}}
{{{a}^{2}}{{({{b}^{2}}{{c}^{2}}+x^{*})}^{k}}+{{({{c}^{2}}x^{*}+{{b}^{2}})}^{k}}}.
\end{equation}
In general, the behavior of solutions to a linear or linearized system
can be predicted based on the eigenvalues of the Jacobian matrix.
Negative eigenvalues cause a solution to tend towards zero,
positive eigenvalues cause the solution to approach infinity,
and imaginary eigenvalues cause a spiral behavior of solutions.%\footnote{www}
\begin{thm}(Liapunov’s Theorem). Let $f :\mathbf{R}^{n}\rightarrow \mathbf{R}^{n}$
be $C^1$ and $x_0\in\mathbf{R}^{n}$ be a fixed point of $x' =
f(x)$ (so $f(x_0) = 0)$. Let $A = Df(x_0)$ be the linearization of
$f$ (so $A_{ij} =\partial f_i/\partial x_j$ is the Jacobian
matrix) and $\lambda_1, \ldots , \lambda_n$ be its eigenvalues.
Then, $x_0$ is
\begin{itemize}
\item asymptotically stable if $Re \lambda_i < 0$ for all $i = 1,
\ldots, n$; \item  unstable if $Re \lambda_i > 0$ for some $i$.
\end{itemize}
\end{thm}
Note that if the real parts of the eigenvalues are all zero, then
further analysis is necessary. Let us consider the equations
\eqref{-IJMPC-rec2a} and \eqref{IJMPC-rec2c}, we have
\begin{eqnarray}\label{JacobiMat}
J &=&\left(\begin{matrix}
\frac{\partial {{F}_{2}}({{x}_{n}},{{y}_{n}},{{z}_{n}})}{\partial {{y}_{n}}} & \frac{\partial {{F}_{2}}({{x}_{n}},{{y}_{n}},{{z}_{n}})}{\partial {{z}_{n}}}  \\
\frac{\partial {{F}_{3}}({{x}_{n}},{{y}_{n}},{{z}_{n}})}{\partial {{y}_{n}}} & \frac{\partial {{F}_{3}}({{x}_{n}},{{y}_{n}},{{z}_{n}})}{\partial {{z}_{n}}}  \\
\end{matrix} \right),
\end{eqnarray}
where
{\small{\small\begin{eqnarray*}\frac{\partial
F_2(x_n,y_n,z_n)}{\partial y_n}&=&
 \frac{4b^2k\left(b^2c^2+x^*\right)^{k-1}\left(b^2+
 c^2x^*\right)^{k-1}\left(2b^2c^2+x^*+c^4x^*\right)}
 {\left(\left(b^2c^2+x^*\right)^k+\left(b^2+c^2x^*\right)^k\right)^2},\\
 \frac{\partial F_2(x_n,y_n,z_n)}{\partial z_n}&=&
 \frac{4k\left(b^2c^2+x^*\right)^{k-1}\left(b^2+
 c^2x^*\right)^{k-1}\left(b^2+b^2c^4+2c^2x^*\right)}
 {\left(\left(b^2c^2+x^*\right)^k+\left(b^2+c^2x^*\right)^k\right)^2},\\
 \frac{\partial F_3(x_n,y_n,z_n)}{\partial y_n}&=&
 \frac{4b^2k\left(b^2c^2+x^*\right)^{k-1}\left(b^2+c^2x^*\right)^{k-1}
 \left(2b^2c^2+x^*+c^4x^*\right)}
 {\left(\left(b^2c^2+x^*\right)^k+\left(b^2+c^2x^*\right)^k\right)^2},\\
 \frac{\partial F_3(x_n,y_n,z_n)}{\partial z_n}&=&
 \frac{4k\left(b^2c^2+x^*\right)^{k-1}\left(b^2+c^2x^*\right)^{k-1}
 \left(b^2+b^2c^4+2c^2x^*\right)}{\left(\left(b^2c^2+
 x^*\right)^k+\left(b^2+c^2x^*\right)^k\right)^2}.
 \end{eqnarray*}}}

It is clear that there exists at most two eigenvalues of the
matrix \eqref{JacobiMat}. In this case, we have two different
situations.\\
Let $\lambda_1, \lambda_2$ be eigenvalues of the matrix
\eqref{JacobiMat}. \begin{itemize}
    \item If $\lambda_1, \lambda_2\in \mathbf{R}$,
    then  the corresponding phase is either paramagnetic or ferromagnetic.
    \item If $\lambda_1=\overline{\lambda_2}\in \mathbf{C}$, then the fixed point is approached in an oscillatory way and the
stability limit (which coincides with the critical limit for the
second order phase transitions) is achieved if we consider the
modulus of $\lambda$ equal to unity \cite{MTA1985a}.
\end{itemize}

\subsubsection{The Para-Ferro Transition}\label{The Para-Ferro Transition}

The variable $x_n$ is unaffected in the first order in $y_n$ and
$z_n$, and in a matrix form, the linearized equations are given by
\begin{equation}\label{stabiliyt2}
\left(
\begin{array}{c}
 y_{n+1} \\
 z_{n+1}
\end{array}
\right)=2L_n\left(
\begin{array}{c}
 y_n \\
 z_n
\end{array}
\right).
\end{equation}
In this case, then the corresponding matrix is given
by
\[L_n=\left(\begin{matrix} \frac{\partial
{{F}_{2}}({{x}_{n}},{{y}_{n}},{{z}_{n}})}{\partial {{y}_{n}}}
& \frac{\partial {{F}_{2}}({{x}_{n}},{{y}_{n}},{{z}_{n}})}{\partial {{z}_{n}}} \\
\frac{\partial {{F}_{3}}({{x}_{n}},{{y}_{n}},{{z}_{n}})}{\partial {{y}_{n}}}
& \frac{\partial {{F}_{3}}({{x}_{n}},{{y}_{n}},{{z}_{n}})}{\partial {{z}_{n}}}\\
\end{matrix} \right)=\left(
\begin{array}{cc}
 \frac{b^2}{b^2+x} & \frac{1}{b^2+x} \\
 -\frac{x}{1+b^2 x} & -\frac{x b^2}{1+b^2 x}
\end{array}
\right).\] For $c=1$, if the eigenvalues of the matrix $L_n$ in
\eqref{stabiliyt2} are real, the situation occurs when this
largest eigenvalue is equal to 1:
$$
3x^2+\left( \frac{5}{b^2}-3b^2\right) x-1=0,
$$
with $x$ positive. As a result, the equation for the para-ferro
transition line is $a=f(b)$, with $a=\exp(J_1/T), b=\exp(J_p/T)$
(see \cite{Vannimenus} for details).
\subsubsection{The Para-Modulated Transition}\label{The Para-Modulated Transition}

The fixed point is approached in an oscillatory manner if the
eigenvalues of the matrix $L_n$ in \eqref{stabiliyt2} with $c=1$
are complex. This is the regular lattice's analogue of a tree of a
periodic perturbation, where $q$ is the modulation wave-vector
\cite{Vannimenus}.
\subsection{Strange Attractor}\label{Strange Attractor}

Why do we investigate a strange attractor and what is the
contribution? As examined in the section \ref{Dynamical behavior
of Ising model on CT}, by means of the recursion relations
associated with a given Hamiltonian we can plot the (numerically)
exact phase diagram of the model by fixing some given parameters
$\alpha, \beta, \gamma$ in the $(\alpha, \beta, \gamma)$ space.
Yokoi et al.~\cite{Yokoi} provided substantial numerical evidence
for chaotic phases linked with strange attractors for the Ising
model with competing interactions on a Cayley tree. One may
undertake a more extensive classification of these modulated
phases by considering the strange attractors of the general
modulated phases. One also computes the Lyapunov exponents to
detect the presence of chaotic structures corresponding to a
strange attractor  of the map \cite{Moreira}.

In addition, each phase is defined by a specific attractor in the
$(X_r,K_r)$ plane, and the phase diagram is created by tracking
the evolution of these attractors and detecting qualitative
changes. Ganikhodjaev and Nawi \cite{GR-AIP2013-SA} investigated
the stange attractors associated with the antiphase with period 4,
the antiferromagnetic phase with period 2 and the paramagnetic
phase for an Ising model. The fractal character of the attractor
associated with the chaotic phase was confirmed by the calculation
of the Hausdorff dimension (see \cite{Tragtenberg}).

By summing over successive shells from the outermost ($r = 1$) to
innermost $(r = N)$ shell, Inawashiro et al. \cite{Inawashiro}
obtained recurrence relations for effective fields $X_r$ and
NN-interactions $K_r$ which are rigorous, but which must in
general be analyzed numerically. In order to plot strange
attractors with a given Ising model, we consider the approach
given in ref. \cite{Inawashiro}.

In this context, graphs of some relevant strange attractors
associated to Ising model on Cayley tree of arbitrary order have
been plotted \cite{GATU2013JPhysConfSer}. 
In \cite{Yokoi}, the occurrence of chaotic phases associated with
strange attractors is supported by extensive numerical evidence.
Moreover, Yokoi et al. \cite{Yokoi} performed calculations to show the
existence of a complete devil's staircase at low temperatures. In
\cite{GATU2013JPhysConfSer}, in the presence of an external
magnetic field, we investigated the Vannimenus model on an
arbitrary order Cayley tree with competing for nearest-neighbor
and next-nearest-neighbor interactions.
%we have studied the Vannimenus model on an arbitrary order Cayley
%tree with competing for nearest-neighbor and next-nearest-neighbor
%interactions in the presence of the external magnetic field.
We obtained robust recurrence relations for effective fields $X_r$
and nearest-neighbor interactions $K_r$, although they must be
studied numerically in general. The role of the external magnetic
field was clarified.

The phase diagram is obtained by watching the evolution of these
attractors and detecting qualitative changes. These changes might
be continuous or abrupt, indicating second- or first-order
transitions, respectively \cite{Yokoi}. Therefore, studying the
strange attractors is very important to characterize the phase
diagrams associated with the relevant models.

In \cite{Tragtenberg}, Tragtenberg and Yokoi  developed some
efficient numerical procedures for this purpose. They  also
investigated the existence of strange attractors on the model in
the presence of a field. In \cite{GR-AIP2013-SA}, 
the phase diagram of the recursive equations corresponding to this
model was investigated using an iterative scheme developed for a
renormalized effective nearest-neighbor coupling $K_r$ and
effective field per site $X_r$ for spins on the $n$-th level of a
Cayley tree with competing one-level $J_0$ and prolonged $J_p$
next-nearest-neighbor interactions between Ising spins on the
tree. Using the approach proposed by Inawashiro et al.~\cite{Inawashiro-T1983}, they were able to get an intermediate
range of $J_p/J_0$ values where $X_r$ and $K_r$ iterate to a
finite attractor in the $X-K$ plane.
\begin{figure} [!htbp]
	%\label{strange-attractor-1aa}
\centering
\includegraphics[width=145mm]{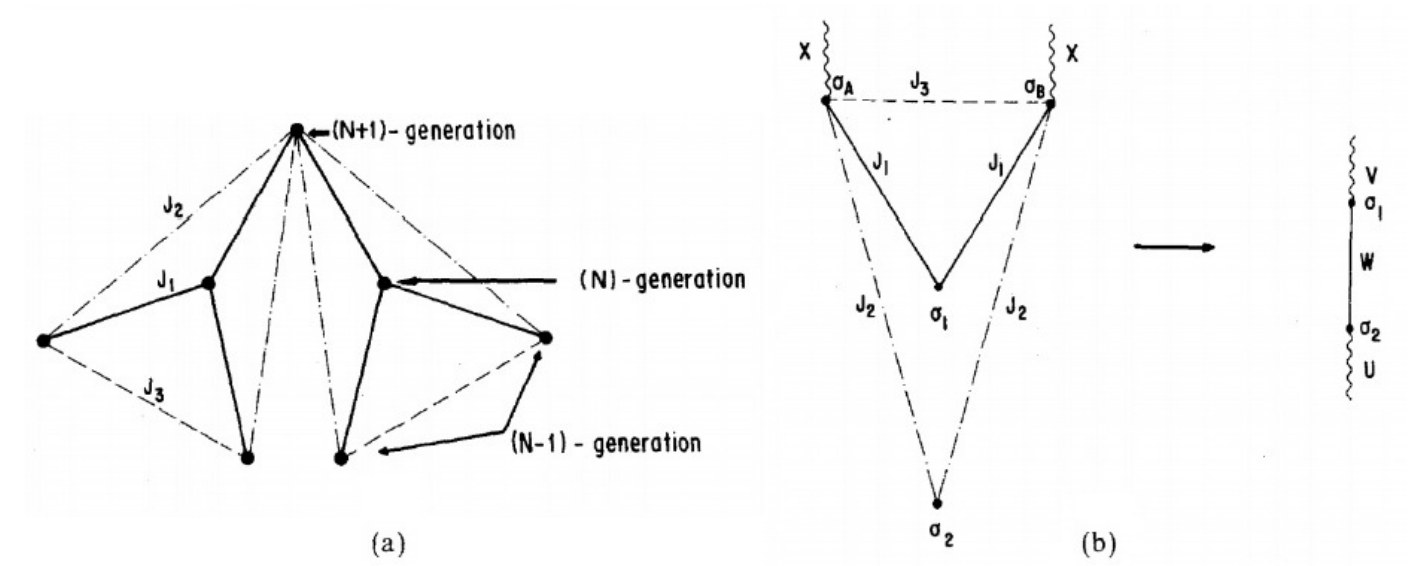}
\caption{(a) Two successive generations of a Cayley tree (dot-dash
line: prolonged second neighbor interactions; double dot-dash
line: one-level second neighbor interactions). (b) Schematic
diagram to illustrate the summation used in equation 2 (see
\cite{GR-AIP2013-SA}). }\label{strange-attractor-1aa}
\end{figure}

Let us consider the following Hamiltonian with spin values in
$\Phi=\{-1,+1\}$, the relevant Hamiltonian with competing
nearest-neighbor and prolonged next-nearest-neighbor binary
interactions
\begin{equation}\label{HM-SA}
H(\sigma )=-J_1\sum _{\langle x,y\rangle} \sigma (x)\sigma
(y)-J_2\sum _{\rangle x,y\langle} \sigma (x)\sigma
(y)-J_3\sum_{x\in V} \prod _{y\in S(x)} \sigma (y)-h\sum _{x\in V}
\sigma (x),
\end{equation}
where the sum in the first term ranges over all nearest neighbors,
the second sum ranges over  all prolonged
next-nearest-neighbors, the third sum ranges over all one-level
$k$-tuple neighbors and the spin variables $\sigma (x)$ assume the
values $\pm 1$. Here, $J_1, J_2, J_3, h\in \mathbf{R}$ are coupling
constants (see figure \ref{strange-attractor-1aa}).

In \cite{GATU2013JPhysConfSer}, we have obtained recursion
relations for effective fields $X_r$ and nearest-neighbor
interactions $K_r$, but they must be numerically investigated. The
role of the external magnetic field has been clarified.
%we have obtained recursion relations for effective fields $X_r$
%and nearest-neighbor interactions $K_r$ which are rigorous, but
%which must be analyzed numerically. We have clarified the role of
%external magnetic field (see Figure \ref{strange-attractor-1aa}).
We sum consecutively over spins to set up the iterative system, as
shown schematically in figure \ref{strange-attractor-1aa} (see~\cite{GATU2013JPhysConfSer}). After performing basic but
sophisticated algebraic operations, we obtained the following
equations;
\begin{eqnarray}
	&&\exp [W\sigma _1\sigma
_2+U\sigma _1+V\sigma _2]=\nonumber\\
&&\frac{1}{C}\sum _{\sigma
_{A_i}=\pm 1} \exp \left[(K_1\sigma _1+K_2\sigma _2)\sum _{i=1}^k
\sigma _{A_i}+K_3\sum _{i=1<j}^k \sigma _{A_i}\sigma _{A_j}+X\sum
_{i=1}^k \sigma _{A_i}\right],
\label{SAttractor-e1}
\end{eqnarray}
where $K_i\equiv J_i/(k_{\text{B}}T), i=1,2,3$ and we get
%\begin{eqnarray}\label{strange3.21}
%U(X,K_1,K_2,K_3)&\equiv& U= \frac{1}{4}\ln
%\frac{\omega(1,1)\omega(1,-1)}{\omega(-1,1)\omega(-1,-1)}\\
%V(X,K_1,K_2,K_3)&\equiv& V= \frac{1}{4}\ln
%\frac{\omega(1,1)\omega(-1,1)}{\omega(1,-1)\omega(-1,-1)}\\
%W(X,K_1,K_2,K_3)&\equiv& W= \frac{1}{4}\ln
%\frac{\omega(1,1)\omega(-1,-1)}{\omega(1,-1)\omega(-1,1)}\\
%C(X,K_1,K_2,K_3)&\equiv& C=
%[\omega(1,1)\omega(-1,-1)\omega(1,-1)\omega(-1,1)]^{
%\frac{1}{4}}\\\nonumber
%\end{eqnarray}
\begin{eqnarray}
\left\{
\begin{array}{l}
 U(X,K_1,K_2,K_3)\equiv U=\frac{1}{4}\ln \frac{\omega (1,1)\omega (1,-1)}{\omega (-1,1)\omega (-1,-1)}, \\
 V(X,K_1,K_2,K_3)\equiv V=\frac{1}{4}\ln \frac{\omega (1,1)\omega (-1,1)}{\omega (1,-1)\omega (-1,-1)}, \\
 W(X,K_1,K_2,K_3)\equiv W=\frac{1}{4}\ln \frac{\omega (1,1)\omega (-1,-1)}{\omega (1,-1)\omega (-1,1)}, \\
 C(X,K_1,K_2,K_3)\equiv C=\sqrt[4]{\omega (1,1)\omega (-1,-1)\omega (1,-1)\omega (-1,1)},
 \label{strange3.21}
\end{array}
\right.
\end{eqnarray}
where the value of  $\omega$ changes according to the oddness or
evenness of the order $k$ of the tree. Therefore, we take $k$ as
an odd  number, and we obtain
$$
\omega (\sigma _1,\sigma _2)=2\sum _{i=1}^{\frac{k-1}{2}}
C(k,i)\exp \left[\frac{k^2-4ki-k+i^2+3i}{2}K_3\right]\cosh
 \left[(k-2i)(K_1\sigma _1+K_2\sigma _2+X)\right],
$$
and if we take $k$ as an even number, we get the following
equation,
\begin{eqnarray*}
\omega (\sigma _1,\sigma _2)&=&2\sum _{i=1}^{\frac{k}{2}-1}
C(k,i)\exp  \left[\frac{k^2-4ki-k+i^2+5i}{2}K_3\right]\cosh
 \left[(k-2i)(K_1\sigma
_1+K_2\sigma _2+X)\right]\\
&+&C(k,k/2)\exp \left[\frac{3k(2-k)}{8}K_3\right].
\end{eqnarray*}
Starting with $X^{(1)} = B = h/(k_\text{B}T)$ and $K^{(1)}_1 = K_1 =
J_1/(k_\text{B}T)$, where $h$ is the initial applied magnetic field per
site and $J_1$ is the initial nearest-neighbor coupling, we obtain
from the above equalities the iteration scheme, for $r = 2,
3,\ldots , N,$
\begin{eqnarray}\label{SAtt3a}
X^{(r)}&=& B + k  U(X^{(r-2)},K^{(r-2)}_1,K_2,K_3) + V
(X^{(r-1)},K^{(r-1)}_1,K_2,K_3),\\
K^{(r)}_1&=&K_1 +W\left( X^{(r-1)},K^{(r-1)}_1,K_2,K_3\right) .\label{SAtt3b}
\end{eqnarray}

Using the $U, V$ and $W$ functions given in \eqref{strange3.21},
we have obtained recursive relations \eqref{SAtt3a} and~\eqref{SAtt3b}. Taking into account the initial conditions
$X^{(0)} = K^{(0)}_1 = 0$ and $X^{(1)} = B,K^{(0)}_1 =K_1$, the
graphs of the strange attractor are plotted.
\begin{figure} [!htbp]\label{strange-attractor-a}
\centering
\includegraphics[width=120mm]{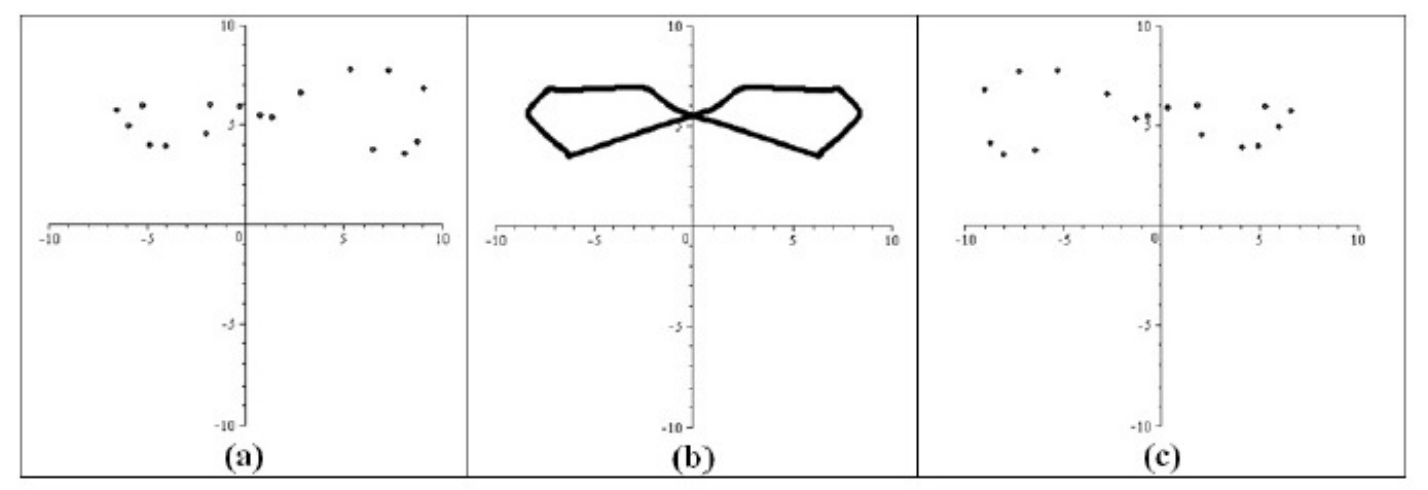}
\caption{The attractors in $X-K_1$ plane for $k = 2$,
 $\gamma=\frac{J_3}{J_1}=-0.38$, $\beta=\frac{J_2}{J_1}= 0.38$, and $\alpha=\frac{k_bT}{J_1}= 0.11$ with
external magnetic for (a) $\delta= \frac{h}{J_1}=+0.1$, (b)
$\delta= \frac{h}{J_1} = 0$ and (c) $\delta= \frac{h}{J_1} =-0.1.$
%(see \cite{GATU2013JPhysConfSer})
}\label{strange-attractor-a}
\end{figure}

We have obtained some attractor graphs in figures
\ref{strange-attractor-a} \cite{GATU2013JPhysConfSer}. The
corresponding phase is incommensurate for $h=0$, but it turns into
a commensurate phase with the presence of an external magnetic
field. Attractors in both figure \ref{strange-attractor-a} (a) and
figure \ref{strange-attractor-a} (c) reflect the commensurate
phase. We refer the readers to~\cite{GATU2013JPhysConfSer} for
detailed information.

\subsection{Notes and remarks}\label{Ising-remarks}
For the Cayley tree of finite order, the system of
nonlinear equations have been successfully derived in \cite{UGAT2012IJMPC,AGUT2012ACTA}. %In this sense, a total of three
%studies have been conducted.
When the prolonged next nearest coupling constant is added to the
Hamiltonian examined by Vannimenus instead of the next nearest
coupling constant \cite{GAUT2011b,AGUT2012ACTA}, a considerable change
is observed in the phase diagrams. In this case, the paramagnetic
(\textbf{P}) phase region is completely lost in the phase diagrams
corresponding to the model in \cite{AGUT2012ACTA} on the
$[-2,2]\times [-2,2]$. As  $k$ increases, it is seen that some
phase regions have narrowed or even disappeared (see
\cite{AGUT2012ACTA} for details). %Unlike our work, the phase
%diagrams of the Ising models that we have described on the
%arbitrary order Cayley tree were plotted.

During the studies obtained for both Ising models and Potts
models, very different phase types were encountered in the form of
narrow islands in the phase diagrams. We cannot determine why
these islands actually exist. These islands may have originated
from the drawing. For this reason, in order to determine the
species of these islands, Lyapunov exponents and drawing of the
wave vectors were needed (see figure~\ref{prolonged-tenary1}).

In contrast to \cite{Vannimenus} and Mariz \cite{MTA1985a}, the
stability analysis for the ferromagnetic phase is more difficult
than the stability analysis for the paramagnetic phase because the
fixed point is not the same for the whole phase. Therefore, in
order to study the stability analysis of the ferromagnetic phase
for the model \eqref{Prolonged-hm}, we need new methods.

In \cite{Inawashiro}, Inawashiro et al have considered
schematic illustration of selective summation of spins from the
outermost to innermost shells to plot the phase diagrams of the
Ising model. They set up an iterative scheme for Vannimenus's
equations \cite{Vannimenus}. They obtained a phase diagram
containing ferromagnetic (\textbf{F}), paramagnetie (\textbf{P}),
chaotic (\textbf{C}), and antiferromagnetic + + -- -- (\textbf{A})
phases. Silva et al. \cite{Silca-C} consider a very general model
on Bethe lattices of arbitrary branching number with arbitrary
couplings between: $(J_1)$ first-, $(J_2)$ second- and $(J_3)$
third-neighbors, and furthermore in the presence of an arbitrary
external field. In particular, in \cite{Silca-C}, for $J_3=0$, the
model reduces to the model analyzed in \cite{UGTA2010AIP}, i.e.,
the generalization of the Vannimenus model to arbitrary order
number. %The novel part of \cite{UGTA2010AIP} is that the authors
%analyzed the periodic phases a bit more in detail, for larger
%periods than the ones considered in \cite{Silca-C}.

When we investigated the phase diagrams given in our recent
articles, very rich and different new phases were obtained.
The structure of the phase diagrams of the generalized systems was
examined and compared
\cite{UGAT2012IJMPC,AGUT2012ACTA,UGTA2010AIP}. Magnetization
graphs were plotted for certain gamma and beta values. Lyapunov
exponents were  successfully plotted for certain values of $k$
and for certain critical coupling constants \cite{UGAT2012IJMPC}.
The stability analysis of the given system was examined in detail.

In addition, the transition lines between the phases were
successfully investigated. This work was published in
\cite{GU2011a}. Finally, for Ising model on a Cayley tree of
arbitrary order, a system of nonlinear equations corresponding to
the Hamiltonian of 3 different interactions was obtained and the
phase analysis was successfully performed and this work was
published in \cite{AGTU2013ACTA}. Since more comprehensive
information and findings are included in the related studies,
details are not given here (see
\cite{UGAT2012IJMPC,AGUT2012ACTA,UGTA2010AIP} for details).

In \cite{SA2011Ankara}, the phase diagrams of an Ising model
involving nearest neighbor $J$, triplet prolonged next nearest
neighbor $J_p$ and one-level next nearest neighbor $J_{1_l}$ on
the third order Cayley tree were plotted. Despite the presence of
paramagnetic regions in the phase diagrams obtained in previous
studies \cite{Vannimenus,MTA1985a,UA2010PhysicaA}, the
paramagnetic phase zones for the low temperatures in the work
\cite{SA2011Ankara} completely disappeared.

We see that ternary prolonged next nearest interaction has
strong effects on the phase diagrams. One of these is the shift
from the zero temperature values of the multi-critical Lifshits
points to the end values. The other effect is the disappearance of
paramagnetic phase zones. For some given coupling constants $J$,
$J_p$ and $J_{1_l}$, the plots of the wave vectors are
interpreted. %There are repetition equations to get these graphs.

%\newpage
\section{\textbf{The dynamical behavior of Potts model on a Cayley tree}}\label{Dynamical behavior of Potts model}
In this section, we  study the phase diagrams of a $q$-state
Potts models having certain characteristics. Notably, we should
mention that most of the results were published in previous
studies. Our goal here is to interpret the results by comparing
them.

It is well-known that the $q$-state Potts models are
generalizations of the 2-state Ising model, which is equivalent to
$q = 2$ in this case. The Potts model can be expressed as a class
of the statistical mechanics models that specifically examine the
long-term behavior of complex systems \cite{Wu}. These models have
a rich structure in order to sample almost every subject area. In
particular, it is at the center of investigations by the
intersection of conformal field theory, infiltration theory,
quantum groups and integrable systems. The Potts model explores
the interactions of system elements based on certain
characteristics of each element \cite{Potts1952}. The Potts model
has a powerful mathematical modelling method that can be used in a
variety of disciplines, including biology, sociology, physics, and
chemistry
\cite{Akin-Ulusoy-2022,Weidlich-1971,Beaudin2007,Beaudin-Ellis2010}.

For all values of the external magnetic field and temperature,
Peruggi et al. \cite{Peruggi-1983} studied the $q$-state
ferromagnetic Potts model (FPM) and antiferromagnetic Potts model
(APM). They derived exact formulas for all thermodynamic functions
of relevance in the FPM and APM, as well as drew entire phase
diagrams for both systems. By using the method of solution of spin
models on Cayley tree, Peruggi et al.~\cite{Peruggi-1987} analyzed
the phases of the ferromagnetic and antiferromagnetic $q$-state
Potts models. Nevertheless, in the literature, the phase diagrams
for the Potts models, except a few publications, have not been
solved and compared using any computer programs
(\cite{GTAU2011a,GTA2009a,TGUA2013ACTA,GTUA2011C,TGAU2010AIP,Gok-Tez}).
For this reason, it has taken some time for the codes to be
created. In general, deriving the recurrence equations associated
to the Potts models is much more difficult than in the Ising
model. Especially, as the number of the branches of the Cayley
tree increases, it becomes even more difficult to derive the
recurrence equations \cite{GNR2013POTTS,UguzGD-Korean-2015}.

In this section, we are going to explain how the phase diagrams
are analyzed, taking into account some Potts models that we 
studied before.
\subsection{The $q$-state Potts model with three different
binary competing interactions on a Cayley tree} Let us consider
the Hamiltonian given by
\begin{equation}\label{PottsHm2}
H(\sigma )=-J_1\sum _{\langle x,y\rangle} \delta _{\sigma
(x)\sigma (y)}-J_p\sum _{\rangle x,y\langle} \delta _{\sigma
(x)\sigma (y)}-J_o\sum _{\rangle\widetilde{x,y}\langle} \delta
_{\sigma (x)\sigma (y)},
\end{equation}
where the coupling constants are referred to as competing
nearest-neighbor (NN) interaction $J_1$, prolonged
next-nearest-neighbor (PNNN) interaction $J_p$ and one-level
next-nearest-neighbor (1LNNN) interaction $J_o$, respectively (see
\cite{GTA2009a} for details). In order to study the behavior of
the recursive relations associated to the Hamiltonian given in
\eqref{PottsHm2}, we established an iterative scheme similar to
that appearing in real space renormalization group frameworks
(RSRGF) \cite{Yesilleten}. In this system, the nearest neighbor
interactions with the triplets on the Cayley tree of the order two
were considered and we observed the presence of the
ferromagnetic and the paramagnetic phase regions in the phase
diagrams, as well as in a new phase region called the
paramodulated phase (see figure \ref{Potts-phase-D2aa}).
Corresponding phase diagrams were plotted by means of the coupling
constants $J_1,J_p$ and $J_o$. At finite temperatures, we
determined several interesting features exhibited for some typical
values of $J_o/J_1$.

After tedious calculations, we obtained the sequence
\begin{equation}\label{Potts-seq1}
[x^{(n+1)}, y_1^{(n+1)}, y_2^{(n+1)}, y_3^{(n+1)}].
\end{equation}
The system of four equations obtained in \eqref{Potts-seq1} has a
more complex structure than the recurrence equations of the
Vannimenus-Ising model and  Potts model associated with the
Hamiltonian \eqref{Potts model}.

By means of the recursive relations given in \eqref{Potts-seq1},
we were interested in the exact phase diagrams in
three-dimensional space $(T /J_1,-J_{p} /J_{1},J_{o} /J_{1})$.
Here, instead of drawing in 3 dimensions, we plotted phase
diagrams in two dimensions without considering one of these
expressions.
%By starting from initial conditions $(x^{(1)}, y_1^{(1)},
%y_2^{(1)}, y_3^{(1)})$ corresponding to boundary condition
%Ѓ$\overline{\sigma}^{(n)}(V\setminus V_n )\equiv 1$, we iterate
%the recursive equations \eqref{Potts-seq1} and observe their
%behavior after a large number of iterations.
We iterate the recursive equations \eqref{Potts-seq1} and analyze
their behavior after a large number of iterations by starting with
initial conditions $(x^{(1)},\,\, y_1^{(1)}, \,\, y_2^{(1)},\,\, y_3^{(1)})$
corresponding to boundary condition
$\overline{\sigma}^{(n)}(V\setminus V_n )\equiv 1$.

In order to plot the phase diagrams associated to a given model, we
try to find a fixed point $(x^{(*)}, y_1^{(*)},\,\, y_2^{(*)},
y_3^{(*)})$. If $y_1^{(*)}=0,\,\,y_2^{(*)}=0, \,\,y_3^{(*)}=0$, it
corresponds to a paramagnetic phase (white colour) or if
$y_1^{(*)}\neq 0, \,\, y_2^{(*)}\neq0, \,\, y_3^{(*)})\neq 0$, then it
corresponds to a ferromagnetic phase \textcolor[rgb]{1.0, 0.0,
0.0}{'FERROMAGNETIC'}.
Secondly, the recursive system of equations may have an iterative
state with period $p$ (see the table \ref{colors}). If we obtain
as $p=2$, the phase corresponds to antiferromagnetic phase (or
\textcolor[rgb]{1.0, 1.0, 0.0}{'PERIOD 2'}) and in the case $p=4$,
the type of phase corresponds to the so-called antiphase
(\textcolor[rgb]{0.0, 1.0, 1.0}{'PERIOD 4'}), that is denoted~$\left\langle 2\right\rangle $
for brevity. If $p=5,6,7,8,9,10,11$, then the relevant phases are
called \textcolor[rgb]{0.0, 0.0, 1.0}{'PERIOD 5'},
\textcolor[rgb]{1.0, 0.0, 1.0}{'PERIOD 6'}, \textcolor[rgb]{0.54,
0.47, 0.36}{'PERIOD 7'}, \textcolor[rgb]{0.5, 0.0, 0.0}{'PERIOD
8'}, \textcolor[rgb]{0.0, 0.5, 0.0}{'PERIOD 9'},
\textcolor[rgb]{0.5, 0.5, 0.0}{'PERIOD 10'} and
\textcolor[rgb]{0.0, 0.5, 0.5}{'PERIOD 11'}, respectively (see
\cite{GTA2009a} for details).

Finally, the system can continue to be aperiodic, that is, in case
of not reaching a certain fixed period. It is difficult to quantitatively 
determine the difference between a really aperiodic example and the
one with a very long period. Having a period of $p$
($p \leqslant 12$) is considered to be a periodic phase. We shall recall all
periodic phases with a period of $p>12$ and aperiodic phase as the
modulated phases \textcolor[rgb]{0.5, 0.0, 0.5}{'MODULATED'} (see
the table \ref{colors} for details).

\begin{figure} [!htbp]
\centering
\includegraphics[width=42mm]{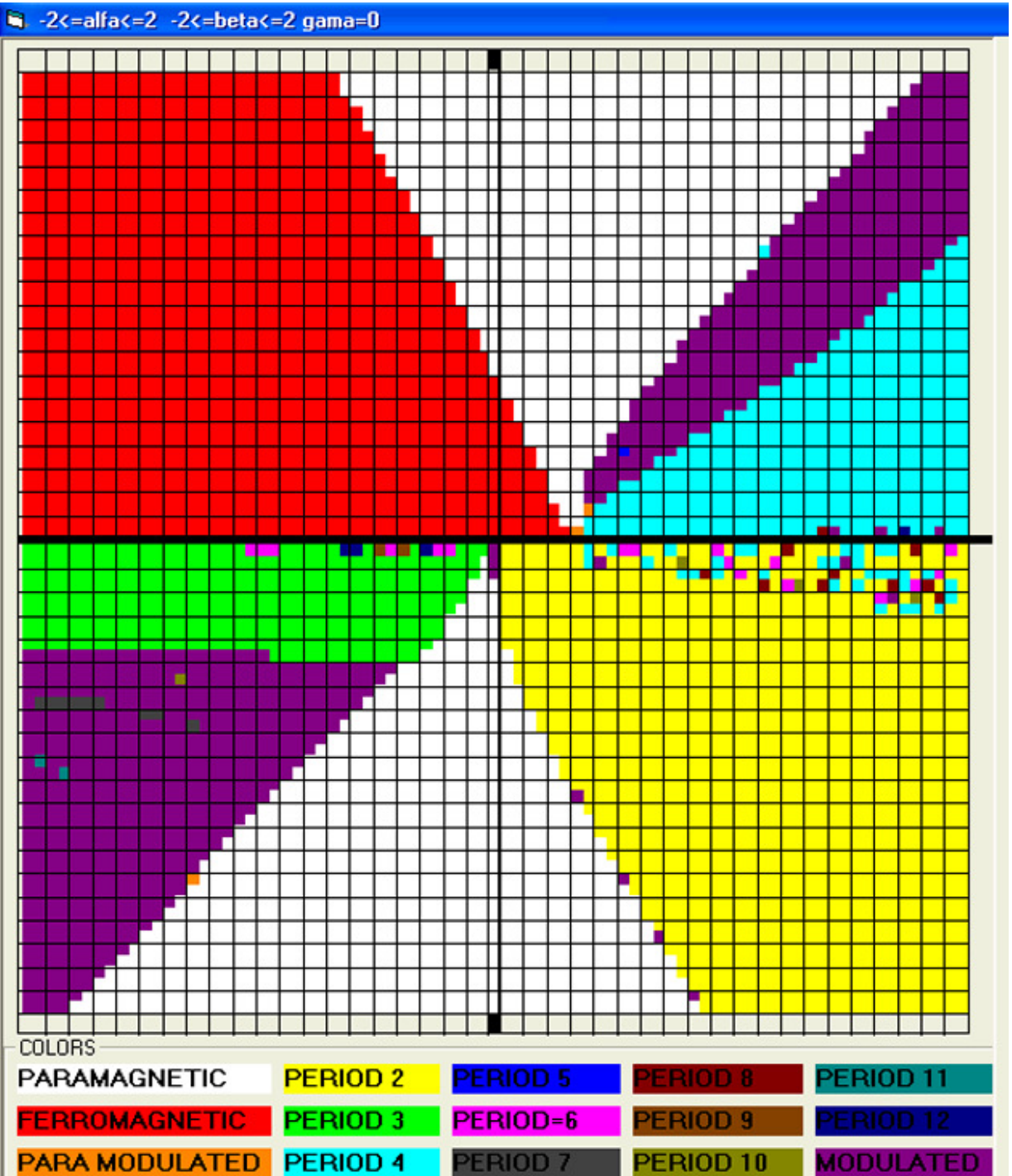}\label{Potts-phase-D2aa}\
\ \ \ \
\includegraphics[width=40mm]{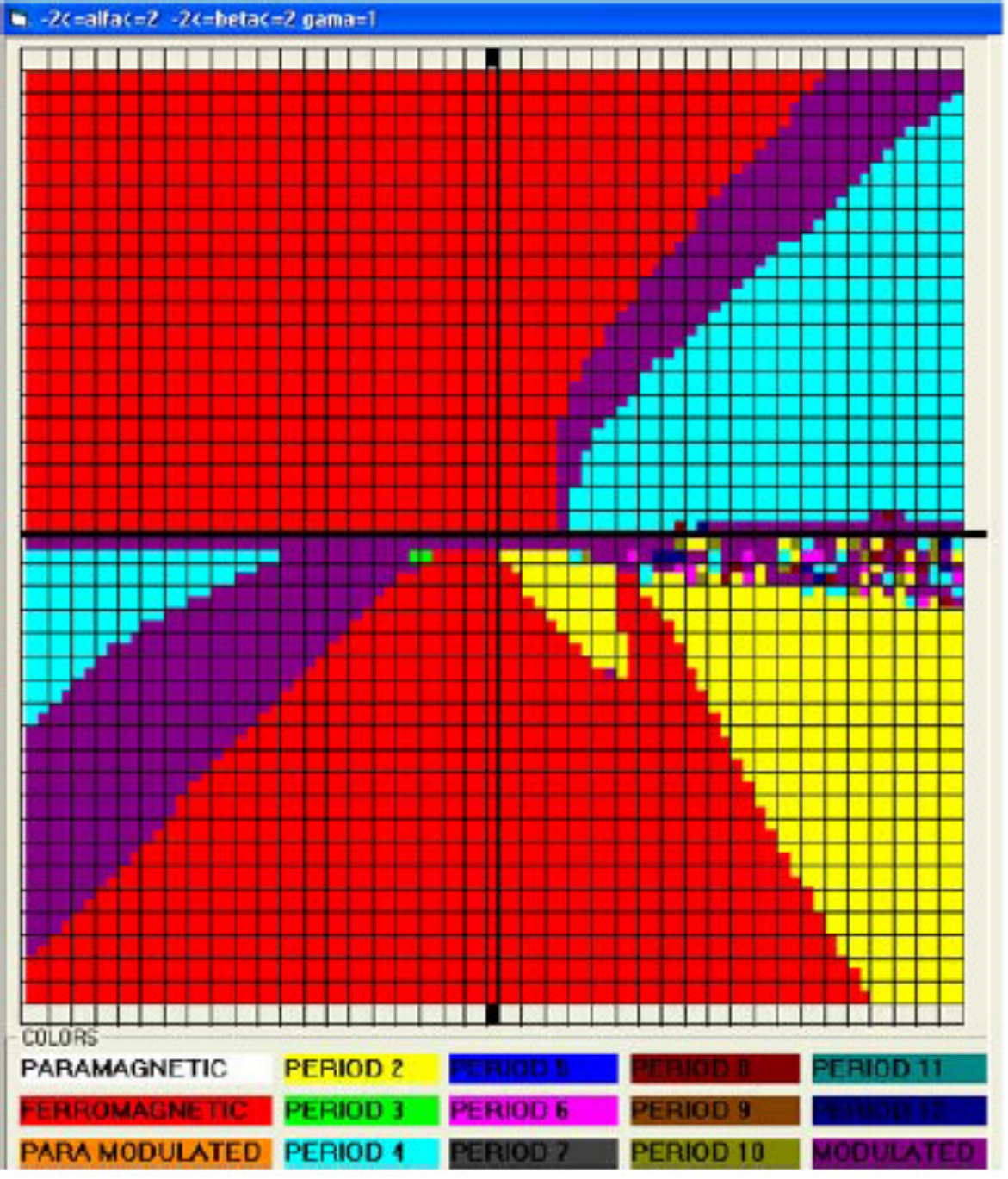}\label{Potts-phase-D2bb}
\caption{(Colour online) Left-hand: The phase diagram of the recursive
relations corresponding to the Hamiltoninan \eqref{PottsHm2} for
$\gamma=0$ on the rectangular region $[-2,2]\times[-2,2]$. Right-hand: Phase diagram of the recursive relations for
$\gamma=1$ on the rectangular region $[-2,2]\times[-2,2]$ (see
\cite{GTA2009a})}
\label{Potts-phase-D2aa}
\end{figure}

Figures given in \ref{Potts-phase-D2aa} show the phase diagrams
corresponding to the Potts model presented in \cite{GTA2009a}. 
The phase regions according to the third parameter variable values
$\gamma$ are seen in  figures given in \ref{Potts-phase-D2aa}.
We observed that for given values of these phase zones, our
diagrams contained the paramagnetic region and we obtained the
multi-critical Lifshitz points for the
increasing values of temperature. 

\subsection{The Potts model with two interactions on a Cayley tree}
Let us consider the following  Hamiltonian as the counterpart of
the Ising model for the Hamiltonian given by Ganikhodjaev et
al. \cite{GAUT2011b}
\begin{equation}\label{Potts model}
H(\sigma)=-{{J}_{p}}\sum\limits_{\langle\widetilde{x,y,z}\rangle}{{{\delta
}_{\sigma (x)\sigma (y)\sigma (z)}}}-J\sum\limits_{\langle
x,y\rangle}{{{\delta }_{\sigma (x)\sigma (y)}}}.
\end{equation}
\begin{figure}[!htbp]
\centering
\includegraphics[width=40mm]{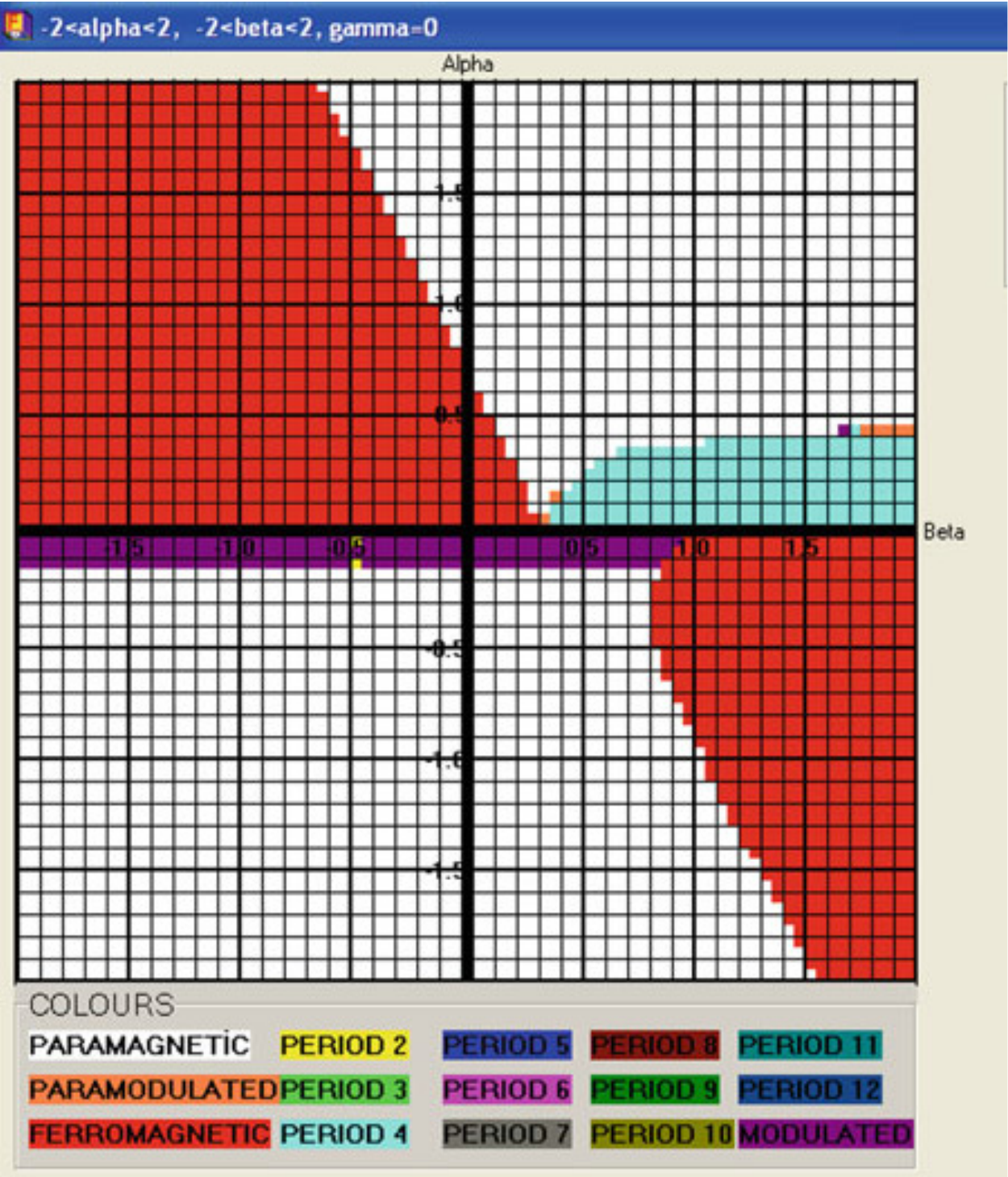}
\caption{(Colour online) Phase diagram of the recursive relations
associated to the Potts model with triple and binary interacting
on the rectangular region
$[-2,2]\times[-2,2]$ (see \cite{GTUA2011C}). %(see \cite{GTUA2011C})
}\label{Potts-phase-D1}
\end{figure}

The dynamical system for the Potts model associated with the
Hamiltonian \eqref{Potts model} on the Cayley tree of the order 2 is
studied in \cite{GTUA2011C}. We analyzed the phase diagram of the
Potts model with competing prolonged ternary and binary NN
interactions and demonstrated that it consisted of the following
phases: ferromagnetic, paramagnetic, and antiphase with period 4,
and modulated phase (figure \ref{Potts-phase-D1})
\cite{GTUA2011C}. Unlike the Vannimenus-Ising model given in
\eqref{Vhm}, the paramagnetic phase for this model \eqref{Potts
model} was observed in a large region (see figure
\ref{Potts-phase-D1}).

The set of spin values $\{1,2,3\}$ associated with the Hamiltonian
\eqref{Potts model} can be replaced by the centered set
$\{-1,0,1\}$ since the form of spins of the Potts model associated
with the Hamiltonian \eqref{Potts model} is unessential.
Therefore, the average magnetization $\tilde{m}$ for the $n\,$th
generation is calculated as follows:
$$
\tilde{m}=-\frac{4(y_1+y_2)(2x+y_2+1)}{(2x+y_1+2y_2+1)^2+2(2x-y_1+1)^2}.
$$

\begin{figure} [!htbp]
\centering
\includegraphics[width=65mm]{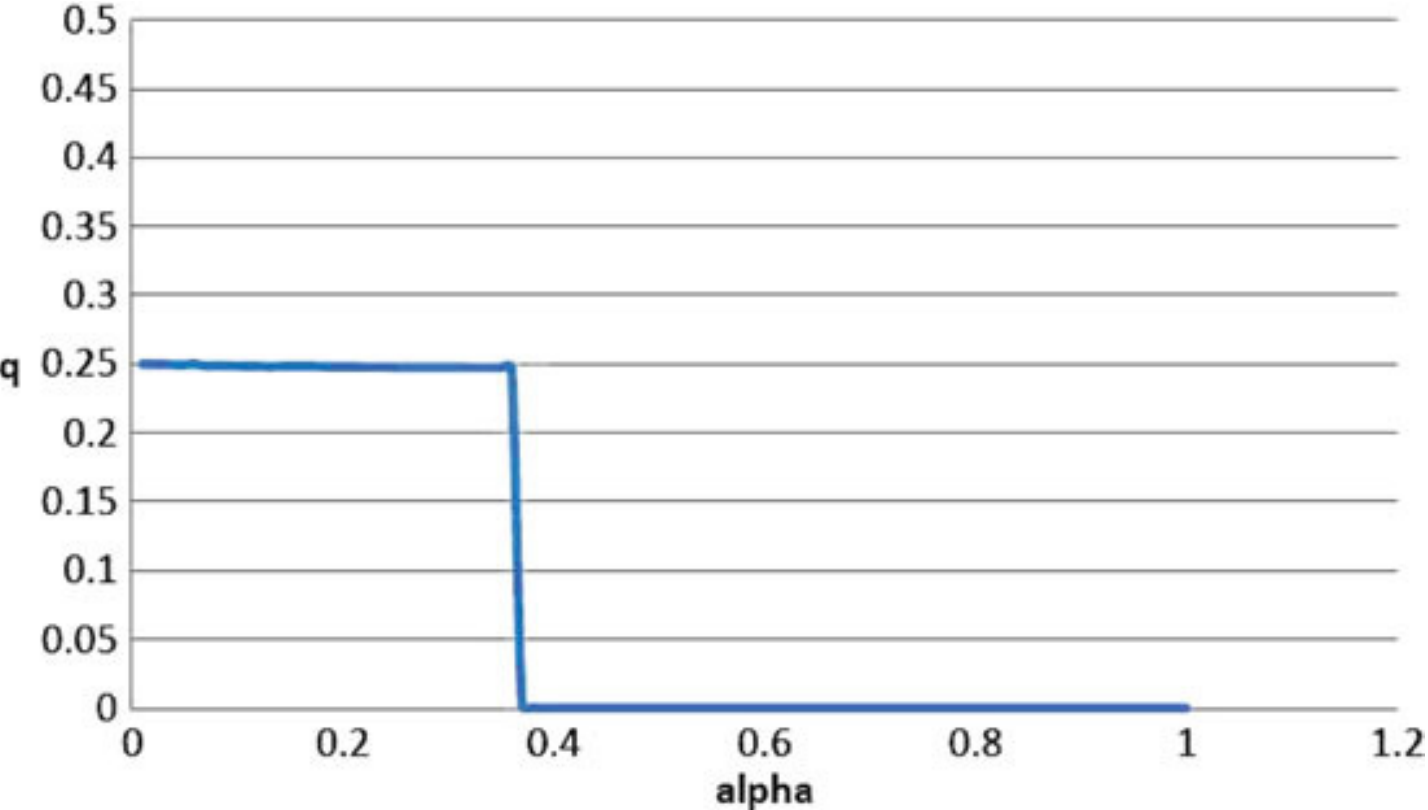}
\caption{(Colour online) Variation of the wave-vector $q$ versus
$T$, for $\beta=-J_p/J_1 =0.8$ (see \cite{GTUA2011C}).
}\label{wave-vectors-Potts}
\end{figure}
Figure \ref{wave-vectors-Potts} shows how the wave-vector changes
as a function of temperature $T$ between its value at the
paramodulated transition and its value in the $\left\langle 2\right\rangle $ phase is equal
to $\frac{1}{4}$ (the figure \ref{wave-vectors-Potts} is borrowed
from~\cite{GTUA2011C}).

If we take into account the Ising model introduced in
\cite{GAUT2011b}, we could not get the fixed point $(x^{*},0,0)$,
but we obtained the fixed points when we considered the Potts
model with the same Hamiltonian. Furthermore, we observed the presence of
the paramagnetic phase regions in the phase diagram of
the Potts model (see figure \ref{Potts-phase-D1}). 
\begin{rem}\label{Potts-4.1}
Note that the same Potts model on a Cayley tree of arbitrary order can be 
numerically  studied. However, deriving the recursive equations for
the same Potts model is a difficult problem. Exact description of
the phase diagrams and recurrence equations is rather tricky.
\end{rem}

\subsection{The 3-state Potts model on Cayley tree of the order 2}
This subsection is based on the paper \cite{GTAU2011a}. Consider
the three-state Potts model with spin values $\Phi=\{1,2,3\}$. The
relevant Hamiltonian takes the form with competing binary
nearest-neighbor and two separate ternary interactions as follows
\cite{GTAU2011a}:
\begin{equation}\label{Potts model3}
H(\sigma
)=-{{J}_{p}}\sum\limits_{\langle\widetilde{x,y,z}\rangle}{{{\delta
}_{\sigma (x)\sigma (y)\sigma (z)}}}-{{J}_{t}}\sum\limits_{\langle
x,\widetilde{y},z\rangle}{{{\delta }_{\sigma (x)\sigma (y)\sigma
(z)}}}-{{J}_{1}}\sum\limits_{\langle x,y\rangle}{{{\delta
}_{\sigma (x)\sigma (y)}}},
\end{equation}
where $J_p,J_t, J_1\in \mathbf{R}$ are coupling constants and
$\delta$ is the Kronecker symbol and defined as
\[{{\delta }_{\sigma (x)\sigma (y)\sigma (z)}}=\left\{
\begin{matrix}
   1,\text{ if }\sigma (x)=\sigma (y)=\sigma (z)\text{ }  \\
   0,\text{ otherwise}\text{. \ \ \ \ \ \ \ \   \ \ \ \ \ \ \ }  \\
\end{matrix} \right.
\]

The system of nonlinear equations associated with the Hamiltonian
given by equation~\eqref{Potts model3} is numerically analyzed in
\cite{GTAU2011a}. %In this sense, we think that we have achieved
By analyzing the recurrence equations numerically, we showed that
the phase diagrams contained ferromagnetic, paramagnetic and $\left<2\right>$
phases only. We also saw that the modulated phases contained only
the $\left<2\right>$ for $J_1 > 0$ and $J_p/J_1 <0$.
%i.e., the set of modulated phases consists of the $\left<2\right>$ phase only
%for $J_1 > 0$ and $J_p/J_1 <0$.
We obtained the transition lines by means of stability
conditions. %By using numerical iterations, we have analyzed
%characteristic points in the phase diagram.  We have plotted the
%wave-vectors versus temperature for some critical points in the
%modulated phases.
We looked at the  characteristic
properties of the phase diagram using numerical iterations. For some critical points in
the modulated phases, we plotted the wave-vectors versus
temperature.

\subsubsection{Recurrence equations}
Here, we  give the recursion equations for the sake of
completeness (see \cite{GTAU2011a} for details). We investigate
the relationship between the partition function on $V_n$ and the
partition function on subsets of $V_{n-1}$ to generate the
recurrent equations. The recurrence equations show how their
impact propagates along the tree having got the initial conditions on
$V_1$. We will look at the partition functions herein below.

\begin{figure} [!htbp]
\centering
\includegraphics[width=30mm]{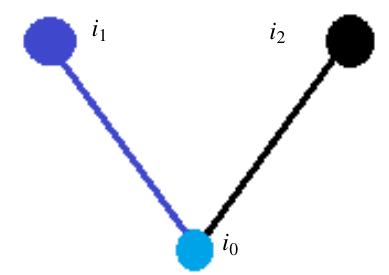}
\caption{(Colour online) A configuration on
$V_1$.}\label{Potts-Partition1}
\end{figure}

Let $Z_i^{(n)}$ be a partition function on $V_n$ such that the
spin $i$ in the root (node) $x^{(0)}$, $i=1,2,3;$ let $Z^{(n)}(i,j)$
be a partition function on $V_n$ with the configuration $(i,j)$ on
an edge $\langle x^{(0)},x\rangle$. To continue this process, we
consider all interactions except interaction of nearest neighbors
$\langle x^{(0)}, x\rangle, \,x\in W_1$ and $i,j=1,2,3;$ let
$Z^{(n)}(i_1,i_0,i_2)$ be a partition function on $V_n$, where
spin $i_0$ is placed at vertex $x^{(0)}$ and spins $i_1$ and $i_2$
are placed at the next
vertices $x_1$ and $x_2$ belonging to $S(x^{(0)})$.\\
If the set of spins is taken as $\Phi=\{1,2,3\}$, the number of
partition functions $Z^{(n)}\left(
\begin{array}{ccc}
 i_1 &   & i_2 \\
 & i_0 &
\end{array}
\right)=Z^{(n)}(i_1,i_0,i_2)$ is 27 (see figure
\ref{Potts-Partition1}) and the total partition function $Z^{(n)}$
in the volume $V_n$ can be written as
\begin{equation}\label{par1}
Z^{(n)}=\sum^{3}_{i_1,i_0,i_2=1}
Z^{(n)}(i_1,i_0,i_2)=\sum^{3}_{i_1,i_0,i_2=1} Z^{(n)}\left(
\begin{array}{ccc}
 i_1 &   & i_2 \\
 & i_0 &
\end{array}
\right).
\end{equation}
\begin{eqnarray}\label{par2}
 Z^{(n)}(i_1,i_0,i_2)%&=&%Z^{(n)}\left(
%\begin{array}{ccc}
% i_1 &  & i_2 \\
% & i_0 &
%\end{array}
%\right)\\\nonumber
&=&\exp \left(\frac{J_t}{2T} \delta
_{i_1i_0i_2}+\frac{J_1}{T}\delta _{i_0i_1}+\frac{J_1}{T}\cdot
\delta _{i_0i_2}\right)Z^{(n)}\left(
\begin{array}{cc}
 i_1 &   \\
 & i_0
\end{array}
\right)Z^{(n)}\left(
\begin{array}{cc}
& i_2 \\
i_0 &
\end{array}
\right) \nonumber\\
 &=&\exp \left(\frac{J_t}{2T}\delta _{i_1i_0i_2}+\frac{J_1}{T}
 \delta _{i_0i_1}+\frac{J_1}{T}\delta _{i_0i_2}\right)Z^{(n)}(i_0,i_1)Z^{(n)}(i_0,i_2),
\end{eqnarray}
where $i_1,i_0,i_2\in\{1,2,3\}.$

If we take into account all partition functions in the volume $V_n$
under the boundary condition $\bar{\sigma}_n\equiv 1,$ then we
assume
\begin{eqnarray}\label{par3}
Z^{(n)}(1,2)&=&Z^{(n)}(1,3),\
Z^{(n)}(2,1)=Z^{(n)}(3,1),\\\label{par4}
Z^{(n)}(2,2)&=&Z^{(n)}(3,3), \ Z^{(n)}(2,3)=Z^{(n)}(3,2).
\end{eqnarray}

For the sake of completeness, we again recall the equations that we
obtained herein below earlier.

Let $a=\exp\left( \frac{J_t}{2T}\right) $; $b=\exp\left( \frac{J_p}{T}\right) $,
$c=\exp\left( \frac{J_1}{T}\right) $.  From \eqref{par2}, we have the partial
partition equations as follows.\\
First, for $\sigma(x^{(0)})=1$ we obtain {\small\begin{eqnarray*}
Z^{(n)}(1,1,1)&=&a^2c^2(Z^{(n)}(1,1))^2,\\
Z^{(n)}(1,1,2)&=&Z^{(n)}(2,1,1)=cZ^{(n)}(1,1)Z^{(n)}(1,2),\\
Z^{(n)}(1,1,3)&=&Z^{(n)}(3,1,1)=cZ^{(n)}(1,1)Z^{(n)}(1,3),\\
Z^{(n)}(2,1,2)&=&Z^{(n)}(3,1,3)=(Z^{(n)}(1,2))^2,\\
Z^{(n)}(2,1,3)&=&Z^{(n)}(3,1,2)=(Z^{(n)}(1,2))^2.
\end{eqnarray*}}
Secondly, for $\sigma(x^{(0)})=2$ we get {\small\begin{eqnarray*}
Z^{(n)}(1,2,1)&=&(Z^{(n)}(2,1))^2,\\
Z^{(n)}(1,2,2)&=&Z^{(n)}(2,2,1)=cZ^{(n)}(2,1)Z^{(n)}(2,2),\\
Z^{(n)}(1,2,3)&=&Z^{(n)}(3,2,1)=Z^{(n)}(2,1)Z^{(n)}(2,3),\\
Z^{(n)}(2,2,2)&=&a^2c^2(Z^{(n)}(2,2))^2,\\
Z^{(n)}(2,2,3)&=&Z^{(n)}(3,2,2)=cZ^{(n)}(2,2)Z^{(n)}(2,3),\\
Z^{(n)}(3,2,3)&=&(Z^{(n)}(2,3))^2.
\end{eqnarray*}}
Thirdly, for $\sigma(x^{(0)})=3$ one gets {\small\begin{eqnarray*}
Z^{(n)}(1,3,1)&=&(Z^{(n)}(3,1))^2,\\
Z^{(n)}(1,3,2)&=&Z^{(n)}(2,3,1)=Z^{(n)}(3,1)Z^{(n)}(3,2),\\
Z^{(n)}(1,3,3)&=&Z^{(n)}(3,3,1)=cZ^{(n)}(3,1)Z^{(n)}(3,3),\\
Z^{(n)}(2,3,2)&=&(Z^{(n)}(3,2))^2,\\
Z^{(n)}(2,3,3)&=&Z^{(n)}(3,3,2)=cZ^{(n)}(3,2)Z^{(n)}(3,3),\\
Z^{(n)}(3,3,3)&=&a^2c^2(Z^{(n)}(3,3))^2.
\end{eqnarray*}}
Due to \eqref{par3} and \eqref{par4}, we can select only five
independent  partition functions $Z^{(n)}(1,1,1)$,
$Z^{(n)}(2,1,2)$, $Z^{(n)}(1,2,1)$, $Z^{(n)}(2,2,2)$,
$Z^{(n)}(3,2,3)$ and therefore we can introduce  new variables
\begin{equation*}
\left\{
\begin{array}{l}
 u_1^{(n)}=\sqrt{Z^{(n)}(1,1,1)},\\
 u_2^{(n)}=\sqrt{Z^{(n)}(2,1,2)},\\
 u_3^{(n)}=\sqrt{Z^{(n)}(1,2,1)},\\
 u_4^{(n)}=\sqrt{Z^{(n)}(2,2,2)},\\
 u_5^{(n)}=\sqrt{Z^{(n)}(3,2,3)}.
\end{array}
\right.
\end{equation*} 
After tedious calculations, we
have the following
equations %
\begin{equation}\label{rec1}
\left\{
\begin{array}{l}
 u_1^{(n+1)}=ac[b^2(u_1^{(n)})^2+4a^{-1}bu_1^{(n)}u_2^{(n)}+4(u_2^{(n)})^2], \\
 u_2^{(n+1)}=[(u_3^{(n)})^2+2a^{-1}u_3^{(n)}u_4^{(n)}+2u_3^{(n)}u_5^{(n)}
 +(u_4^{(n)})^2+2a^{-1}u_4^{(n)}u_5^{(n)}+(u_5^{(n)})^2], \\
 u_3^{(n+1)}=[(u_1^{(n)})^2+4a^{-1}u_1^{(n)}u_2^{(n)}+4(u_2^{(n)})^2], \\
 u_4^{(n+1)}=ac[(u_3^{(n)})^2+2a^{-1}bu_3^{(n)}u_4^{(n)}+2u_3^{(n)}u_5^{(n)}
 +b^2(u_4^{(n)}){}^2+2a^{-1}bu_4^{(n)}u_5^{(n)}+(u_5^{(n)})^2], \\
 u_5^{(n+1)}=[(u_3^{(n)})^2+2a^{-1}u_3^{(n)}u_4^{(n)}+2u_3^{(n)}u_5^{(n)}
 +(u_4^{(n)})^2+2a^{-1}u_4^{(n)}u_5^{(n)}+(u_5^{(n)})^2].
\end{array}
\right.
\end{equation}
Since $u_2^{(n+1)}=u_5^{(n+1)}$, we can rewrite the system of
equations given in \eqref{rec1} as
{\small\begin{equation}\label{rec2} \left\{
\begin{array}{l}
 u_1^{(n+1)}=ac[b^2 (u^{(n)}_1)^2+4a^{-1}bu^{(n)}_1 u^{(n)}_2+4(u^{(n)}_2)^2 ], \\
 u_2^{(n+1)}=[(u^{(n)}_2+{u^{(n)}_3})^2+2a^{-1}(u^{(n)}_2+u^{(n)}_3)u^{(n)}_4+{u^{(n)}_4}^2], \\
 u_3^{(n+1)}=[(u^{(n)}_1)^2+4a^{-1}u^{(n)}_1u^{(n)}_2+4(u^{(n)}_2)^2], \\
 u_4^{(n+1)}=ac[({u^{(n)}_2}+{u^{(n)}_3})^2+2a^{-1}b(u^{(n)}_2+u^{(n)}_3)u^{(n)}_4+b^2(u^{(n)}_4)^2].
 %u_5^{(n+1)}=[(u_3^{(n)})^2+2a^{-1}u_3^{(n)}u_4^{(n)}+2u_3^{(n)}u_5^{(n)}+(u_4^{(n)})^2+2a^{-1}u_4^{(n)}u_5^{(n)}+(u_5^{(n)})^2],
\end{array}
\right.
\end{equation}}

In order to draw the phase diagram, we prefer the following
choice;
$$
x=\dfrac{u_2+u_3}{u_1+u_4}, \quad
y_1=\dfrac{u_1-u_4}{u_1+u_4},\quad y_2=\dfrac{u_2-u_3}{u_1+u_4}.
$$
If we substitute  as follows;
\begin{equation*}
 u_1=\frac{(1+y_1)A}{2},\quad u_2=\frac{(x+y_2)A}{2}, \quad u_3=\frac{(x-y_2)A}{2}, \quad u_4=\frac{(1-y_1)A}{2},
\end{equation*}
where $A=u_1+u_4$. Then, equations \eqref{rec2} yield:
\begin{equation}\label{rec3}
\begin{array}{lll}
x^\prime &=&\dfrac{1}{acD}[4x^2+4xy_2+2y_2^2+2a^{-1}(2x+y_2+y_1y_2)+y_1^2+1] ;\\[3mm]
y^\prime_1&=&\dfrac{2}{D}[b^2y_1+a^{-1}b(2xy_1+y_2+y_1y_2)+2xy_2+y_2^2];\\[3mm]
y^\prime_2&=&-\dfrac{2}{acD}[y_1+2xy_2+y_2^2+a^{-1}(2xy_1+y_2+y_1y_2)],
\end{array}
\end{equation}
where $D=b^2(1+y_1^2)+2a^{-1}b(2x+y_2+y_1y_2)+4x^2+4xy_2+2y_2^2.$

In order to plot the phase diagram we have to look at the local
properties, namely the local magnetization or the magnetization of
the root $x^{(0)}.$
Since the  spin form of the Potts model is unessential, the set of
spin values $\{1,2,3\}$ can be replaced by the centered set
$\{-1,0,1\}$. The average magnetization $\tilde{m}$ for the $n$
generation is then calculated as:

\begin{equation}\label{mag1}
\tilde{m}=-\frac{(2x+y_2)(a^{-1}y_1+y_2)+y_1+a^{-1}y_2}{12x^2+3y_1^2+4y_2^2+8xy_2-2y_1+4a^{-1}(3x+y_2+y1y_2-xy_1)}.\\
\end{equation}
Thus, a situation where $y_1^*$, $y_2^*\neq 0$ but $\tilde{m}=0$ can
occur. Considering the system of nonlinear equation~\eqref{rec3}
and the average magnetization equation \eqref{mag1} together, we
draw the phase diagrams of the given Potts model using numerical
calculations (see \cite{GTAU2011a} for details).

\subsubsection{The phase diagrams of the 3-state Potts model associated with the Hamiltonian \eqref{Potts model3}}
Before delving deeper into the various transitions, it is helpful
to understand the main aspects of the phase diagram. As we
mentioned in detail in previous sections, the numerically precise
phase diagrams are plotted by the recursive relations \eqref{rec3}
in 3-dimensional space $(T/J_1, -J_p/J_1, J_t/J_1)$.
Based on the previous work, we make the following variable
replacement;
$(-J_p/J_1)=\beta $, $T/J_1=\alpha$ and $J_t/J_1=\gamma$ and
respectively $b=\exp(-\alpha^{-1}\beta)$, $c=\exp(\alpha^{-1})$
and $a=\exp((2\alpha )^{-1}\gamma).$ Then, we have
$a=c^{\frac{\gamma}{2}}.$ For some given values of $\gamma$, under
boundary condition $\bar{\sigma}^{(n)}\equiv 1,$ we obtain the
initial conditions of the system of equations \eqref{rec3} as
follows,
$$
x^{(1)}=\dfrac{1+a^2c^2}{ac(a^2c^2b^2+1)},\quad
y_1^{(1)}=\dfrac{a^2c^2b^2-1}{a^2c^2b^2+1},\quad
y_2^{(1)}=\dfrac{1-a^2c^2}{ac(a^2c^2b^2+1)}.
$$
The initial boundary conditions are  iterated in the system of
equations \eqref{rec3}. By doing a sufficiently large number of
iterations, we observe the behavior of the system of iterative
equations \eqref{rec3} in the plane
$(\beta=-\frac{J_p}{J_1},\alpha=\frac{T}{J_1}).$ As stated in the
previous sections, it is aimed at reaching the fixed point
$(x^*,y_1^*,y_2^*)$, and phase diagrams are obtained according to
the status of these fixed points (see table \ref{colors}).

In order to investigate the limit behavior of Potts models
associated to the Hamiltonian with different coupling constants on
a Cayley tree of 2nd order, we have studied the systems of
nonlinear equations corresponding to dynamical systems in order to realize
the analysis of the dynamical systems of Hamiltonian equations for
different interactions according to 3-valued ($q=3$) Potts model
\cite{GTAU2011a}. We should immediately say  that to obtain
recursive equations is becoming more difficult and time-consuming.
\begin{figure} [!htbp]\label{Korean-potts}
\centering
\includegraphics[width=40mm]{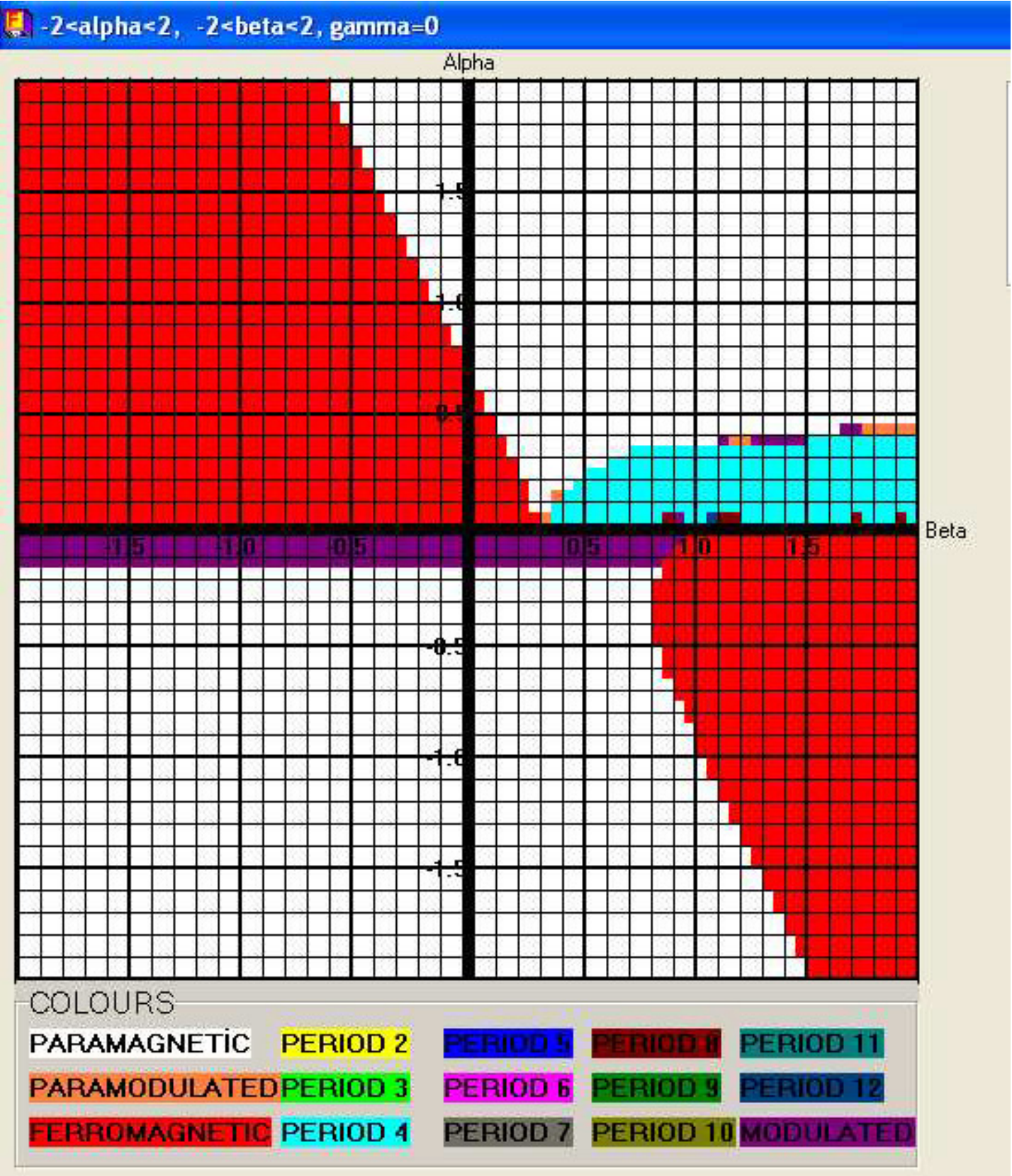}\ \ \ \ \ \ \ \ \ \ \ \
\includegraphics[width=40mm]{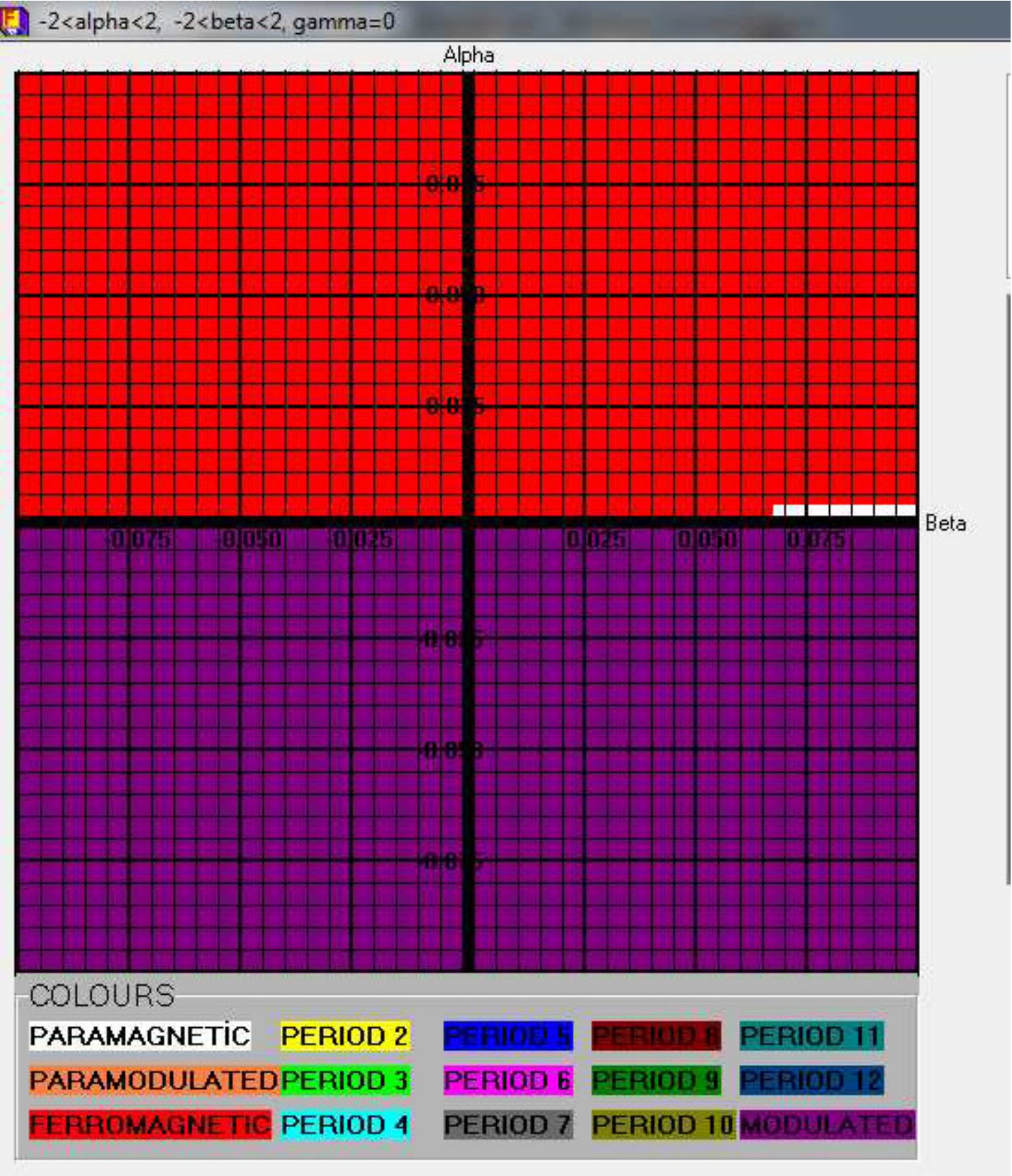}
\caption{(Colour online) Left-hand: Phase diagram of the Potts model
\cite{GTAU2011a} for $\gamma=0$ on the rectangular region
$[-2,2]\times [-2,2]$. Right-hand: The phase diagram of
Potts model \cite{GTAU2011a} for $\gamma=0$ on the rectangular
region $[-0.1,0.1]\times [-0.1,0.1]$ (see \cite{GTAU2011a} for the
figure) (see \cite{GTAU2011a} for the figure).
}\label{Korean-potts}
\end{figure}

By examining the variation of the wave vectors, it is possible to
determine the commensurate phases in the modulated regions.
Looking at the related work, the regions in the considered phase
diagrams are examined in smaller intervals and the transitions
from one phase to the other are seen more clearly.

Let us use the fixed value $\gamma=0$ to see the diagrams. A two
phase diagram is shown in figure \ref{Korean-potts}. The
modulated phase is reached in a very narrow interval $-0.11<\alpha
<0$ with $\beta < 0$ [see figure \ref{Korean-potts} (left)]. In
order to investigate the phases with a very small mesh width of
0.005, in a more narrow square $[-0.1,0.1]\times [-0.1,0.1]$, the
phase diagrams of the Potts model associated with the Hamiltonian
\eqref{Potts model3} are plotted [see figure \ref{Korean-potts}
(right)].
\begin{rem}
In the results that we have obtained as a team in the last decade, we
followed the above method to investigate the dynamical behaviour of
$q$-state Potts models and to plot the phase diagrams in detail (see
\cite{GTA2009a,GTAU2011a}).
\end{rem}

\subsection{The phase diagrams of 3-state Potts model on the Cayley tree of the  order 3} 

As emphasized earlier, getting the
recurrence equations associated to the Potts model is much more
difficult compared to the Ising model. When the number of branches
increases, it becomes even more difficult to find the recursive
equations. The systems of nonlinear equations have been obtained
for some Potts models with the competing interaction on the Cayley
tree of arbitrary order. The behavior of the nonlinear dynamical
systems corresponding to the Potts models obtained for
Hamiltonians with various coupling  constants on the Cayley tree
has been studied.

Note that this subsection is based on the paper
\cite{TGUA2013ACTA}. The appropriate Hamiltonian for the 3-state
Potts model has the form with competitive NN, PNNN, and two-level
triple interactions as:
%For 3-state Potts model, the relevant Hamiltonian with competing
%nearest-neighbor, prolonged next-nearest-neighbor and two-level
%triple interactions has the form
%\begin{small}\langle\rangle
\begin{equation}\label{Potts-hm4}
H(\sigma )=-{{J}_{tt}}\sum\limits_{\langle
\overline{x,y,z}\rangle}{{{\delta }_{\sigma (x)\sigma (y)\sigma
(y)}}}-
{{J}_{pn}}\sum\limits_{\rangle\widetilde{x,y}\langle}{{{\delta
}_{\sigma (x)\sigma (y)}}}- {{J}_{nn}}\sum\limits_{\langle
x,y\rangle}{{{\delta }_{\sigma (x)\sigma (y)}}},\end{equation}
%\end{small}
where ${{J}_{tt}},{{J}_{pn}},{{J}_{nn}}\in \mathbb{R}$ are
coupling constants and $\delta$ is the Kronecker symbol. Here, the
function
\[{{\delta }_{\sigma (x)\sigma (y)\sigma (y)}}=\left\{ \begin{matrix}
   1; & \sigma (x)=\sigma (y)=\sigma (y)  \\
   0; & \text{otherwise}, \\
\end{matrix} \right.
\]
is expressed as the generalized Kronecker's symbol.
\begin{figure} [!htbp]\label{Configuration-v1-3-tree}
\centering
\includegraphics[width=30mm]{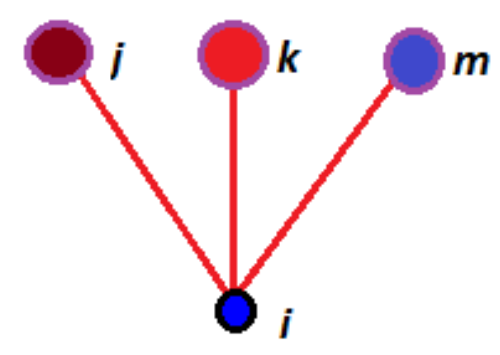}%\ \ \ \ \ \ \ \ \ \ \ \
\caption{(Colour online) Representation of the configurations in
$V_1$ on a Cayley tree of the order three, $i,j,k,m\in \{1,2,3\}$. For
$n=1$, the number of configurations is 81.}\label{Potts-phase-D4}
\end{figure}

If we take into account the 3-state Potts model associated with
the Hamiltonian given in \eqref{Potts-hm4} on a Cayley tree of the
order 3, we obtain the following partition equation:
\begin{eqnarray}\label{Partition-Function-Seyit}
Z^{(n)}\left(
\begin{array}{ccc}
 j & k & m \\
  & i &
\end{array}
\right)&=&\re^{A}\re^{B}\re^{C}Z^{(n)}(i,j)Z^{(n)}(i,k)Z^{(n)}(i,m),
\end{eqnarray}
where $A=\beta J_{nn}(\delta _{ij}+\delta _{ik}+\delta _{im}),
B=\beta J_{tt}(\delta _{ijk}+\delta _{ikm}+\delta _{ijm}), C=\beta
J_{pn}(\delta _{jk}+\delta _{jm}+\delta _{km})$.

Due to $i,j,k,m\in \{1,2,3\}$ there are $3^{4}=81$ different
partial partition functions
$$
Z^{(n)}\left(
\begin{array}{ccc}
 i_1 & i_2& i_3 \\
  & i_0 &
\end{array}
\right).
$$
Moreover, the total partition function $Z^{(n)}$ in volume $V_n$ is
obtained as;
$$
Z^{(n)}=\sum _{i_0,i_1,i_2,i_3=1}^3 Z^{(n)}\left(
\begin{array}{ccc}
 i_1 & i_2& i_3 \\
  & i_0 &
\end{array}
\right).
$$
However, it is reasonable to assume that the different branches
are comparable, as is common in the tree models. We have only
chosen five independent variables in \cite{TGUA2013ACTA} as
follows;
% It is reasonable, though, to assume that the different branches are
%equivalent, as is usually done for models on trees. In
%\cite{TGUA2013ACTA}, we have only chosen  five independent
%variables,
$$
z_1^{(n)}=Z^{(n)}\left(
\begin{array}{ccc}
 1 & 1 & 1 \\
  & 1 &
\end{array}
\right), \quad z_2^{(n)}=Z^{(n)}\left(
\begin{array}{ccc}
 1 & 1 & 1 \\
  & 2 &
\end{array}
\right), \quad z_3^{(n)}=Z^{(n)}\left(
\begin{array}{ccc}
 2 & 2 & 2 \\
  & 1 &
\end{array}
\right),
$$
$$\quad z_4^{(n)}=Z^{(n)}\left(
\begin{array}{ccc}
 2 & 2 & 2 \\
  & 2 &
\end{array}
\right), \quad z_5^{(n)}=Z^{(n)}\left(
\begin{array}{ccc}
 3 & 3 & 3 \\
  & 2 &
\end{array}
\right).
$$
%We introduce five  new variables as:
%$$
%Z^{(n)}=\sum _{i_0,i_1,i_2,i_3=1}^3 Z^{(n)}\left(
%\begin{array}{ccc}
% i_1 & i_2& i_3 \\
%  & i_0 &
%\end{array}
%\right).
%$$
%It is reasonable, though, to assume that the different branches
%are equivalent, as is usually done for models on trees. In
%\cite{TGUA2013ACTA}, we have only chosen  five independent
%variables, namely $$z_1^{(n)}=Z^{(n)}\left(
%\begin{array}{ccc}
% 1 & 1 & 1 \\
%  & 1 &
%\end{array}
%\right), z_2^{(n)}=Z^{(n)}\left(
%\begin{array}{ccc}
% 1 & 1 & 1 \\
%  & 2 &
%\end{array}
%\right),z_3^{(n)}=Z^{(n)}\left(
%\begin{array}{ccc}
% 2 & 2 & 2 \\
%  & 1 &
%\end{array}
%\right),
%$$
%$$z_4^{(n)}=Z^{(n)}\left(
%\begin{array}{ccc}
% 2 & 2 & 2 \\
%  & 2 &
%\end{array}
%\right),z_5^{(n)}=Z^{(n)}\left(
%\begin{array}{ccc}
% 3 & 3 & 3 \\
%  & 2 &
%\end{array}
%\right).
%$$
We investigate the relationship between the partition function on
$V_n$ and the partition function on subsets of $V_{n-1}$ to get
the recursive equations \cite{TGUA2013ACTA}. If we add the
following five new variables to the equation:
$$
u_i^{(n)}=\sqrt[3]{z_i^{(n)}},\ \text{for} \ i=1,2,...5,
$$
after some very long and tedious calculations, we derive the
corresponding recursive equations:
{\small\begin{eqnarray}\label{3-s-3-order-dynamic}
\left\{\begin{array}{l} u_1^{(n+1)}=a^3c\left[\sum\limits _{i=0}^3
C_i^{3}\left(bu_1^{(n)}\right)^{3-i}+\left(2u_2^{(n)}\right)^ia^{\frac{3i(i-3)}{2}}\right],\\
u_2^{(n+1)}=\sum\limits_{i=0}^3 C_i^{3}(u_3^{(n)})^ib^i
\left[\sum\limits _{j=0}^{3-i} C_j^{3-i}\left(u_4^{(n)}\right)^j\left(u_5^{(n)}\right)^{3-i-j}a^{\frac{3j(j-3)}{2}}\right],\\
u_3^{(n+1)}=\sum\limits _{i=0}^3
\left[\left(u_1^{(n)}\right)^{3-i}\left(u_2^{(n)}\right)^i(b+1)^ia^{\frac{3i(i-3)}{2}}\right],\\
u_4^{(n+1)}=a^3c\left[\sum\limits _{i=0}^3
C_i^{3}(bu_4^{(n)}){}^ia^{\frac{3i(i-3)}{2}}\sum\limits
_{j=0}^{3-i} \left(C_j^{3-i}\left(u_3^{(n)}\right)^j\left(u_5^{(n)}\right)^{3-i-j}\right)\right],\\
u_5^{(n+1)}=\sum\limits _{i=0}^3
C_i^{3}(u_5^{(n)})^ib^i\left[\sum\limits _{j=0}^{3-i}
C_j^{3-i}\left(u_4^{(n)}\right)^j\left(u_3^{(n)}\right)^{3-i-j}a^{\frac{3j(j-3)}{2}}\right],\end{array}
\right.
\end{eqnarray}}
where $C_i^{3}=\left(
\begin{array}{c}
 3 \\
 i
\end{array}
\right)$.
Let $\alpha=T/J_{nn}$, $\beta=-J_{pn}/J_{p}$ and
$\gamma=J_{tt}/J_{nn}$. In \cite{TGUA2013ACTA}, to investigate
the phase diagrams, as it is a common approach in the literature,
we  obtained the recursive relations:
\begin{eqnarray}\label{3-s-3-order-dynamic1}
\left\{\begin{array}{l}
x^{(n)}=\frac{2u_2^{(n)}+u_3^{(n)}+u_5^{(n)}}{u_1^{(n)}+u_4^{(n)}},
y_1^{(n)}=\frac{u_1^{(n)}-u_4^{(n)}}{u_1^{(n)}+u_4^{(n)}},\\
y_2^{(n)}=\frac{u_2^{(n)}-u_3^{(n)}}{u_1^{(n)}+u_4^{(n)}},
y_3^{(n)}=\frac{u_2^{(n)}-u_5^{(n)}}{u_1^{(n)}+u_4^{(n)}}.\end{array}
\right.
\end{eqnarray}
If we substitute
\begin{eqnarray*}
u_1&=&\frac{(1+y_1)A}{2},\quad u_2=\frac{(x+y_2+y_3)A}{2}, \quad u_3=\frac{(x+y_3-3y_2)A}{2},\\ 
u_4&=&\frac{(1-y_1)A}{2}, \quad u_5=\frac{(x+y_2-3y_3)A}{2},
\end{eqnarray*}
where $A=u_1+u_4$, we finally obtain a system of the following
recursive relations;
\begin{equation}\label{3-s-3-order-dynamic1a}
\left\{\begin{array}{l}
x^{(n+1)}=F(x^{(n)},y_1^{(n)},y_2^{(n)},y_3^{(n)}),\\
y_1^{(n+1)}=G(x^{(n)},y_1^{(n)},y_2^{(n)},y_3^{(n)}),\\
y_2^{(n+1)}=H(x^{(n)},y_1^{(n)},y_2^{(n)},y_3^{(n)}),\\
y_3^{(n+1)}=K(x^{(n)},y_1^{(n)},y_2^{(n)},y_3^{(n)}).
\end{array}
\right.
\end{equation}
Note that since the equations are excessively long, we do not
write the complete result of the iterative system of equations
\eqref{3-s-3-order-dynamic1a} here. As a result, we recommend
consulting the reference \cite{TGUA2013ACTA}.

We obtain the initial conditions on $V_{1}$ under boundary
conditions $\overline{\sigma}^{(n)}(V\setminus V_n)\equiv 1$ as
follows:
$$
x^{(1)}=\frac{2b^2+a^2c^2+1}{a^3c^3b^2+ac}, \quad
y_1^{(1)}=\frac{a^2c^2b^2-1}{a^2c^2b^2+1}, \quad
y_2^{(1)}=\frac{b^2-a^2c^2}{a^3c^3b^2+ac}, \quad
y_3^{(1)}=\frac{a^2c^2-1}{a^3c^3b^2+ac}.
$$
The recursive equations \eqref{3-s-3-order-dynamic1} show how
their impact propagates along the tree.
\begin{figure} [!htbp]
\centering
\includegraphics[width=40mm]{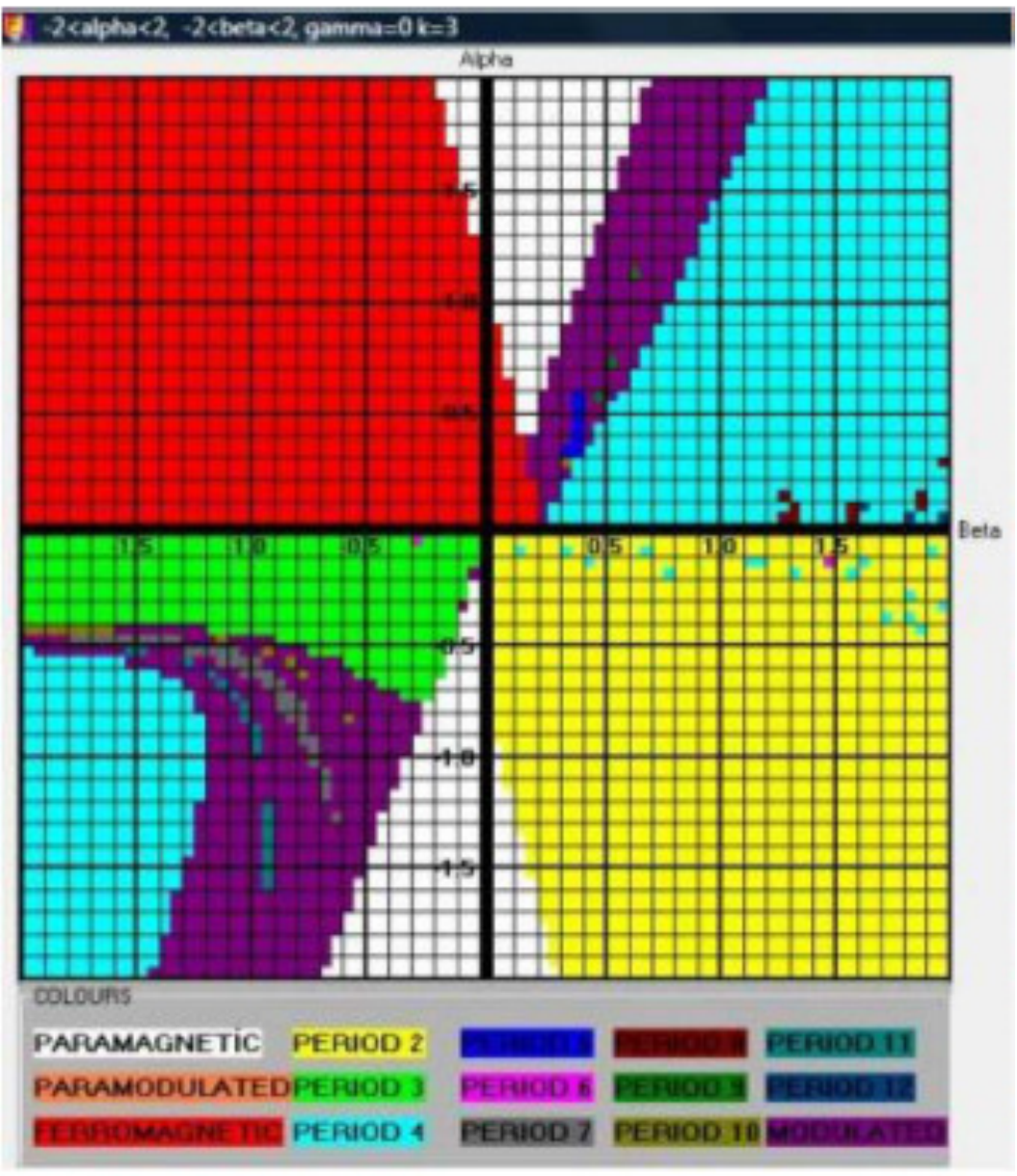}\ \ \ \ \ \ \ \ \ \ \ \
\includegraphics[width=42mm]{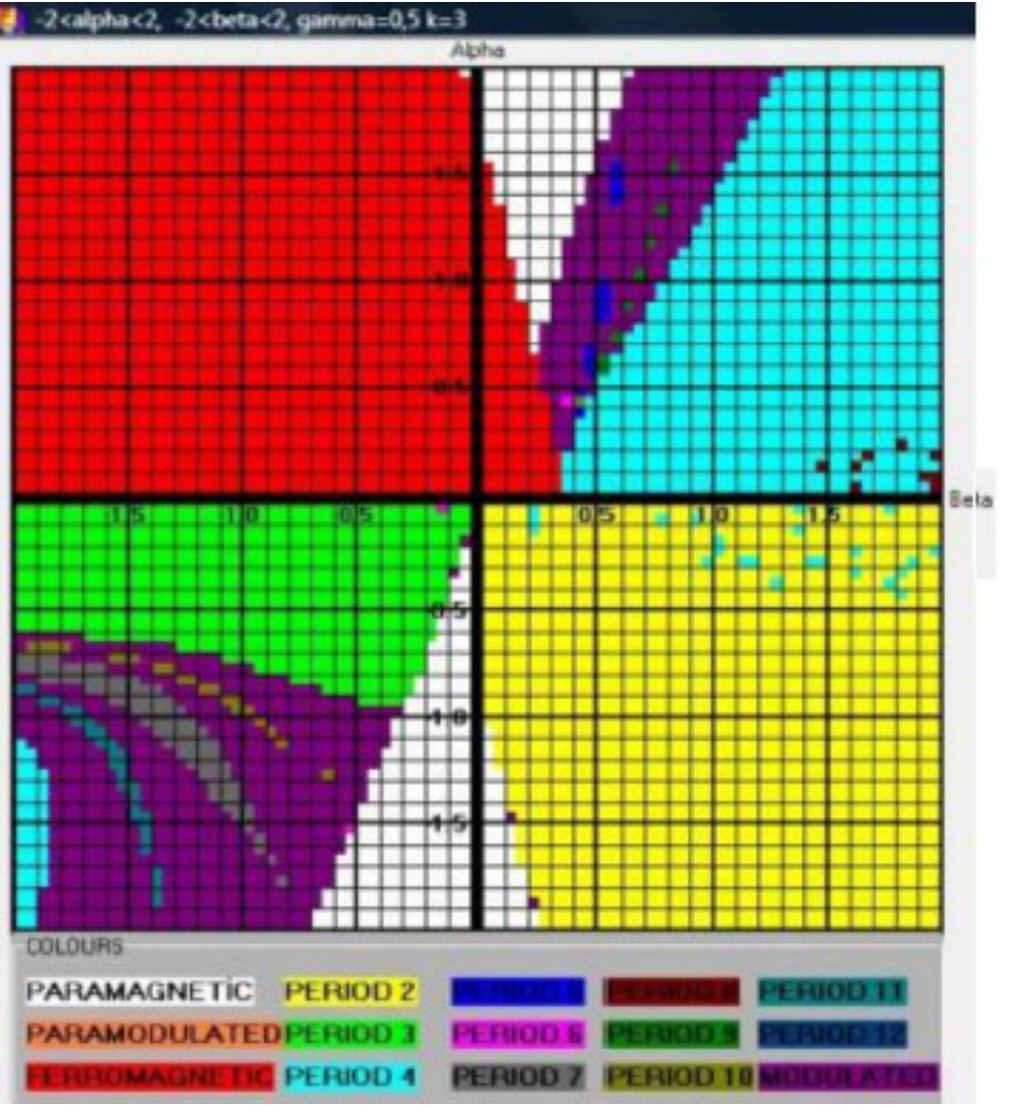}
\caption{(Colour online) The phase diagrams of the model for $k =
3$, $\gamma= 0$ (left-hand) and  $\gamma= 0.5$ (right-hand), respectively
(see \cite{TGUA2013ACTA}). }
\label{Potts-phase-D4}
\end{figure}
As mentioned in \cite{TGUA2013ACTA}, in the simplest situation a
fixed point $\left(x^*,\,\, y_1^*, \,\, y_2^*, \,\,y_3^*\right)$ is reached. After
a large number of iterations, if the fixed point is satisfied as
$y_1^*=0,\,\, y_2^*=0, \,\, y_3^*=0$, then it corresponds to a paramagnetic
phase. If $y_1^*, \,\, y_2^*, \,\, y_3^*\neq 0$, the phase diagram corresponds
to the ferromagnetic phase. 
When we look at the figures~\ref{Potts-phase-D4} here, we can see that there are many
different and diversified phases compared to the previous ones.

In \cite{Gok-Tez}, the phase diagrams of $q=5, q=6$ and $q=15$
spin states Potts models were obtained. Additionally, Gok
\cite{Gok-Tez} analyzed the phase diagrams of 3-state Potts model
with competing for nearest neighbor, next nearest neighbor and one
level triple of neighbors interactions.

Ganikhodjaev et al. \cite{GNR2013POTTS} defined a single-trunk
Cayley tree, obtained recursive equations for the model with
competing interactions on the Cayley tree and for the same model
on the single-trunk Cayley tree, and demonstrated how to reduce
the recursion equations on the Cayley tree to the simpler
recursive equations on the single-trunk Cayley tree, 
extended the results of \cite{GAUT2011a} to the Potts model with
competing interactions on a Cayley tree of arbitrary order $k$.
The role of the order $k$ was clarified. They  studied 
the differences between both types of the model on Cayley tree
order 3 and order 4 and order 10. They extended the findings of
\cite{GAUT2011a} to the Potts model with competing interactions on
a Cayley tree of any order $k$. The role of the order $k$ was
described. They investigated the differences between the two types
of models on Cayley tree of orders 3, 4, and 10. They demonstrated
that there were six phases in the phase diagrams: ferromagnetic,
paramagnetic, antiferromagnetic, period 3, antiphase, and
modulated phase.
\subsection{Comments and remarks}
Note that the results in the section \ref{Dynamical behavior of
Potts model} are based on the studies
\cite{GTAU2011a,GTA2009a,GMM2006,TGUA2013ACTA,GTUA2011C,TGAU2010AIP,Gok-Tez,GNR2013POTTS}.
In this section, the system of nonlinear equations for 3-state
Potts model on Cayley tree of the finite-order is derived.
Secondly, the structure of the phase diagrams of given $q$-state
Potts models is examined and compared and the wave vectors are plotted.

In the case $k=3$, the recursive equations are obtained. When the
phase diagrams are compared with the results in the Potts model
given in the previous sections, it is found that new phase
regions appear in the phase diagrams. Thus, we show that the
coupling constants have considerable effects on the phase diagrams.

In our works, we  also see that the number of branches in the
Cayley tree causes considerable changes in the phase diagrams. The
phase types obtained in some regions are either completely lost or
shrinking. In this case, in the regions where the modulated phases
are observed, the differences are also determined by drawing the
wave vectors corresponding to the magnetization in order to
distinguish between the superimposed phases and the long period
phases. Since more comprehensive information and findings are
included in the related studies, details are not given here.

This is not rewritten here because of the length of the recursive
equations. In \cite{TGUA2013ACTA}, we generalized the results of
Ganikhodjaev et al. \cite{GTA2009a} to the 3-state Potts model
with competing NN, PNNN and two-level triple neighbor interactions
on a a third-order Cayley tree. We studied the modulated phases
arising from the frustration effects introduced by NN, PNNN and
two-level triple neighbor interactions.

For $(x^{*}, 0,0,0)$, we
performed the stability of the dynamical system in our previous
work. However, the problem of linearization around $(x^{*},
y_1^{*}, y_2^{*}, y_3^{*})$ (where at least one of $y_1^{*},
y_2^{*}, y_3^{*}$ is different from zero) is extremely difficult.

In \cite{GNR2013POTTS}, Ganikhodjaev et al. defined a single-trunk
Cayley tree. By deriving the recursion equations for the same
model on the single-trunk Cayley tree, they showed how to reduce
the recursion equations on a Cayley tree to the simpler recursion
equations on the single-trunk Cayley tree. There are many open
problems to be solved using this approach.

A detailed examination of $q$-state Potts models on Cayley tree of
arbitrary order has not yet been done. In order to fully
understand the overlapping phase regions in modulated phases, we
should investigate the Lyapunov exponents, the magnetization, and
the length of wave-vectors. In addition, as the number of spins
increases, it becomes more difficult for us to analyze the
corresponding $q$-state Potts models. Therefore, we need new
methods to dissect the $q$-state Potts models  with
arbitrary spin values $q$.

\begin{rem} If there is no paramagnetic phase, then the transition between
the phases and the analysis of the stability problem become a more
complex issue which leads to many challenging problems. We will
investigate these unresolved problems in the future.  
\end{rem}

%\newpage
\section{\textbf{The dynamical behaviors of the Ising models on chandelier networks}}\label{Cayley-like lattice}
I have an interesting story regarding how the chandelier networks
model came to be. During the days when I was focused on obtaining the
phase diagrams of the lattice models on the Cayley tree as part of
our project's scope, I noticed the chandelier in our living room's
ceiling while playing on the floor with my two-year-old son. When
I thought of the Cayley tree's structure, I imagined that by
duplicating the tree's structure, a Cayley-like lattice, which was
the reverse of the tree, could be identified. Thus, since that date,
I have been working on the concept of \textbf{chandelier network}.

Take, for example, a chandelier network with $k$ bulbs suspended
from the ceiling. Assume that the identical $k$ quadrants are strung on
each light of the first chandelier. We get a chandelier set that
looks like a semi-infinite Cayley tree in this case. We 
presume that the nearest lighting on the same level are linked. By
computing the internal, external, and entire energies
corresponding to a Hamiltonian on the chandelier network that we
have established, we can analyze the themes investigated in
statistical physics.

As stated in the introduction, the Cayley tree is an unrealistic
lattice. Many problems in statistical physics have been studied in
the Cayley tree (or Bethe lattice), because the various operations
and computations on this lattice are easier to grasp than the
d-dimensional $\mathbf{Z}^{\text{d}}$ lattice. Hence, the results
of the Cayley tree have served as inspiration for the
d-dimensional $\mathbf{Z}^{\text{d}}$ lattice.

Then, we considered a matching chandelier network, which we
labelled triple, quadruple, quintuple, and so on. Furthermore, we looked
at how Ising models behaved dynamically on these chandelier
networks. A limited number of studies have been carried out on
chandelier networks so far
\cite{AUT2010AIP,UA2010PhysicaA,UGAT2012ACTA,UA2011CJP}. Although
the results are similar to those of lattice models on the Cayley
tree, I believe that in the future we will be able to find more
realistic results about the lattice models (Ising and Potts)
defined on a chandelier network. Now, let us briefly present our
findings obtained by elaboration. For more comprehensive
information, we encourage the readers to consult 
\cite{AUT2010AIP,UGAT2012ACTA}.

\subsection{Structure of the chandelier networks}

In this subsection, we are going to introduce a chandelier network
with $k$ branches which is similar to a semi-infinite Cayley tree
of order $k$ ($k\geqslant 3$).

In a chandelier network with $k$ branches (order), the starting
vertex of the chandelier network is connected to $k$ vertices,
while all the other vertices are related to $k+3$ vertices.
Let $C^k=(V, E, i)$ be order $k$ chandelier network  with a root
vertex $x^{(0)}\in V$. The set of vertices in the lattice is
denoted by $V$, and the set of edges is denoted by $E$. The
notation $i$ represents the incidence function corresponding to
each edge $e\in E$, with end points $x_1,x_2\in V$. Except for the
initial vertex $x^{(0)}$, each vertex has $(k+3)$ nearest
neighbors belonging to $V$ as the set of vertices and the set of
edges. It is obvious that the root vertex $x^{(0)}$ has $k$
nearest neighbors (see figure \ref{ternary-chandiler}).
Consider a three-lamp chandelier that hangs from a ceiling. Assume that
the same three quadrants from the first chandelier were added to
each bulb (see figure \ref{ternary-chandiler}).

In this case, we get a network resembling a semi-infinite Cayley
tree. We  suppose that each lamp is linked to the
lamps in its immediate vicinity. 
For $x,y\in V$, the number of edges in the shortest path from $x$
to $y$ on the chandelier network $C^{k}$ ($k>2$) is denoted by
$d(x,y)$. The $0$-th level refers to the fixed vertex $x^{(0)}$,
while the $n$-th level refers to the vertices in $W_n$. We set
$|x|=d(x,x^{(0)})$, $x\in V$ for the sake of simplicity.
\begin{figure} [!htbp]\label{ternary-chandiler}
\centering
\includegraphics[width=60mm]{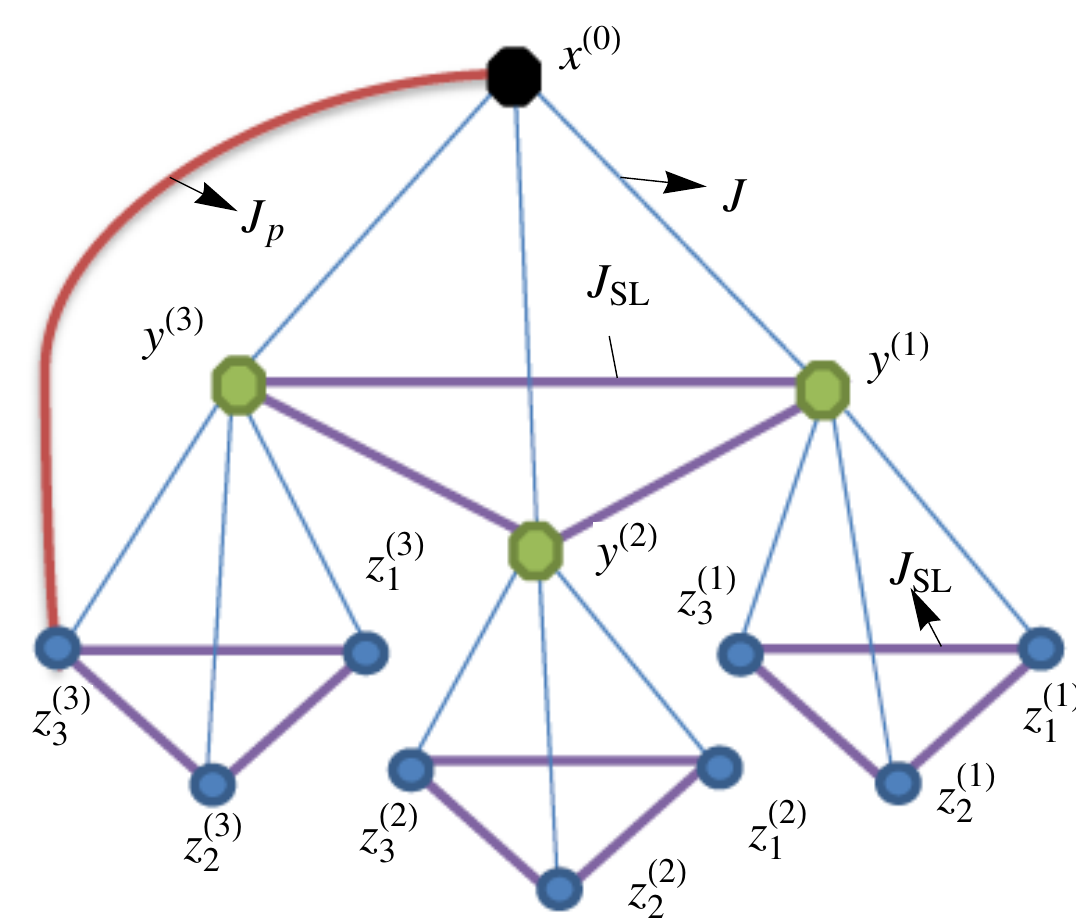}
\caption{(Colour online) third order chandelier network of two
level.}\label{ternary-chandiler}
\end{figure}

The sphere with radius $n$ on $V$ is denoted by
$$
W_n^{(P)}=\{x\in V: d(x,x^{(0)})=n \}.
$$
For example, the sphere with radius $2$ on $V$ is denoted by $
W_2^{(P)}=\{z^{(u)}_v: u,v=1,2,3 \}$ (see figure \ref{ternary-chandiler}).\\
The ball with radius $n$ is denoted by
$$
V_n^{(P)}=\{x\in V: d(x,x^{(0)})\leqslant
n\}=\bigcup_{k=0}^{n}W_k^{(P)}.
$$
The set of direct prolonged successors of any vertex $x\in W_n$ is
denoted by
$$
S_k(x)=\{y\in W_{n+1}:d(x,y)=1 \}.
$$
The set of the same-level neighborhoods of any vertex $x\in W_n$
will be denoted by
$$
SL_k(x)=\{y\in W_{n}:d(x,y)=1 \}.
$$
It is clear that $|SL_k(x)|=2$, for all vertices $x\in W_n$.
\\ Now, let us give the definitions of the neighborhood types that
we considered in the chandelier networks.
\begin{defin}\label{neighborhoods}
\begin{enumerate}
    \item Two vertices $x$ and $y$, $x,y \in V$ are called {\it
\textbf{nearest-neighbors (NN)}} if there exists an edge $e\in E$
connecting them, which is denoted by $e=\langle x,y\rangle$.
    \item The nearest-neighbor vertices $x,y\in V$ that are not
prolonged are called {\it \textbf{same-level nearest-neighbors
(SLNN)}} if $|x|=|y|$ and are denoted by $\widetilde{\rangle
x,y\langle}$.
    \item Two vertices $x,y\in V$ are called {\it \textbf{the
next-nearest-neighbors (NNN)}} if there exists a vertex $z \in V$
such that $x, z$ and $y, z$ are NN, that is if $d(x,y)=2$.
    \item The next-nearest-neighbor
vertices $x\in W_n^{(P)}$ and $y\in W_{n+2}^{(P)}$ are called {\it
\textbf{prolonged next-nearest-neighbors}} if $|x|\neq |y|$,
$d(x,y)=2$ and is denoted by $\rangle x,y\langle$ (see figure
\ref{Triangular-Chandelier}).
%%\item Three vertices $x, y$ and $z$ are called a \emph{\textbf{triple of neighbors}}
%%if $<x, y>$, $<y, z>$ are nearest neighbors. they are denoted by
%%$< x, y, z >.$
%%\item The triple of vertices $x,y,z$ is called {\it \textbf{prolonged}}
%%if $x\in W_n,y\in S(x)$ and $z\in S(x)$ ($x\in W_n,y\in W_{n+1}$
%%and $z\in W_{n+2}$) for some nonnegative integer $n$  and is
%%denoted by $\widetilde{<x,y,z>}$.
%%\item The triple of vertices $x,y,z\in V$ that are not prolonged is
%%called {\it \textbf{two-level} } since $|x|=|z|$ and is denoted by
%%$\bar{<x,y,z>}$.
\end{enumerate}
\end{defin}
\subsection{An Ising model on triangular chandelier network (TCN)}
The phase diagrams for the Ising model on a third order triangular
chandelier network (TCN, shortly) are studied in
\cite{UA2010PhysicaA}, with competing nearest-neighbor
interactions $J_1$, prolonged next-nearest-neighbor interactions
$J_p$, and one-level next-nearest-neighbor quadruple interactions
$J_{l_1}$ (figure \ref{Triangular-Chandelier}).

The phase diagrams show the nonzero temperature multicritical
points (Lifshitz points) as well as various modulated phases. An
iterative technique similar to that found in real space
renormalization group frameworks is used to carry out this study.
%
%In  \cite{UA2010PhysicaA}, the phase diagrams for the Ising model
%on a Cayley tree-like lattice, called Triangular Chandelier
%(Figure \ref{Triangular-Chandelier}), with competing
%nearest-neighbor interactions $J_1$, prolonged
%next-nearest-neighbor interactions $J_p$ and one-level
%next-nearest-neighbor quadruple interactions $J_{l_1}$ are
%studied. The phase diagrams display the multicritical points (the
%Lifshitz points)  that are at nonzero temperature and many
%modulated phases. To perform this study, an iterative scheme
%similar to that appearing in real space renormalization group
%frameworks is established;

Let us consider the following Hamiltonian;
\begin{eqnarray}\label{3order-CL-Ham1}
H(\sigma )&=&-J_1\sum _{\langle x,y\rangle} \sigma (x)\sigma
(y)-J_p\sum _{\widetilde{\rangle x,y\langle}} \sigma (x)\sigma
(y)\nonumber \\ 
&&-J_o\sum _{\rangle \widehat{x,y}\langle} \sigma
(x)\sigma (y)-J_{l_1}\sum _{\rangle x,\widehat{y,z},t\langle}
\sigma (x)\sigma (y)\sigma (z)\sigma (t),
\end{eqnarray}
where we have competing nearest-neighbor $J_1$, prolonged
next-nearest-neighbor binary interactions $J_p$, one-level
next-nearest-neighbour binary interactions $J_o$, and one-level
next-nearest-neighbor quadruple interactions $J_{l_1}$.

%It retrieves, as an example, Ganikhodjaev and Uguz's previous work
%\cite{GU2011PhysicaA} on Vannimenus extension results for $k>3$.
In \cite{UA2010PhysicaA}, for some given values and signs of
$J_1,J_p,J_o$ and $J_{l_1}$, the phase diagrams were fully
determined at vanishing temperature. For typical values of
$J_{l_1}/J_1$ and $J_p/J_1$ at finite temperatures, numerous
interesting aspects were ovserved.
\begin{figure} [!htbp]\label{Triangular-Chandelier}
\centering
\includegraphics[width=70mm]{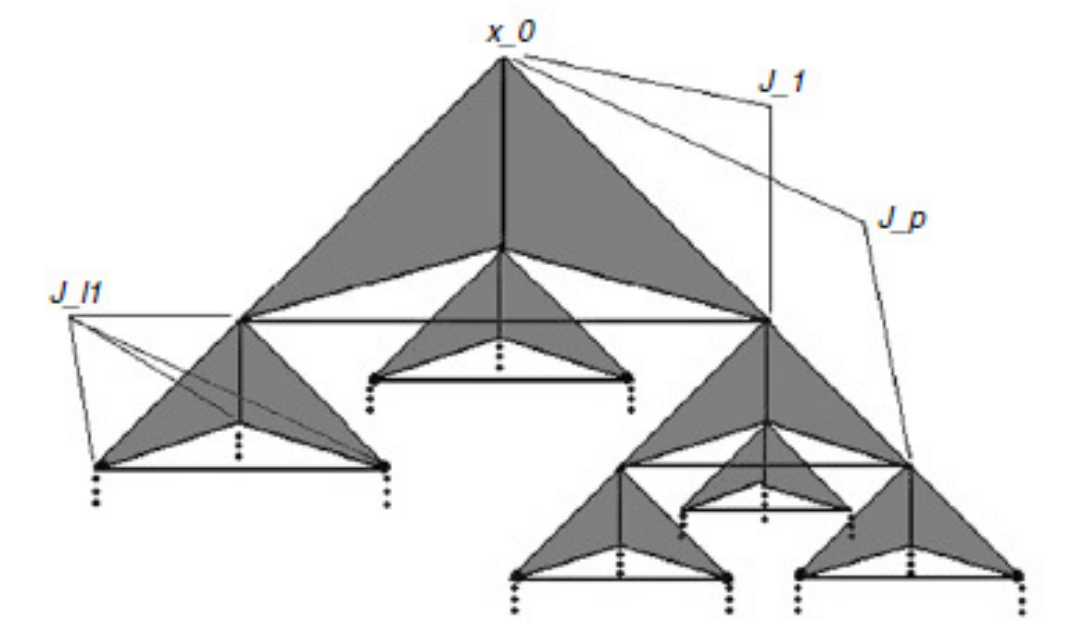}
\caption{An infinite TCN.  %(see \cite{UA2010PhysicaA}).
}\label{Triangular-Chandelier}
\end{figure}

\subsection{The phase diagrams of an Ising model on rectangular chandelier network (RCN)}
On a new kind of lattice which we called the rectangular
chandelier, the authors of \cite{AUT2010AIP} investigated the
phase diagrams of the Ising model with competing ternary and
nearest-neighbor binary interactions, as well as one-level quinary
interactions. Moreover, we observed the existence of multicritical
Lifshitz points at non-zero temperature. The same methods can be
used to obtain the phase diagrams on tetragon, pentagon,
hexagon, ..., and generally to understand the role of $k$ for
triangular $(k=3)$ and rectangular ($k=4$) chandelier networks as
in the case of a Cayley tree \cite{GU2011PhysicaA}.

To plot the phase diagrams of Ising models on chandelier networks,
we first derived the systems of nonlinear equations by applying
the procedure in the previous sections. We investigated relevant
phase diagrams in detail with the help of computer code by
calculating the initial conditions (see
\cite{AUT2010AIP,GU2011PhysicaA,UA2011CJP} for details).

%This can be accomplished mathematically in a simple manner.
\begin{figure} [!htbp]
\centering
\includegraphics[width=79mm]{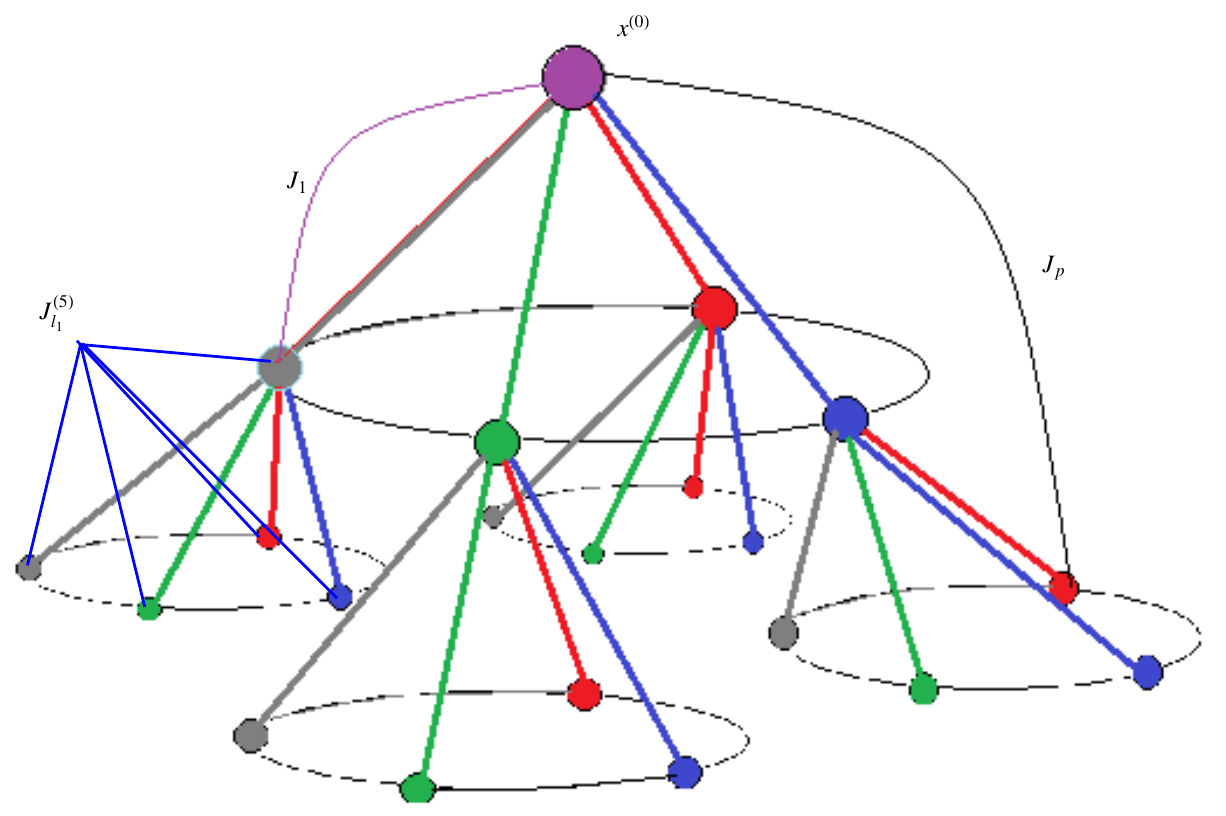}
\caption{(Colour online) A infinite RCN. $x^{(0)}$ is the root of this lattice,
and four new branches (edges) are formed from the root of the
lattice. We take into account competing interactions $J_1$, $J_p$,
and $J^{(5)}_{l_1}$, in this model.}
\label{Rectangular-Chandelier}
\end{figure}
\begin{figure} [!t]
\centering
\includegraphics[width=45mm]{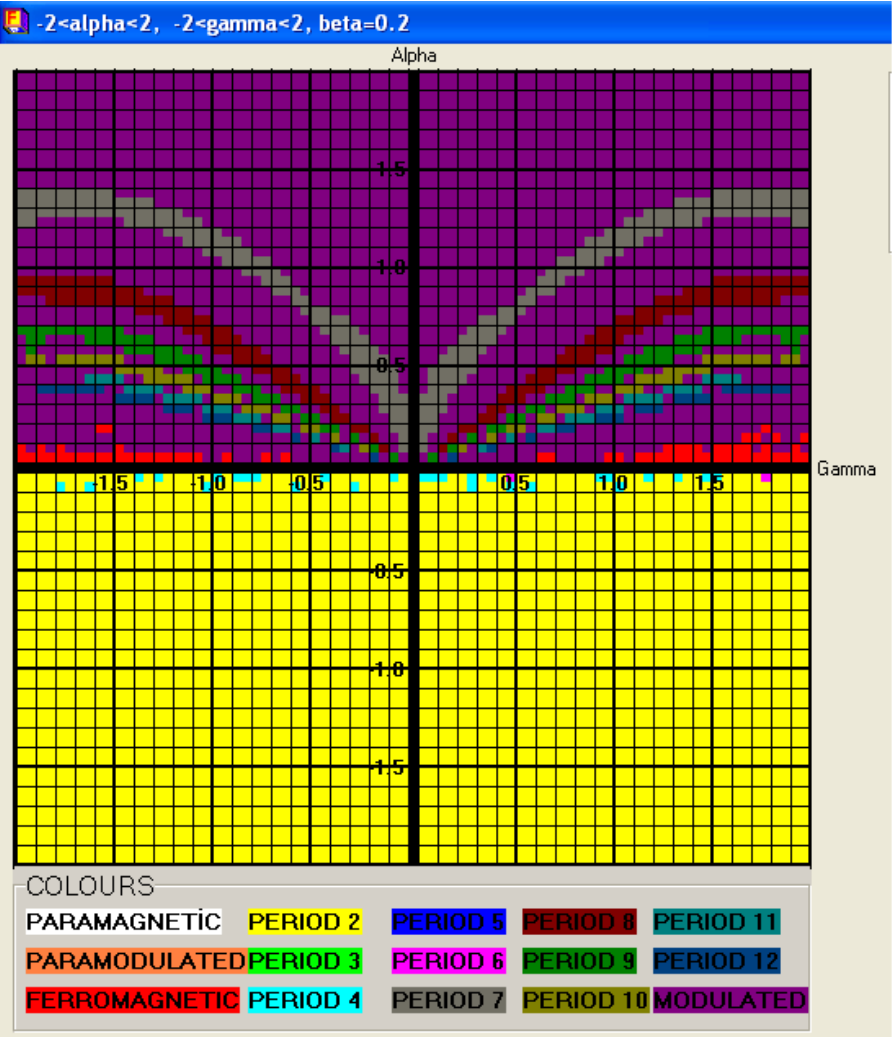}\ \ \ \ \ \ \ \ \ \ \ \
\includegraphics[width=45mm]{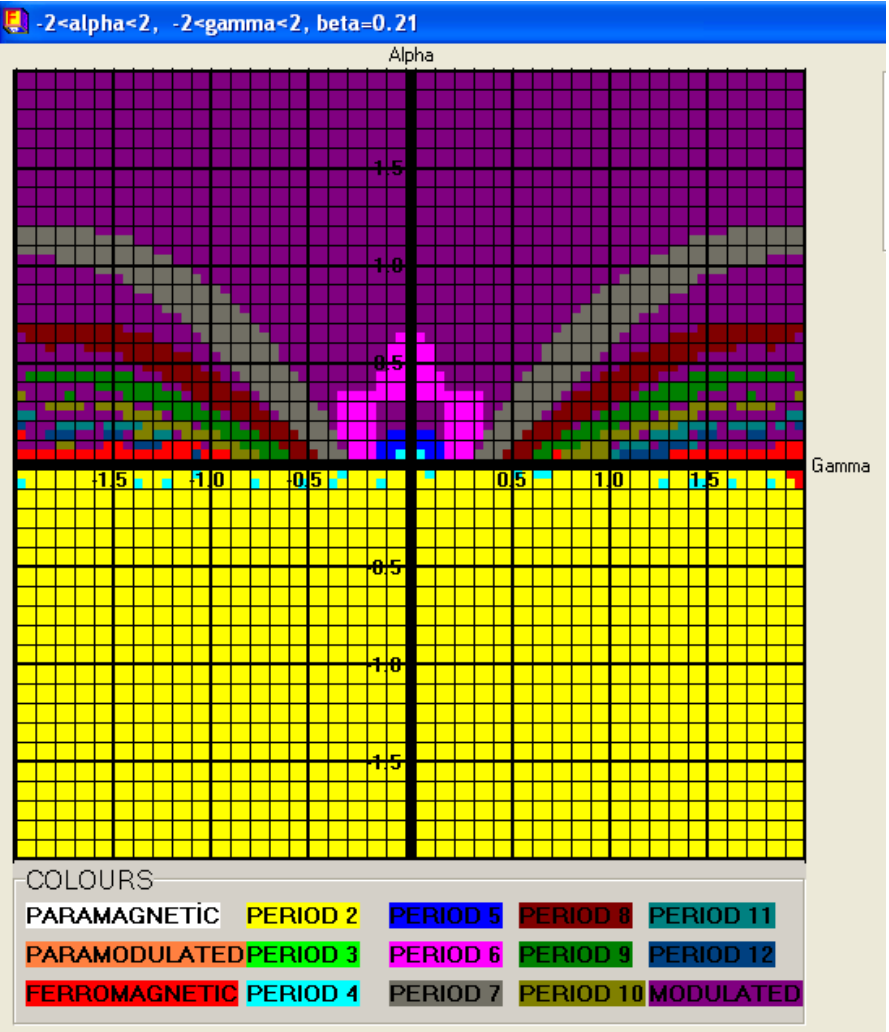}
\caption{(Colour online) The phase diagrams of an Ising model
associated with the Hamiltonian \eqref{Ham-RCN} on RCN, for
$\beta=0.2$ (left-hand) and $\beta=0.21$ (right-hand) (see
\cite{UA2011CJP}).}\label{RCN-beta=0.21}
\end{figure}
\begin{figure} [!t]
\centering
\includegraphics[width=45mm]{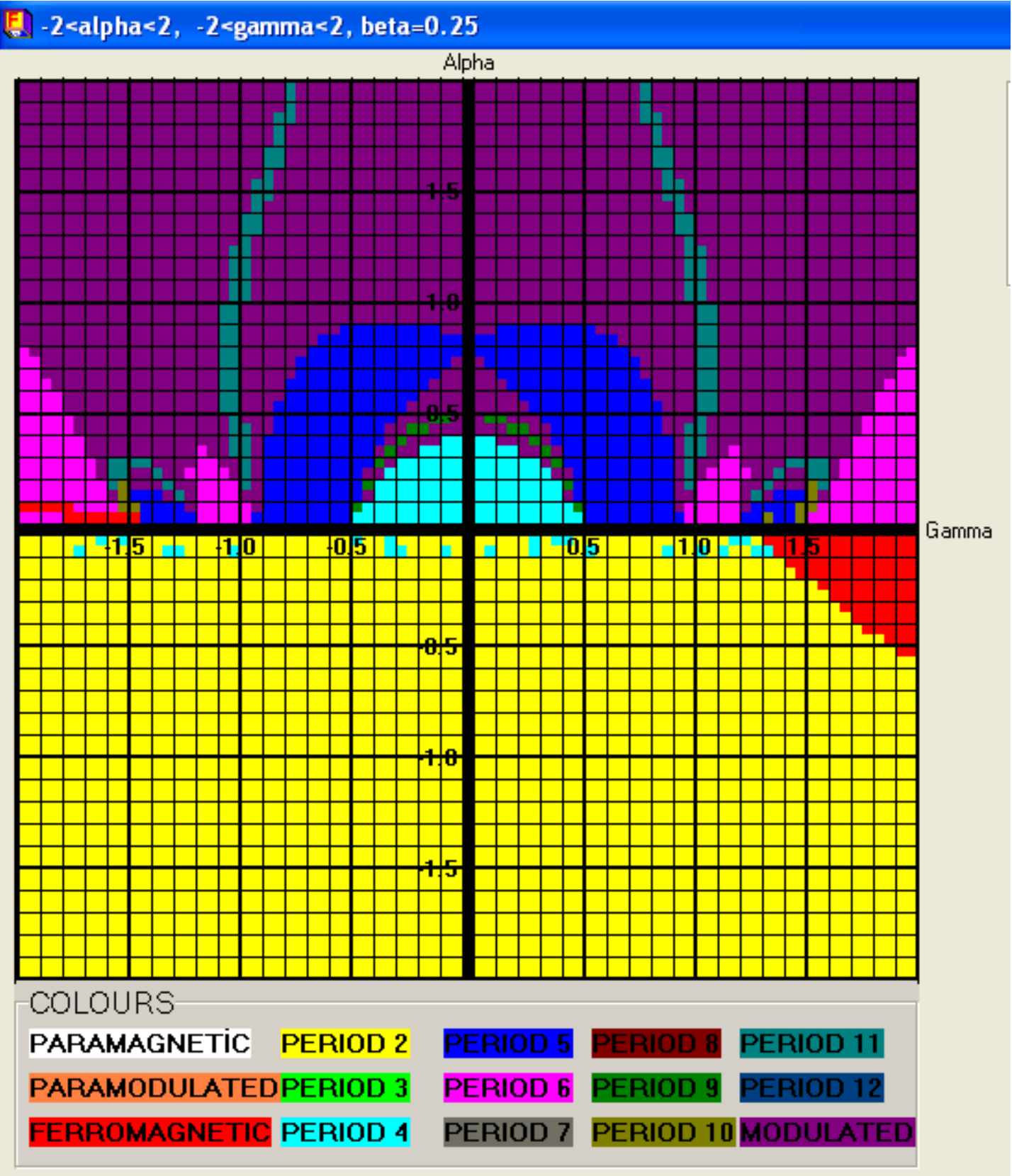}\ \ \ \ \ \ \ \ \ \ \ \
\includegraphics[width=45mm]{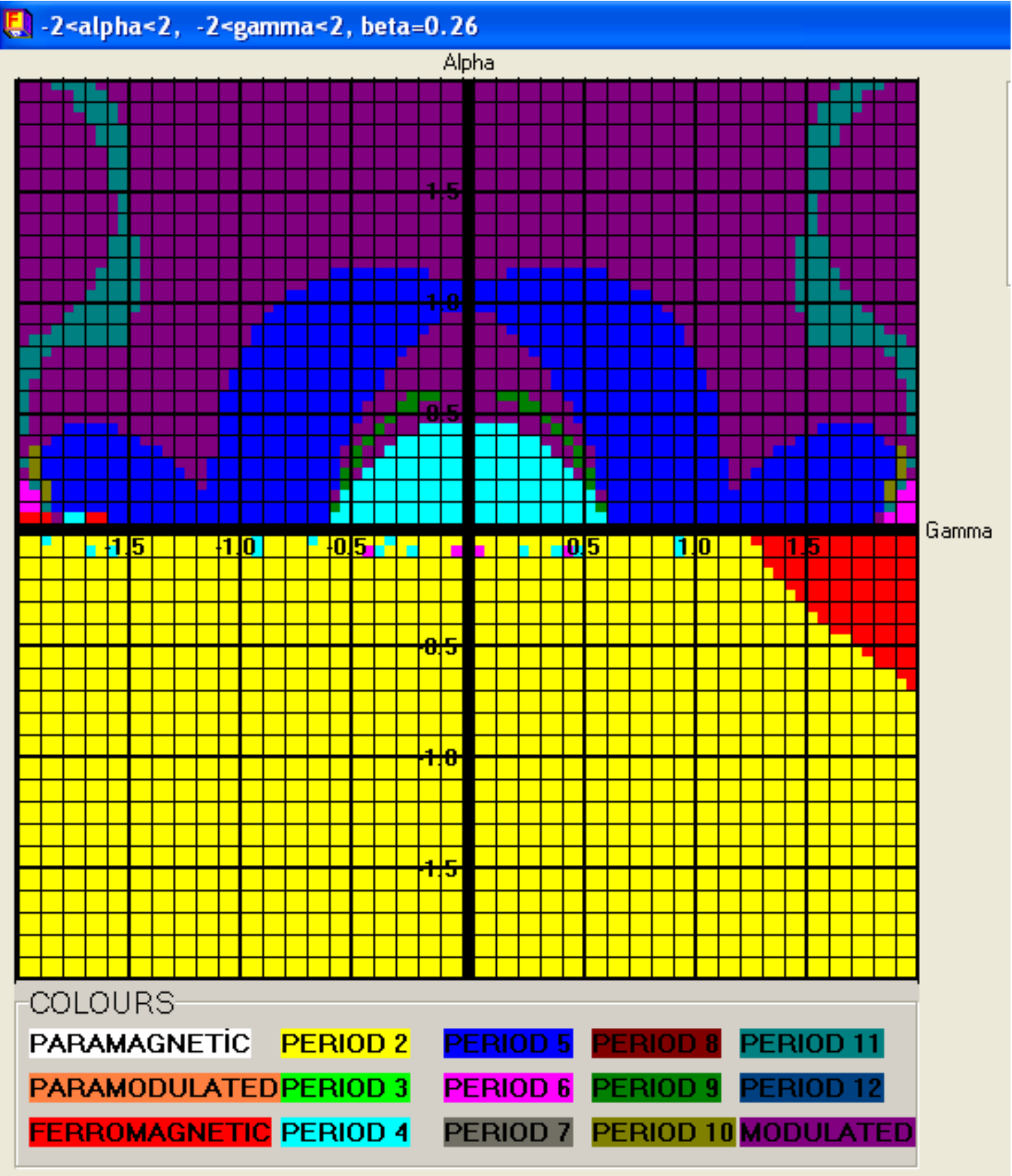}
\caption{(Colour online) The phase diagrams of an Ising model
associated with the Hamiltonian \eqref{Ham-RCN} on RCN, for
$\beta=0.25$ (left-hand) and $\beta=0.26$ (right-hand) (see
\cite{UA2011CJP}).}\label{RCN-beta=0.25}
\end{figure}

In \cite{UA2011CJP}, we considered the following Hamiltonian:
\begin{eqnarray}\label{Ham-RCN}
H(\sigma )=-J_1\sum _{\langle x,y\rangle} \sigma (x)\sigma
(y)-J_p\sum _{\rangle \overset{\tilde{ }}{x,y}\langle} \sigma
(x)\sigma (y)-J_{l_1}^{(5)}\sum _{\langle
x,y,\overline{z},t,w\rangle} \sigma (x)\sigma (y)\sigma (z)\sigma
(t)\sigma (w).
\end{eqnarray}
We studied the Ising model associated with Hamiltonian
\eqref{Ham-RCN} on the RCN (see figure
\ref{Rectangular-Chandelier}). The recursion relations numerically
provided us the phase diagrams in $(T/J_1=\alpha; -J_p/J_1=\beta;
J^{(5)}_{l_1}/J_1=\gamma)$ space. In addition, in
\cite{UA2011CJP}, the authors presented the variation of the
wave-vector with temperature in the modulated phase for some
critical points. For the case $\beta= 0.22$, we observed a
distinctive feature of the phase diagram. We detected new
interesting phases containing geometric figures, i.e., the phases
with \textbf{P5, P6, P7}, and \textbf{P8} in the first and second
quadrants (see the figures given in \ref{RCN-beta=0.21} and
\ref{RCN-beta=0.25}). As can be seen in these figures, small
changes in the value of $\beta$ completely change the phase
diagrams.

Note that when the phase diagrams obtained in \cite{UA2011CJP} and
\cite{AUT2010AIP} are compared, the paramagnetic phase 
completely disappears in the phase diagrams obtained for the
model associated with the Hamiltonian in~\cite{AUT2010AIP}.

\subsection{The phase diagrams of an Ising model on pentagonal chandelier network (PCN)}
We referred to the lattice in figure \ref{Pentegonal2}  as pentagonal
chandelier network (PCN, shortly).
\begin{figure} [!htbp]\label{Pentegonal2}
\centering
\includegraphics[width=55mm]{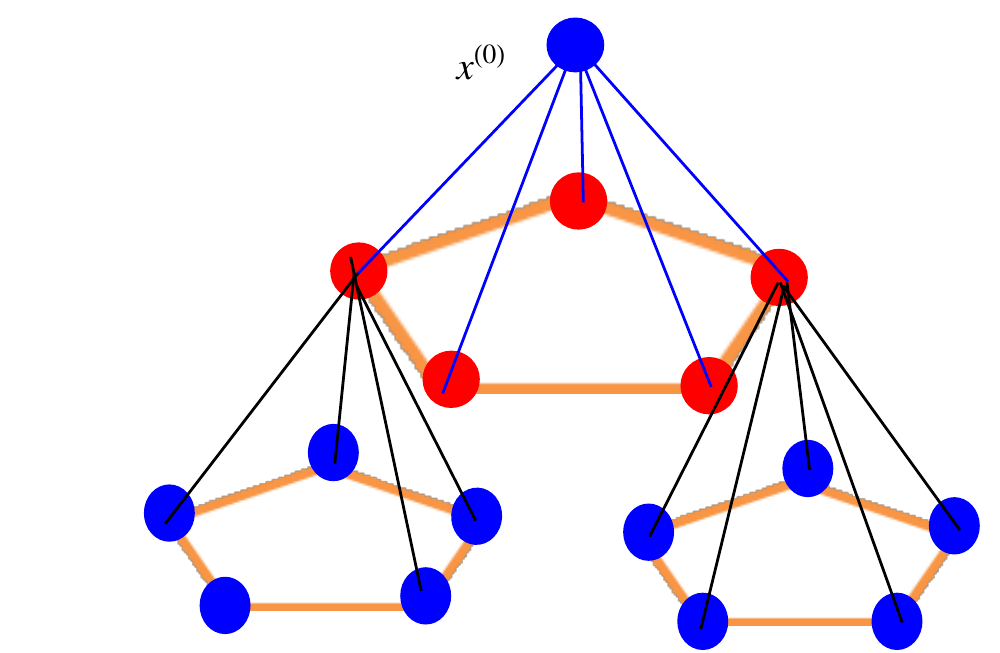}
\caption{(Colour online) The first and second-generation branch of a
PCN.}\label{Pentegonal2}
\end{figure}
On this network shown in figure \ref{Pentegonal2}, we consider a
Hamiltonian with spin values in $\Phi=\{-1,+1\}$ defined by
%
%\begin{eqnarray}\label{Pentegonal-hm} H(\sigma)&=&-J_1\sum
%_{\langle x_1,x_2\rangle} \sigma (x_1)\sigma (x_2)-J_p\sum
%_{\rangle\widetilde{x_1,x_2}\langle} \sigma (x_1)\sigma
%(x_2)-J_{l_1}^{(6)}\sum _{\rangle x_1,x_2,\cdots,x_2\langle}
%\sigma (x_1)\sigma (x_2)\cdots \sigma (x_6),
%\end{eqnarray}
%
\begin{eqnarray}\label{Pentegonal-hm} 
	H(\sigma)=&-&J_1\sum
	_{\langle x_1,x_2\rangle} \sigma (x_1)\sigma (x_2)-J_p\sum
	_{\rangle\widetilde{x_1,x_2}\langle} \sigma (x_1)\sigma
	(x_2)\nonumber\\
	&-&J_{l_1}^{(6)}\sum _{\rangle x_1,x_2,\ldots,x_2\langle}
	\sigma (x_1)\sigma (x_2)\ldots \sigma (x_6),
\end{eqnarray}
where $J_1, J_p, J_{l_1}^{(6)}\in \mathbf{R}$ are  the coupling
constants. Similar to the previous models, we derived the
recurrence equations corresponding to the Hamiltonian given in
\eqref{Pentegonal-hm} as: 
\begin{eqnarray}
\label{2}
\nonumber x^{\prime}&=&\frac{1}{a^2D}\left\{\sum_{r=0}^5
C_{r}^{5}\left[ b^{5-2r}c^{(-1)^{r+1}}(x-y_2)^{5-r}(1- y_1)^r +
b^{-5+2r}c^{(-1)^r}(1+y_1)^{5-r}(x+y_2)^r \right]
\right\},\\\nonumber 
y^{\prime}_1&=&\frac{1}{D}\left\{\sum_{r=0}^5
C_{r}^{5} \left[ b^{5-2r}c^{(-1)^r}(1+y_1)^{5-r}(x+y_2)^r -
b^{-5+2r}c^{(-1)^{r+1}}(x-y_2)^{5-r}(1- y_1)^r \right]
\right\},\\\nonumber 
y^{\prime}_2&=&\frac{1}{a^2D}\left\{
\sum_{r=0}^5 C_{r}^{5} \left[
b^{5-2r}c^{(-1)^{r+1}}(x-y_2)^{5-r}(1- y_1)^r   -
b^{-5+2r}c^{(-1)^r}(1+y_1)^{5-r}(x+y_2)^r \right]
\right\},\nonumber
\end{eqnarray}
and {\small\begin{eqnarray*} D(x,y_1,y_2)&=& \sum_{r=0}^5
C_{r}^{5}\left[ b^{5-2r}c^{(-1)^r}(1+y_1)^{5-r}(x+y_2)^r+
b^{-5+2r}c^{(-1)^{r+1}}(x-y_2)^{5-r}(1- y_1)^r \right],
\end{eqnarray*}} 
where  $a=\exp(J_1/T), \quad b=\exp(J_p/T), c=\exp(J^{(k+1)}_{l_1}/T) $ and
$C_{r}^{5}=\left(\begin{array}{c}
  5 \\r
\end{array}
\right)$.
\begin{figure}[!htbp]\label{Avize-phase-D1}
	\centering
\rotatebox{270}{\scalebox{0.30}{\includegraphics{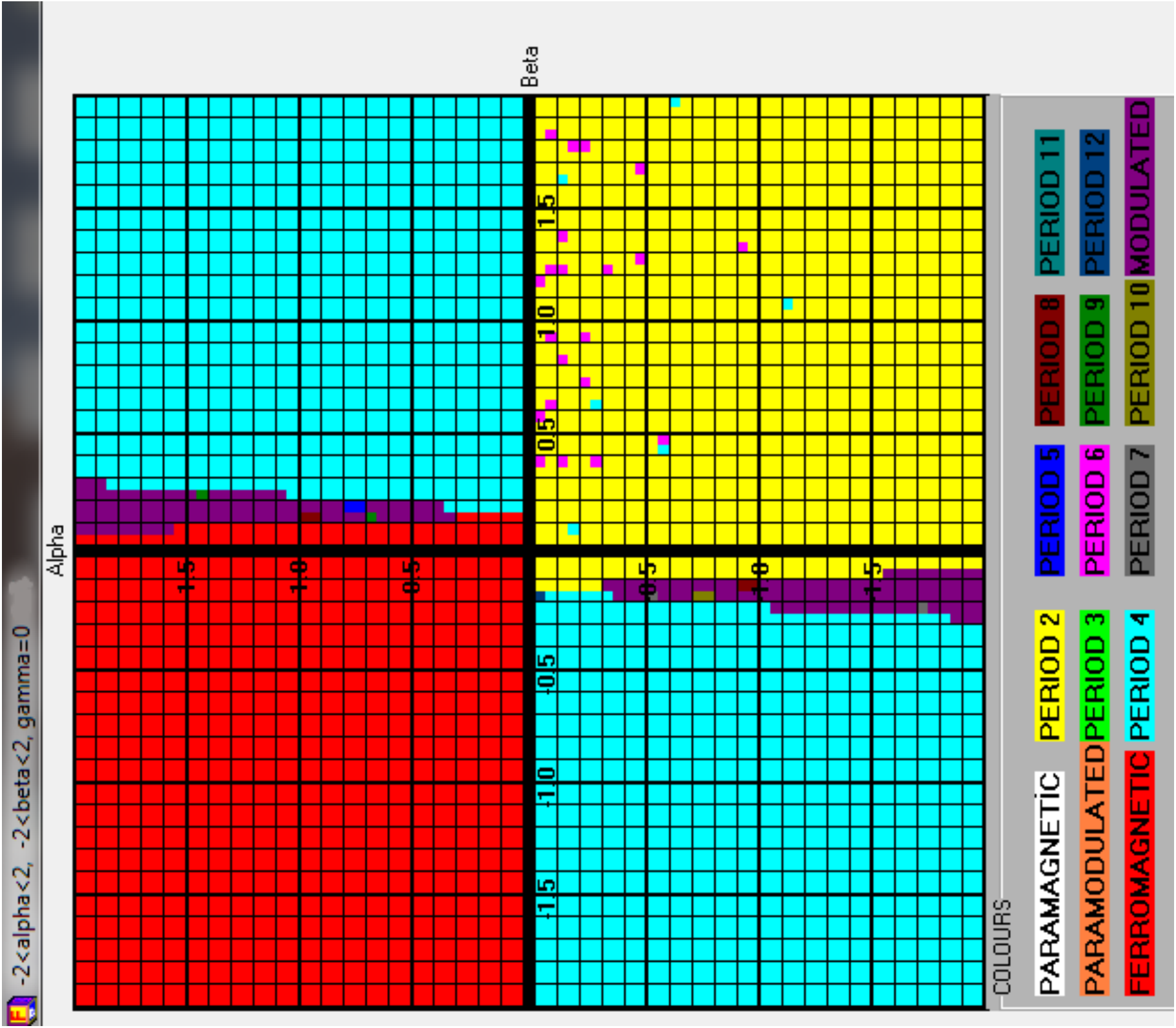}}}\
\ \ \ \ \
\rotatebox{270}{\scalebox{0.30}{\includegraphics{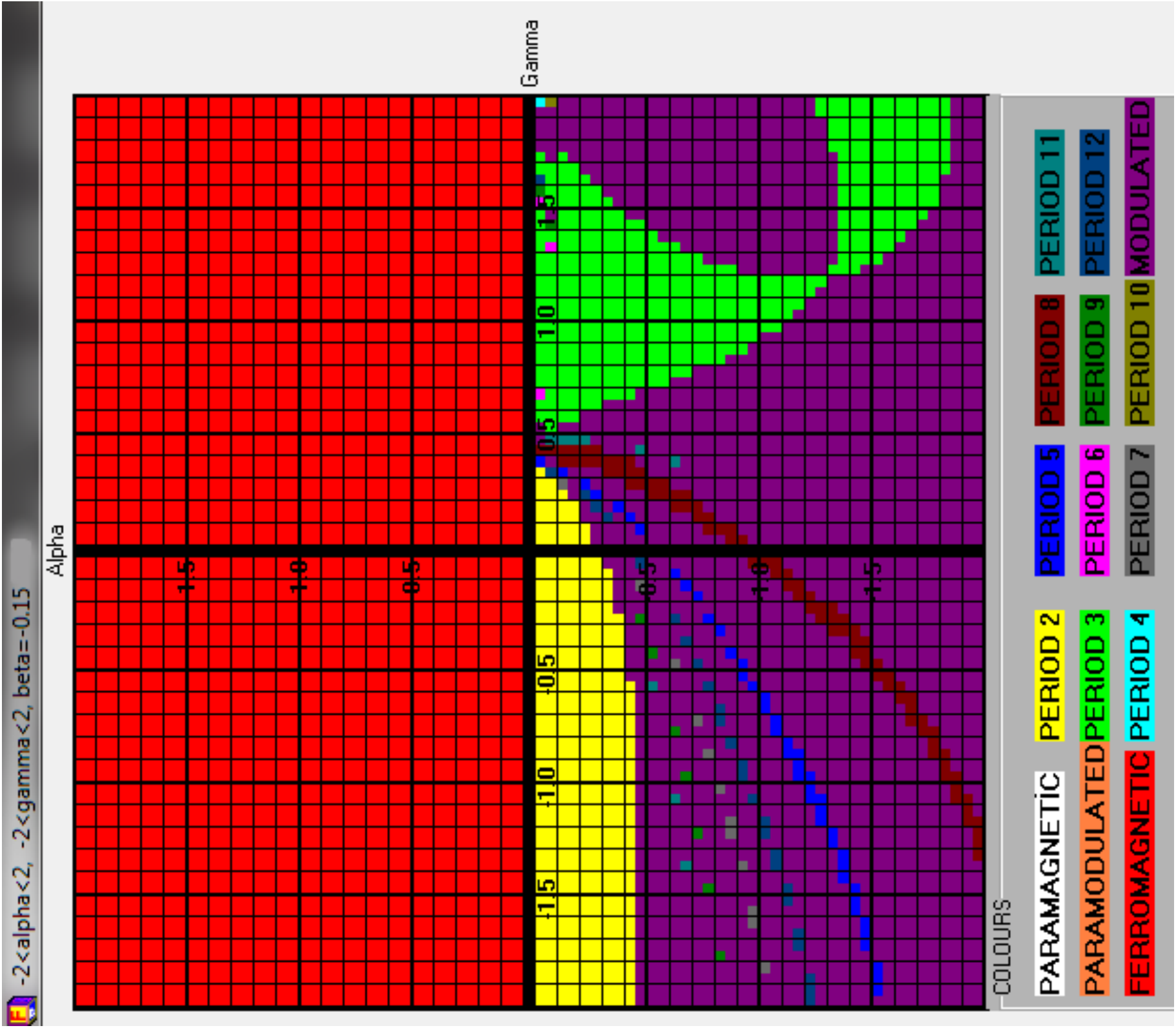}}}
\caption{(Colour online) The phase diagrams of the Ising model
associated to Hamiltonian \eqref{Pentegonal-hm} on the PCN for
$(\alpha,\beta)\in [-2,2]\times[-2,2]$, $\gamma=0$ (left-hand) and
$\beta=-0.15$ (right-hand).}
\label{Avize-phase-D1}
\end{figure}
\begin{figure}[!htbp]\label{Avize-Lyapunov2}
\centering
\includegraphics[width=75mm]{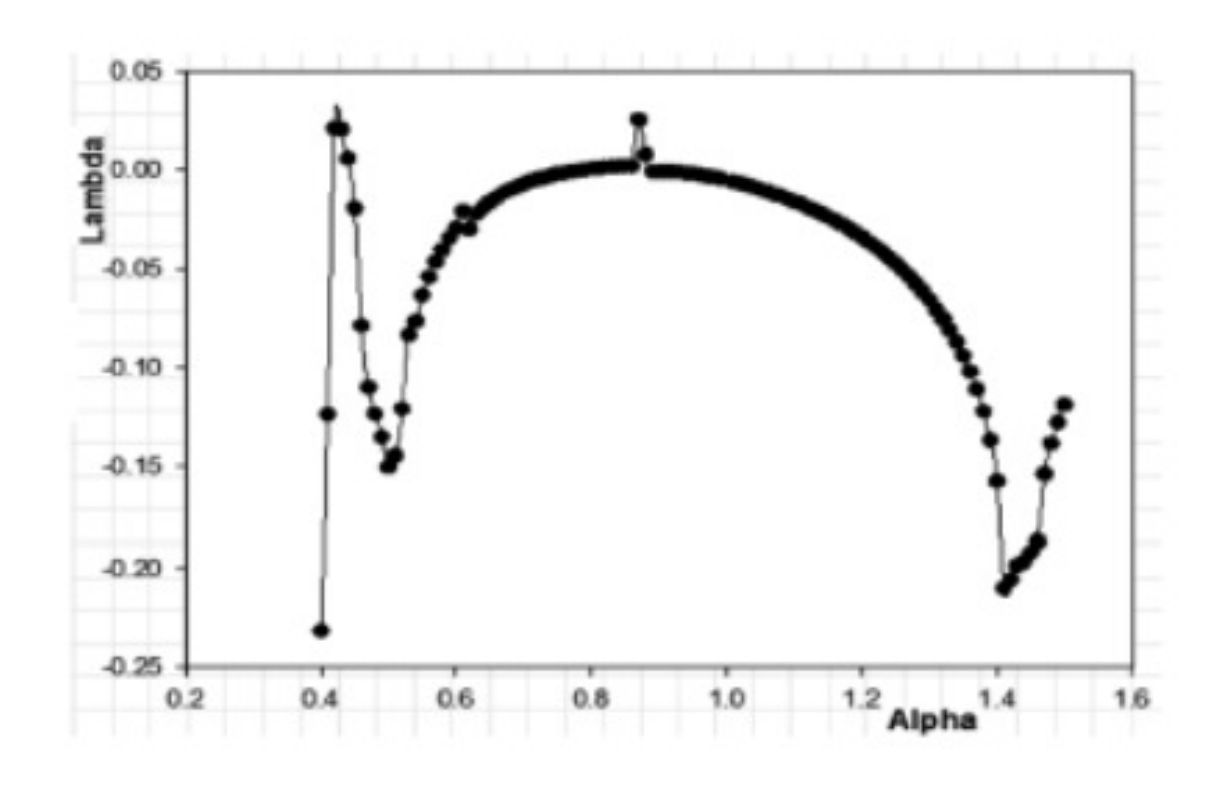}\ \ \ \ \
\includegraphics[width=60mm]{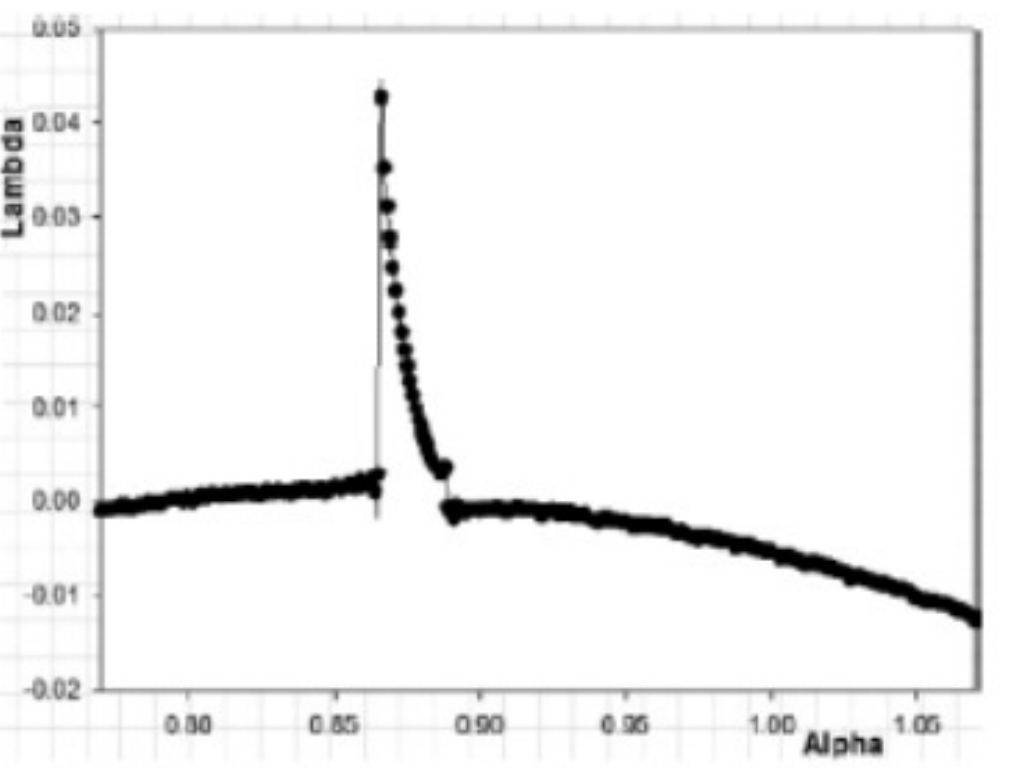}
\caption{Lyapunov exponents of the Ising model associated with
\eqref{Pentegonal-hm} for some values 
.}\label{Avize-Lyapunov2}
\end{figure}
For the Ising model corresponding to Hamiltonian in
\eqref{Pentegonal-hm}, by using the method given in the previous
sections, we can plot the phase diagrams. For example, for
$\gamma=0$ and $\alpha \in [-2,2], \beta\in [-2,2]$ we have
plotted the phase diagram of the Ising model associated with the
Hamiltonian \eqref{Pentegonal-hm} in figure \ref{Avize-phase-D1}.
In the diagram, \textbf{P2}, \textbf{P4}, \textbf{Ferromagnetic}
and \textbf{modulated} phases mainly appear. Phases in different
types of islands are also observed (see figure
\ref{Avize-phase-D1}). Figure \ref{Avize-phase-D1} (right-hand) shows a
phase diagram of the Ising model associated to the Hamiltonian
\eqref{Pentegonal-hm} for $(\alpha,\gamma)\in [-2,2]\times[-2,2]$
and $\beta=-0.15$. When figures given in \ref{Avize-phase-D1} are
compared with each other, it is seen that the phase diagrams
change completely.

As known, the negative Lyapunov exponent shows that an
infinitesimal perturbation in the beginning conditions has an
infinitesimal effect. The negative Lyapunov exponent $\lambda$
coincides with a particular cycle phase. The positive Lyapunov
exponent leads to a totally different trajectory
\cite{GR2014AIP-CL}. Therefore, we calculated the Lyapunov
exponents for certain critical parameter values to reveal the
details in the phases.

In this way, for $\beta =0.33$ value in different intervals of
$\alpha$, in figure \ref{Avize-Lyapunov2}, we plotted the Lyapunov
exponents of the Ising model associated with the Hamiltonian given
in \eqref{Pentegonal-hm}. The figure on the right is the enlarged
version of the Lyapunov exponent in the narrower range of the
values. As seen in the figures, the value of the Lyapunov exponent
is positive between 0.86 and 0.89. This implies that the model is
chaotic in the range $(0.86,0.89)$. We look at the relevant works
for details. Considering figure \ref{Avize-phase-D1}, in 
figure \ref{Avize-Lyapunov2}, we plot two variations of the
Lyapunov exponent $\lambda$, $\beta= 0.33$ with $\alpha\in [0.4,
1.5]$ (left) and  $\alpha\in [0.75, 1.1]$ (right-hand). Note that the
figures given in \ref{Avize-phase-D1} are borrowed from
\cite{UGAT2012ACTA}.
\subsection{Comments and remarks}
The results in this section are based on the papers
\cite{AUT2010AIP,UA2010PhysicaA,UGAT2012ACTA,UA2011CJP,GR2014AIP-CL}.
As we have seen in the relevant studies, the morphology of phase
diagrams associated with a given Hamiltonian on the chandelier
networks is completely different from the phase diagrams
corresponding to the Ising model on the Cayley tree, if different
coupling constants are added to the same Hamiltonian.

Many problems in statistical physics have lately been considered
on the Cayley tree \cite{Dang2015,Akin2016}, because other
operations and computations based on this lattice are considerably
simpler than the d-dimensional $\mathbf{Z}^{\text{d}}$ lattice. As a
result, the Cayley tree's results provide a source of inspiration
for the d-dimensional $\mathbf{Z}^{\text{d}}$ lattice. The Ising model
has relevance to physical, chemical, and biological systems
\cite{Georgi,Iof1996,Lebowitz1}. Compared with $\mathbf{Z}^{\text{d}}$
lattice, we think that the chandelier networks are more realistic
than the Cayley tree~\cite{AUT2010AIP,UA2010PhysicaA,UGAT2012ACTA,UA2011CJP,Moraal}.

The phase diagrams for the Ising model defined on a pentagonal
chandelier network with competing one-level pentagon interactions
were studied by Ganikhodjaev and Rodzhan \cite{GR2014AIP-CL}. They
studied the variation of wavevector with temperature in the modulated
phase to detect narrow commensurate steps between incommensurate
regions. They also investigated the Lyapunov exponent associated
with the trajectory of the system.

\subsubsection{Open problem in chandelier networks} 
Let us consider the following Hamiltonian
\begin{equation}\label{Hm-C3}
H(\sigma)=-J\sum\limits_{\langle x,y\rangle}{\sigma (x)\sigma
(y)}-{{J}_{p}}\sum\limits_{\rangle x,y\langle}{\sigma (x)\sigma
(y)}-{{J}_{SL}}\sum\limits_{\langle \widetilde{x,y}\rangle}{\sigma
(x)\sigma (y)},
\end{equation}
where the first sum encompasses all nearest neighbors, the second
sum encompasses all prolonged next-nearest-neighbors, and the
third sum encompasses all same-level nearest-neighbors, and
$J,J_{p},J_{\text{SL}}\in \mathbf{R}$ are coupling constants (see figure
\ref{ternary-chandiler}). If $J_{\text{SL}}=0$, the Hamiltonian
\eqref{Hm-C3} coincides with the Hamiltonian of Vannimenus
\cite{Vannimenus}.

In this case, unlike the symmetry of a Cayley tree of arbitrary
order \cite{UGAT2012IJMPC}, the $k$-order chandelier networks do
not have symmetry when $k>3$. As a result, to derive the recursive
equations associated with the Hamiltonian \eqref{Hm-C3} for $k>3$
is much more complicated.

\section{\textbf{Conclusions and recommendations}}\label{CONCLUSIONS-RECOMMENDATIONS}

I should emphasize that many of the figures given in this review
paper are borrowed from the related articles.

Vannimenus \cite{Vannimenus} and Mariz et al. \cite{MTA1985a}
investigated the phase diagrams of the Ising model associated with
given Hamiltonian on the Cayley tree of 2{th} order.

Considering Vannimenus's approach, we have determined the Lifshitz
points (critical points) in the phase diagrams for certain models.
The system of nonlinear equations corresponding to the Ising model
was analyzed. Then, the phase diagrams of a given model were plotted
and the phase diagram of Vannimenus's model was plotted in four
regions. The results and the suggestions associated with the Ising
and Potts models are expressed in the discussion parts of the
publications which we have accomplished, so we will not discuss
them here in detail. The literature concerning the phase diagrams
of the lattice models is enormous. We have tried to be as complete
as possible in the bibliography, so the reader is invited to study
further this interesting theory.

We have also cited some papers concerning Ising and Potts on the
Cayley tree, for which many of the results presented here hold as
well.

Let us summarize some of the results given in this review paper
now;
\begin{enumerate}
 \item The phase diagram of the Ising model associated with the Hamiltonian
\[
H(\sigma )=-{{J}_{p}}\sum\limits_{\langle x,y,z\rangle}{\sigma
(x)\sigma (y)\sigma (z)} -J\sum\limits_{\langle x,y\rangle}{\sigma
(x)\sigma (y)}
\]
was studied and the fixed points were searched to
obtain the phase transition curve. The relevant phase diagram was
published in the study \cite{GAUT2011b}.
    \item The limiting behaviors of the new Ising systems with three
competing interactions were studied. This phenomenon was
published in the study of the corresponding phase diagrams
\cite{GAUT2011Chaos}. These phase diagrams are much richer 
compared to the previous ones, and the Lifshitz points exhibited
different properties.
    \item The phase diagrams of the 3-state Potts model associated with the relevant Hamiltonian with
competing nearest-neighbor, prolonged next-nearest-neighbor and
two-level triple interactions were analyzed. The system of
equations corresponding to the $q$-state Potts model is
essentially more complicated than the similar basic equations of the
Ising model \cite{GAUT2011Chaos}.

\item Ganikhodjaev et al. \cite{GMP2008} investigated the
different $q$-state Potts models associated with given
Hamiltonians on the Cayley tree. In this case, different and
richer phase diagrams were obtained for some coupling constants
and temperatures. The computer codes obtained in the study
\cite{GMP2008} were further improved by using computer programming
languages such as $C^{++}$, MATLAB, MAPLE, DELPHI, VISUAL BASIC.
In this context, we  examined in detail the phase diagrams of
the $q$-state Potts models on the Cayley tree by comparing the
results obtained in the references
\cite{GTAU2011a,GTUA2011C,TGAU2010AIP}.

\item The problem given in 5 is generalized for the Potts model,
and both the phase diagrams and the linearization problem were studied. However, the Potts model  encountered a longer
and more complicated situation than in the investigation of these
problems. The analysis of this model was examined for $k=3$.

\item We have talked about the emergence of chandelier networks.
We have presented the results obtained about the lattice  models
described on chandelier networks so far.
\end{enumerate}

\section{Open problems}\label{Openproblems}

\begin{itemize}
\item When the prolonged ternary next nearest-neighbor interaction
is added to the Hamiltonian, although paramagnetic phase regions
for the Ising system disappear, the paramagnetic phase regions
appear for the Potts model associated to similar Hamiltonian. This
cannot be explained at the moment.

\item 
On a Cayley tree of order three, Ak\i n \cite{Akin2017-arXiv} has
recently analytically  analysed the recurrence equations of an Ising model with
three competing interactions. He  precisely
described the Ising model's paramagnetic and ferromagnetic phases.
To analytically derive the phases associated with similar models on
higher order Cayley trees is a difficult task.

\item For both Ising and Potts models, we have obtained multiple
critical points in non-zero temperatures, contrary to the works of
Vannimenus \cite{Vannimenus} and Mariz et al. \cite{MTA1985a}. In
the phase diagrams drawn for a model, we could perform the
linearization process around the multi-critical point in the
presence of the paramagnetic phase. If there is no \textbf{P}
phase for the model, then different methods are needed to examine
the stability of the system. 

\item It is possible to study the strange attractors of the given
systems  more in detail. In this context, we studied the strange
attractor of an Ising system on the $k$-branch Cayley tree
\cite{GATU2013JPhysConfSer}.

\item Ganikhodjaev \cite{G2011Chaos}  studied the strange attractor
in the Potts model on a Cayley tree in the presence of competing
interactions. We  tried to draw ``Strange attractors'' of Potts
models for the 3rd and 4th order Cayley tree
generalization of this work \cite{G2011Chaos}. Relevant equations
were obtained. In order to draw the graphs of these recursive
equations, we need computer code studies.

\item On the Cayley tree of arbitrary order, we continue to
investigate Potts models with various coupling constants. The
system of nonlinear equations corresponding to the Ising models
was obtained as a consequence of a long calculation in the
previous chapter, and new results for the Potts model associated
with the Hamiltonian are expected to be investigated in the near
future.
\item 
We considered the Potts models with arbitrary spin valence ($q>3$)
and came up with some interesting results. We are first referred
to as phase diagrams of 4-state Potts model with mutual
interaction in this manner (see \cite{Gok-Tez}).

\item As already mentioned, the models we have studied in this
paper indicate the existence of a chaotic structure because of the
diversity of the wave vectors of the phase diagrams obtained and
the positive values of the Lyapunov exponents
\cite{UGAT2012IJMPC,GATU2013JPhysConfSer}. We think that these
calculations could be a source of inspiration for investigating
more realistic models studied in physics, chemistry and other
basic sciences as a result of adjusting these calculations to be
more accurate with Monte Carlo and other computational techniques.

\item Silva {et al}. \cite{Silca-C} investigated the Ising
model in the presence of an external magnetic field on a Cayley
tree of any order with competitive interactions between the
first-, second-, and third-nearest-neighbor interactions spins
adhering to the same branch. Ganikhodjaev and Rodzhan
\cite{GR2014AIP-CL} investigated phase diagrams of the Ising model
on the Cayley tree with competing interactions up to the
third-nearest-neighbor generation for critical variables in their
paper \cite{GanMohd2016}. There are numerous issues with this
model. In the future, we will pursue these unsolved issues.

\item Note that investigation of the stability of the Potts model
is much more difficult than the stability of Ising model (see
\cite{GTAU2011a}). Moreover, we could not examine the quantity of
Lyapunov exponent and the strange attractors associated to the
Potts models in our works. We are going to investigate these
concepts in the future works.

\item I should note that, as far as I know, the phase diagrams of
the $q$-state Potts models on $k$-order chandelier networks have
not yet been studied. In \cite{Akin-Chang-2020}, we investigated the
reversibility of a family of linear cellular automata (LCAs) on
Cayley tree of order $k$ over the field $\mathbb{F}_p$ and
computed the measure theoretic entropy of the family of these
LCAs. We will examine similar problems such as entropy and ergodic
properties of these LCAs on the chandelier network. In
\cite{A-CMP-2019}, we investigated the Gibbs measures of an Ising
model with competing interactions on the TCL. Gibbs measures on the
other chandelier networks have not yet been explored.
\end{itemize}

\section*{Acknowledgements}

I dedicate this work to Prof. Dr. Nasir Ganikhodjaev from whom I
learned a lot of new topics. 
The author thanks ICTP for providing financial support and all
facilities. The author is supported by the Simons Foundation and
IIE.
%-SRF.

 \vspace{10mm}
{\it  Note added in proof}
 \vspace{2mm}

The Ising model's phase transition phenomena on Cayley trees have
recently been studied in [{\it Ak\i n~H., Physica B, 2022, 645, 414221}]. In [{\it Ak\i n~H., Mukhamedov F., J. Stat. Mech., 2022, 053204}], using the Kolmogorov consistency
theorem, we established the translation invariant splitting Gibbs measures
(TISGMs) connected to the Ising model with mixed spin (1,1/2) on the
second-order Cayley tree. We showed that the mixed spin Ising model
comprises three TISGMs in both the ferromagnetic and anti-ferromagnetic
regions, in contrast to the classic Ising model [\cite{Akin2017-arXiv}, {\it Ak\i n, H., Physica B, 2022, 645, 414221}]. In this context,
considering the theoretical results, detailed analysis of the transitions
between phases will be carried out by drawing phase diagrams.

\newpage
\ukrainianpart

\title{Фазові діаграми ґраткових моделей на дереві Кейлі та
	люстроподібні мережі: огляд
}
\author[Х. Акин]{Х. Акин\refaddr{label1}\refaddr{label2}}
\addresses{
	\addr{label1}Міжнародний центр теоретичної фізики ім. Абдуса Салама
	(ICTP), Strada Costiera, 11, I - 34151 Тріест, Італія 
\addr{label2}	Кафедра математики, факультет науки та мистецтв Харранського університету, 63290, Шанлиурфа, Туреччина}

\makeukrtitle

\begin{abstract}
	Основна мета цієї оглядової статті --- систематичний виклад усіх відомих результатів стосовно фазових діаграм ґраткових моделей (Ізинга та Поттса) на дереві Кейлі (або гратці Бете) і
	люстроподібних мережах. Здійснено детальний огляд різноманітних застосувань ґраткових моделей. З використанням підходу Ваніменуса, представлено і проаналізовано рекурсивні рівняння для моделей Ізинга і Поттса, пов'язаних з конкретним гамільтоніаном на дереві Кейлі. Наведено відповідні фазові діаграми, а також алгоритми для їх обчислення на різних мовах програмування. Для виявлення фазових переходів у модульованій фазі детально досліджується залежність хвильового вектора $q$ від температури і показник Ляпунова, пов'язаний з траєкторією конкретної рекурсивної системи. Визначено перехід між впорядкованою ($C$) і невпорядкованою  ($I$) фазами, використовуючи показники Ляпунова, хвильовий вектор і дивний атрактор для ґрунтовного порівняння. Здійснено огляд динамічної поведінки моделі Ізинга на люстроподібних мережах. Досліджено фазові діаграми моделі Ізинга, що відповідає даному гамільтоніану на ґратках нового виду типу дерева Кейлі, таких як
	{\it трикутні, чотирикутні, п'ятикутні люстроподібні мережі
	}. Окрім того, обговорюється багато задач, які ще очікують на свої розв'язки.
	\keywords дерево Кейлі, люстроподібна мережа, модель Ізинга,
		модель Поттса, хвильові вектори, показник Ляпунова, дивні атрактори, фазовий перехід
\end{abstract}

%\listoffigures
\lastpage
\end{document}